\documentclass[]{aastex631}
\shorttitle{Molecular Gas in the Mon OB1 Region}
\shortauthors{ZHUANG et al.}
\graphicspath{{./}{figures/}}

\begin{document}

\title{Distribution and Properties of Molecular Gas toward the Monoceros OB1 Region}

\author[0000-0002-7413-7574]{Zi Zhuang}
\affiliation{Purple Mountain Observatory and Key Laboratory of
Radio Astronomy, Chinese Academy of Sciences, Nanjing 210034, People's Republic of China}
\affiliation{School of Astronomy and Space Science, University of Science and Technology of China, 96 Jinzhai Road, Hefei 230026, People's Republic of China}

\author[0000-0002-0197-470X]{Yang Su*}
\affiliation{Purple Mountain Observatory and Key Laboratory of
Radio Astronomy, Chinese Academy of Sciences, Nanjing 210034, People's Republic of China}
\affiliation{School of Astronomy and Space Science, University of Science and Technology of China, 96 Jinzhai Road, Hefei 230026, People's Republic of China}

\author[0009-0002-2379-4395]{Shiyu Zhang}
\affiliation{Purple Mountain Observatory and Key Laboratory of
Radio Astronomy, Chinese Academy of Sciences, Nanjing 210034, People's Republic of China}
\affiliation{School of Astronomy and Space Science, University of Science and Technology of China, 96 Jinzhai Road, Hefei 230026, People's Republic of China}

\author[0000-0003-3151-8964]{Xuepeng Chen}
\affiliation{Purple Mountain Observatory and Key Laboratory of
Radio Astronomy, Chinese Academy of Sciences, Nanjing 210034, People's Republic of China}
\affiliation{School of Astronomy and Space Science, University of Science and Technology of China, 96 Jinzhai Road, Hefei 230026, People's Republic of China}

\author[0000-0003-4586-7751]{Qing-Zeng Yan}
\affiliation{Purple Mountain Observatory and Key Laboratory of
Radio Astronomy, Chinese Academy of Sciences, Nanjing 210034, People's Republic of China}

\author[0000-0003-1714-0600]{Haoran Feng}
\affiliation{Purple Mountain Observatory and Key Laboratory of
Radio Astronomy, Chinese Academy of Sciences, Nanjing 210034, People's Republic of China}
\affiliation{School of Astronomy and Space Science, University of Science and Technology of China, 96 Jinzhai Road, Hefei 230026, People's Republic of China}

\author[0000-0003-2732-0592]{Li Sun}
\affiliation{Purple Mountain Observatory and Key Laboratory of
Radio Astronomy, Chinese Academy of Sciences, Nanjing 210034, People's Republic of China}
\affiliation{School of Astronomy and Space Science, University of Science and Technology of China, 96 Jinzhai Road, Hefei 230026, People's Republic of China}

\author[0009-0001-3487-1870]{Xiaoyun Xu}
\affiliation{Center for Astrophysics, Guangzhou University, Guangzhou 510006, People's Republic of China}

\author[0000-0002-3904-1622]{Yan Sun}
\affiliation{Purple Mountain Observatory and Key Laboratory of
Radio Astronomy, Chinese Academy of Sciences, Nanjing 210034, People's Republic of China}
\affiliation{School of Astronomy and Space Science, University of Science and Technology of China, 96 Jinzhai Road, Hefei 230026, People's Republic of China}

\author[0000-0003-2418-3350]{Xin Zhou}
\affiliation{Purple Mountain Observatory and Key Laboratory of
Radio Astronomy, Chinese Academy of Sciences, Nanjing 210034, People's Republic of China}

\author[0000-0003-0746-7968]{Hongchi Wang}
\affiliation{Purple Mountain Observatory and Key Laboratory of
Radio Astronomy, Chinese Academy of Sciences, Nanjing 210034, People's Republic of China}
\affiliation{School of Astronomy and Space Science, University of Science and Technology of China, 96 Jinzhai Road, Hefei 230026, People's Republic of China}

\author[0000-0001-7768-7320]{Ji Yang}
\affiliation{Purple Mountain Observatory and Key Laboratory of
Radio Astronomy, Chinese Academy of Sciences, Nanjing 210034, People's Republic of China}

\correspondingauthor{yangsu@pmo.ac.cn}



\begin{abstract}

We perform a comprehensive CO study toward the Monoceros OB1 (Mon OB1) region based on the Milky Way Imaging Scroll Painting survey at an angular resolution of about $50''$.  
The high-sensitivity data, together with the high dynamic range, show that molecular gas in the $\rm 8^{\circ}\times4^{\circ}$ region displays complicated hierarchical structures and various morphology (e.g., filamentary, cavity-like, shell-like, and other irregular structures).
Based on Gaussian decomposition and clustering for $\mathrm{^{13}CO}$ data, a total of 263 $\mathrm{^{13}CO}$ structures are identified in the whole region, and 88\% of raw data flux is recovered.
The dense gas with relatively high column density from the integrated CO emission is mainly concentrated in the region where multiple $\rm ^{13}CO$ structures are overlapped.
Combining the results of 32 large $\mathrm{^{13}CO}$ structures with distances from Gaia DR3, we estimate an average distance of $\rm 729^{+45}_{-45}~pc$ for the giant molecular cloud (GMC) complex.
The total mass of the GMC complex traced by $\mathrm{^{12}CO}$, $\mathrm{^{13}CO}$, and $\mathrm{C^{18}O}$ is $1.1\times10^5~M_\odot$, $4.3\times10^4~M_\odot$, and $8.4\times10^3~M_\odot$, respectively.
The dense gas fraction shows a clear difference between Mon OB1 GMC East (12.4\%) and Mon OB1 GMC West (3.3\%).
Our results show that the dense gas environment is closely linked to the nearby star-forming regions. On the other hand, star-forming activities have a great influence on the physical properties of the surrounding molecular gas (larger velocity dispersion, higher temperatures, more complex velocity structures, etc.). We also discuss the distribution/kinematics of molecular gas associated with nearby star-forming activities.

\end{abstract}



\section{Introduction} \label{sec:intro}

Molecular clouds (MCs) are the birthplace of stars.
Research on MCs helps us gain useful information on their distribution and properties, the relation between MCs and the star-forming process, and the feedback of star activities on the surrounding molecular gas.
The most abundant molecule in MCs, hydrogen, is hard to observe because it radiates ineffectively in cold and dense interstellar environments. The next most abundant molecule, CO, has served as the most important tracer of MCs since its detection in the 1970s \citep{wilson1970carbon}. The emission of low-$J$ transition (i.e., $J$ = 1$\rightarrow$0) of $\rm^{12}CO$ is strong and extended; therefore, it is suitable for studying the distribution and properties of MCs. 
$\rm^{13}CO$ and $\rm C^{18}O$ lines usually trace high-density regions owing to their relatively low optical depth.
In the past decades, large-scale CO surveys have provided us with abundant information about the morphological and kinematic properties of MCs in the Milky Way (e.g., \citealt{dame2001milky, burton2013mopra, su2019milky}). 

Mon OB1 giant molecular cloud (GMC) is likely located in the Local arm toward the Galactic anticenter. Toward this direction, the Mon OB1 GMC suffers from less contamination from foreground and background emission, making it a good target for studying properties of molecular gas and the connection between star formation and MCs. The distance of the GMC complex is at about 720 -- 790 pc (e.g., \citealt{yan2019molecular, chen2020large, 2020A&A...633A..51Z}). \citet{oliver1996new} performed a detailed CO study toward the Mon OB1 region based on the CfA 1.2 m telescope. 
The distribution of column density and the abundance ratio of molecular gas ($\rm C^{18}O / ^{13}CO$) in the region were also investigated recently (\citealt{ma2022gas, wang2023molecular}). 

Mon OB1 GMC East (hereafter the east cloud) harbors the famous young open cluster NGC 2264  \citep{2008hsf1.book..966D}. The cluster contains more than 1000 members and several substructures, for example, the Cone Nebula, the Fox Fur Nebula, and two star-forming clusters, NGC 2264C and NGC 2264D. Among the stars, the massive O7 V star S Mon and several B-type stars in NGC 2264 form the Mon OB1 association \citep{de1999hipparcos}. NGC 2264C and NGC 2264D are associated with two famous infrared sources, IRS 1 \citep{allen1972infrared} and IRS 2 \citep{margulis1989young}, respectively. 
The magnetic field was studied for the east cloud as well as the filamentary structures toward NGC 2264C and NGC 2264D (\citealt{alina2022large, wang2024filamentary}).
The average age of NGC 2264 is estimated as $\sim$ 3 Myr with an age dispersion of at least 5 Myr (\citealt{sung2004initial, 2008hsf1.book..966D, wright2023gaia}). Recent work found that some clusters in NGC 2264 are young with an age of $\sim$ 1 -- 2 Myr (\citealt{venuti2019deep}). 
NGC 2264 is an active star-forming region based on the detection of many outflows, Herbig-Haro (HH) objects, and shock $\rm H_2$ jets (\citealt{margulis1988molecular, schreyer1997molecular, wolf2003star, reipurth2004deep, peretto2006probing, buckle2012structure}), including  the famous NGC 2264G \citep{lada1996structure}. 
The distribution of many young stellar object candidates is tightly associated with the surrounding dense gas in NGC 2264 (\citealt{rapson2014spitzer, zhang2023distances}). In addition, some studies also have been done to investigate the kinematic connection between stellar populations/activities and associated molecular gas (e.g., \citealt{tobin2015kinematic, montillaud2019multib,flaccomio2023spatial}).
The molecular gas toward NGC 2264 was proposed to undergo a global collapse and then drive the dense parts to concentrated ridges\citep{2021A&A...645A..94N}. 

The Mon R1 loop ($l\sim201\fdg6$, $b\sim0\fdg2$) is located about $2^{\circ}$ away from the center of the Mon OB1 association and is visually connected to Mon OB1 GMC West (\citealt{2008hsf1.book..966D}; hereafter the west cloud). The Mon R1 loop contains several reflection nebulae, such as IC 446, NGC 2245 and NGC 2247. 
Mon R1 loop exhibits a semiring structure (\citealt{kutner1979ring, bhadari2020star}) at a distance of $\sim$ 660 -- 715 pc (\citealt{movsessian2021new, lim2022gaia}).  
Outflows and $\rm H_2$ jets are also identified to be associated with Mon R1 loop (\citealt{movsessian2021new, magakian2022near}), indicating the ongoing star-forming activities therein. 

Many studies have been conducted toward the Mon OB1 region in multiwavelength. Based on the Milky Way Imaging Scroll Painting (MWISP) survey in CO and its isotopic transitions (see Section \ref{subsec:MWISP}), we aim to investigate the large-scale distribution and detailed physical properties of molecular gas toward the whole Mon OB1 region. The large-scale CO data with high resolution and sensitivity are important to reveal the global picture of the Mon OB1 GMC complex.
The paper is structured as follows. The observational data set used in this study is presented in Section \ref{sec: observation}. In Section \ref{sec:result}, we first analyze the basic distribution and physical properties of molecular gas in the whole Mon OB1 region. We then identify $\rm ^{13}CO$ structures from the MWISP data and determine the distances of large $\rm ^{13}CO$ structures based on the Gaia DR3 data. We further study the 3-D distribution and properties of the identified $\rm ^{13}CO$ structures. In Section \ref{sec:discuss}, we discuss the connection between the molecular gas and star-forming activities near NGC 2264.
Finally, we summarize the main results in Section \ref{sec:summary}.

\section{Observation and Data Reduction} \label{sec: observation}
\subsection{MWISP CO Data}\label{subsec:MWISP}
The observation is a part of the MWISP project (see \citealt{su2019milky}) using the PMO-13.7 m telescope at Delingha in China.
The half-power beamwidth (HPBW) of the telescope is $\sim50''$ at 115 GHz, while the pointing accuracy is $\sim5''$.
The nine-beam Superconducting Spectroscopic Array Receiver (\citealt{shan2012development}) is employed to observe the $\rm ^{12}CO$, $\rm ^{13}CO$, and $\rm C^{18}O$ ($J=1\rightarrow 0)$ lines simultaneously under the on-the-fly (OTF) mode. 
A total of 18 fast Fourier transform spectrometers are used as the back end of the receiver. The spectrometers have a bandwidth of 1 GHz and 16,384 channels, yielding channel separations of $\sim\rm0.16~km~s^{-1}$ for the $\rm ^{12}CO$ line ($\sim \rm 115~GHz$) and $\sim\rm0.17~km~s^{-1}$ for $\rm ^{13}CO$ and $\rm C^{18}O$ lines ($\sim \rm 110~GHz$).

The entire data region of $\rm \sim91~deg^2$ (i.e., $l=[195^{\circ},208^{\circ}],~b=[-2^{\circ},-5^{\circ}],~v=[-20, 50]~\mathrm{km~s^{-1}}$) is divided into 364 cells of $30'\times30'$; each is made with a grid spacing of $30''$.
A first-order baseline is applied for all three CO lines' spectra. After removing the bad channels and abnormal spectra, the reduced data cubes have typical rms noise of $\rm 0.45~K$ for $\rm ^{12}CO$ and $\rm0.25~K$ for $\rm ^{13}CO$ and $\rm C^{18}O$, respectively. The data has been released on ScienceDB (DOI: https://doi.org/10.57760/sciencedb.17451).

\subsection{Gaia DR3}
We use the astrophysical data and extinction data from Gaia DR3. Gaia DR3 was released in 2022 \citep{2023A&A...674A...1G}, which includes data on about 1.8 billion objects. The data products of Gaia include the astrophysical parameter catalogs produced by the General Stellar Parameterizer from Photometry (GSP-Phot; \citealt{2023A&A...674A..27A}) module based on Gaia astrometry, photometry, and low-resolution BP/RP spectra. GSP-Phot contains parallax and extinction of 471 million sources with apparent magnitude $G$ $<$ 19 mag. 

We select Gaia DR3 data with $A_G>0$, and the ratio of parallax errors ($\Delta\varpi$) to parallaxes ($\varpi$) is less than 20\%. The distances ($D$) become unreliable when $\Delta\varpi/\varpi\geqslant20\%$ \citep{bailer2015est}. In the Mon OB1 region that we studied, 187,204 stars are included. On the other hand, $A_G$ errors for each star are not a direct measurement but are derived from $A_G$. The process may lead to a small unreliable $A_G$ error for some star samples (\citealt{yan2019molecular, zhang2024multilayer}). Therefore, we set a cutoff of 0.01 mag for $A_G$ errors. After discarding those stars with very small $A_G$ error, 160,905 stars are kept for the following study.


\section{Result} \label{sec:result}
\subsection{Molecular Gas toward the Whole Region} \label{subsec: distribution}
\begin{figure}[ht!]
  \centering
  \includegraphics[scale=0.9]{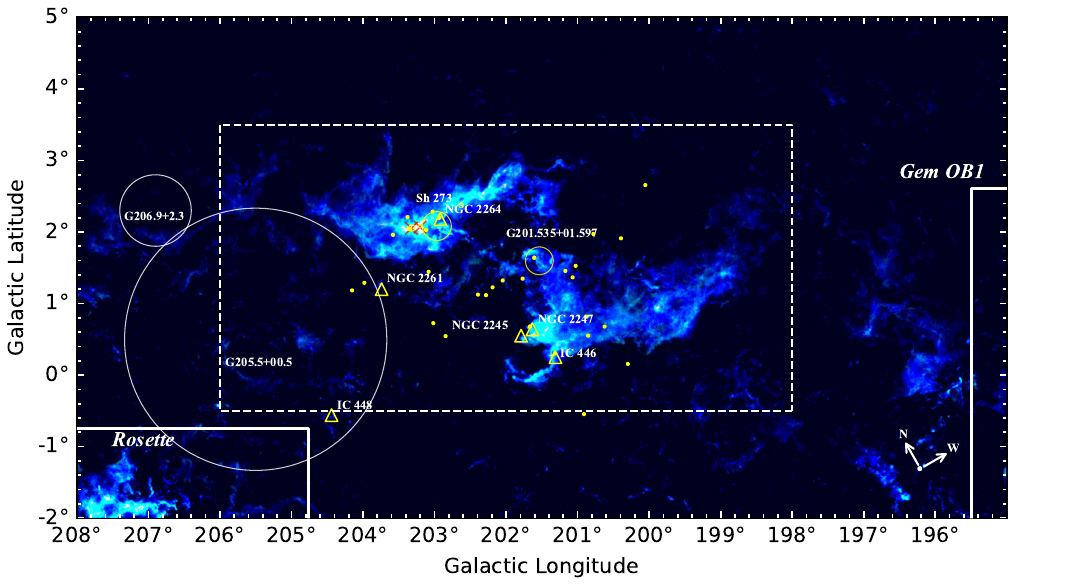}
  \caption{Integrated intensity map of $\rm ^{12}CO$ (blue), $\rm ^{13}CO$ (green), and $\rm C^{18}O$ (red) toward the whole data region.
  The $l,b,v$ ranges are $l$=[$195^{\circ},208^{\circ}$], $b$=[$-2^{\circ},5^{\circ}$], $v$ = $\rm[-20, 50]~km~s^{-1}$. Regions framed out by white solid lines are the two surrounding GMCs, Rosette and Gem OB1. The white circles symbolize SNRs \citep{green2019revised}. The red crosses indicate the two famous infrared sources, IRS 1 \citep{allen1972infrared} and IRS 2 \citep{margulis1989young}. The yellow circles indicate HII region candidates \citep{anderson2014wise}. Bright nebulae \citep{lynds1965catalogue} are represented by yellow triangles, and yellow dots indicates OB stars (spectral type earlier than B2, \citealt{xu2021local}). The region surrounded by white dashed rectangle is where we mainly focus on in the paper.
  \label{fig:1}}
  \end{figure}

Figure \ref{fig:1} shows the integrated intensity map of three CO isotopologues' emission. The Mon OB1 GMC complex is framed out by white dashed rectangle in the map.
The Gem OB1 GMC complex (see details in \citealt{wang2017molecular}, \citeyear{wang2019molecular}; \citealt{2018ApJS..235...15L}) is situated in the southwest part.
The Rosette GMC in the southeast part has been studied in detail by \citet{li2018large}. Two supernova remnants (SNRs) in the east of the GMC were investigated by \citet{su2017molecular}.

\begin{figure}[ht!]
  \centering
  \includegraphics[scale=0.6]{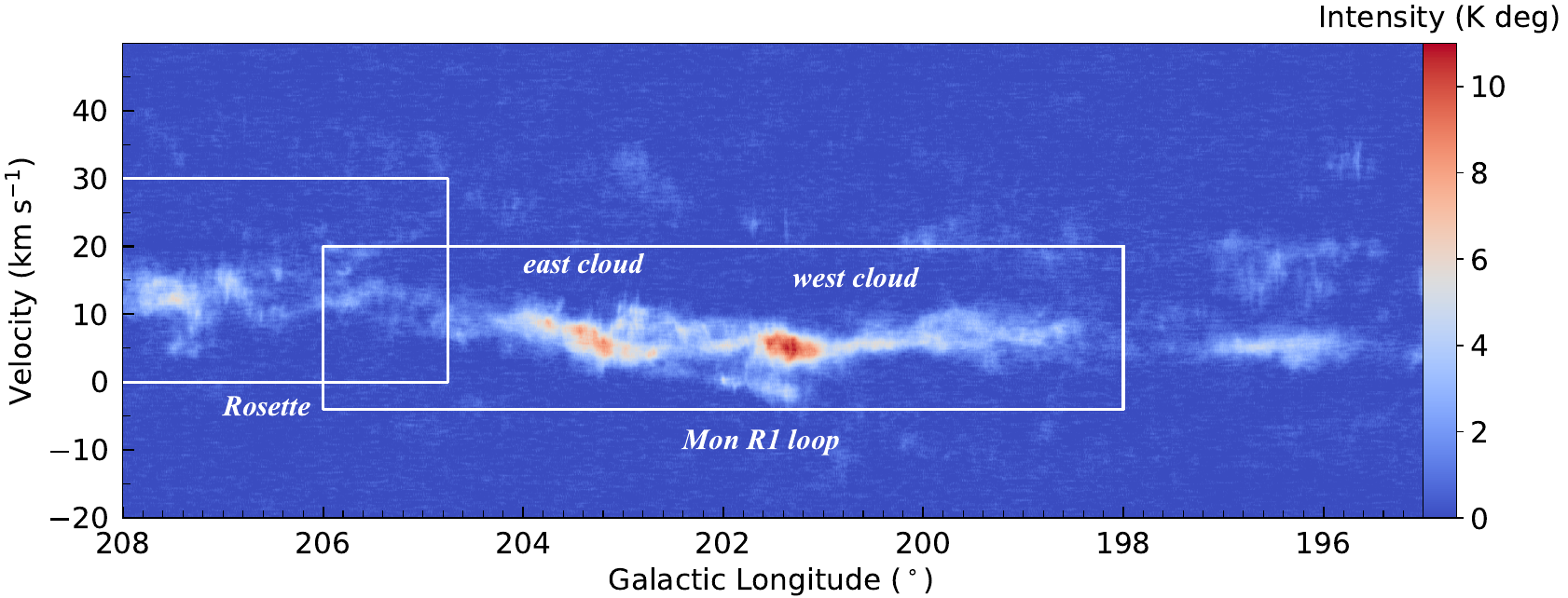}
  \caption{Longitude-velocity map of $\rm ^{12}CO$ emission toward the Mon OB1 region. The rectangles indicate the velocity range of molecular gas in the Mon OB1 GMC and the Rosette GMC. The east cloud, the west cloud and the Mon R1 loop are also labeled in the map.
  \label{fig:2}}
  \end{figure}

A longitude-velocity ($l-v$) map of $\rm^{12}CO$ emission is shown in Figure \ref{fig:2}. The emission of MCs in the Mon OB1 region is mostly confined to $\rm[-4,20]~km~s^{-1}$, with Mon OB1 in $\rm[2,20]~km~s^{-1}$ and Mon R1 loop in $\rm[-4,2]~km~s^{-1}$. 
Generally, the velocity distribution along the longitude is complicated. Some large-scale velocity structures can be seen toward the east and west clouds. For example, multiple velocity components are found toward NGC 2264 (see details in Section \ref{subsec:cloud-identification}).
In addition, a parallelogram-like structure ($l\sim[201\fdg0,~203\fdg2]$, $v\sim[-5, 12]~\mathrm{km~s^{-1}}$) is discerned in the $l-v$ map between the east and west clouds. Many high-velocity features (HVFs) are widely distributed in the region, indicating the local outflows associated with accreting protostars (\citealt{margulis1988molecular, wolf2003star, movsessian2021new}).

In the map, some weak CO emission with a negative velocity of $\sim\rm[-15,-5]~km~s^{-1}$ is from local molecular gas.
In addition, CO emission in the velocity range of $\sim\rm[20,35]~km~s^{-1}$ is very likely from the Perseus arm based on distance measurements of MCs (see Section \ref{subsubsec:distances}). 
We will not discuss the CO gas outside $\rm[-5,20]~km~s^{-1}$ unless otherwise specified.

\begin{figure}[ht!]
  \centering
  \includegraphics[scale=1.3]{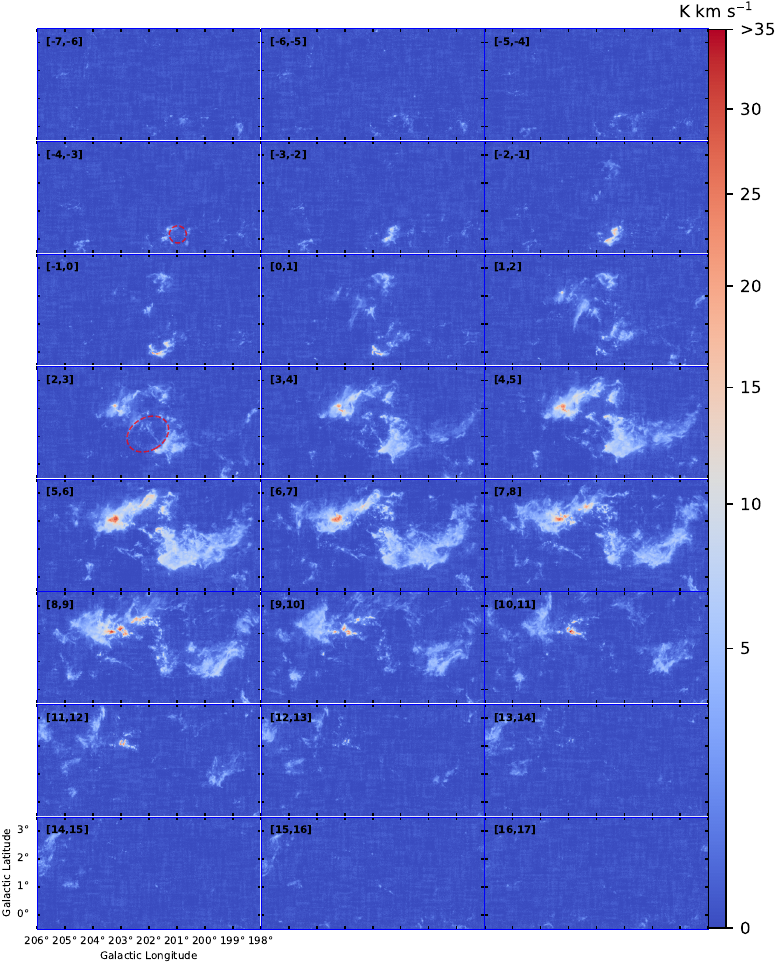}
  \caption{Velocity channel maps of $\rm ^{12}CO$ emission in the Mon OB1 region. The $l,b$ range is $l=[198^\circ,206^\circ]$, $b=[-0\fdg5, 3\fdg5]$, which is consistent with the area framed by the white dashed rectangle in Figure \ref{fig:1}. The velocity range is between $\rm -7~km~s^{-1}$ and $\rm 17~km~s^{-1}$. Each panel was integrated over $\rm 1~km~s^{-1}$. The corresponding velocity range is marked in the upper left corner of each panel.
  \label{fig:3}}
\end{figure}

\begin{figure}[ht!]
  \centering
  \includegraphics[scale=1.3]{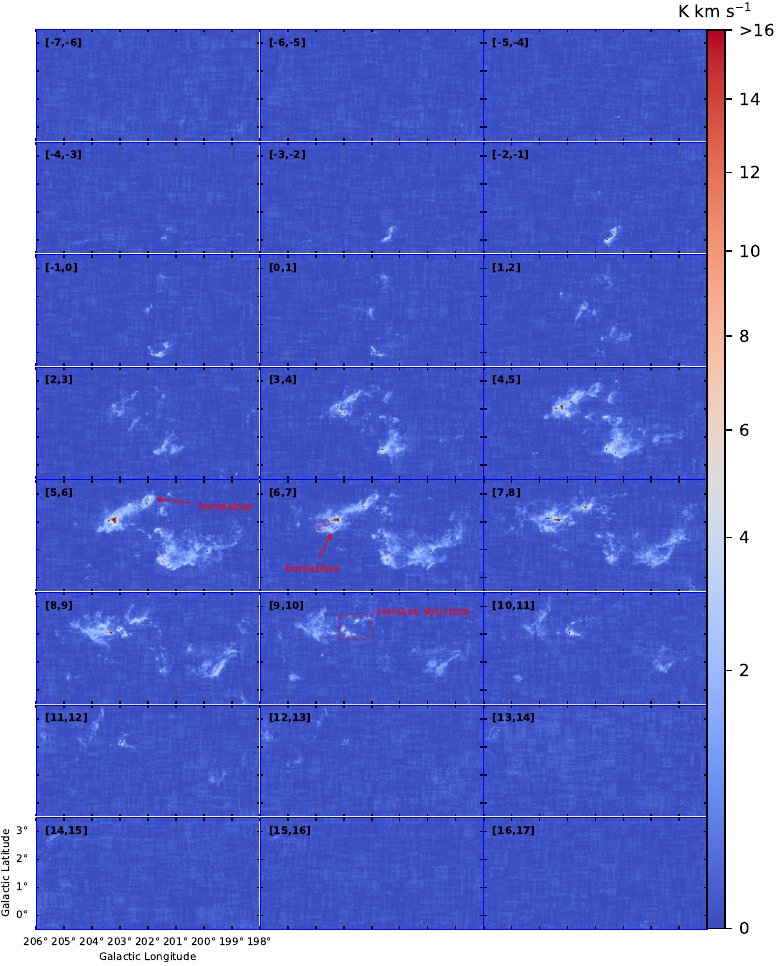}
  \caption{Velocity channel maps of $\rm ^{13}CO$ and $\rm C^{18}O$ (black contours) emission.
  The contours of $\rm C^{18}O$ emission are [0.75, 1.5, 2.25] $\rm K~km~s^{-1}$ (i.e., $5\times, 10\times, \mathrm{and} ~15\times$ of background noise levels).
  \label{fig:4}}
\end{figure}

The velocity channel maps of $\rm ^{12}CO$ and  $\rm ^{13}CO$ ($\rm C^{18}O$) are shown in Figure \ref{fig:3} and Figure \ref{fig:4}, respectively. The velocity bin of each panel is $\rm 1~km~s^{-1}$. The channel maps of CO emission reveal that the Mon OB1 GMC first appears in [-2,-1] $\rm~km~s^{-1}$ and disappears roughly in [15,16] $\rm~km~s^{-1}$.
A cavity structure with weak $\rm ^{12}CO$ emission is revealed at ($l\sim200\fdg95,~b\sim0\fdg12, ~r\sim0\fdg30$) in the velocity interval of [-5,-3] $\rm~km~s^{-1}$ (see the red dashed circle in [-4,-3] panel of Figure \ref{fig:3}). This structure is next to the Mon R1 loop.
Based on Figure \ref{fig:4}, we find that $\rm C^{18}O$ emission mainly occurs in the east cloud, while there is relatively weak $\rm C^{18}O$ emission in the west cloud and Mon R1 loop. In the channel maps, $\mathrm{^{12}CO}$ emission shows extended and diffuse molecular gas distribution in the whole region, and $\mathrm{^{13}CO}$ emission delineates the main structures of the GMC complex, while $\mathrm{C^{18}O}$ emission traces the densest gas.

Compared to the east cloud, the west cloud is more diffuse. Molecular gas near NGC 2264 shows high integrated intensity from $\rm 2~km~s^{-1}$ to $\rm 12~km~s^{-1}$, and Mon R1 loop also exhibits relatively high intensity from $\rm -3~km~s^{-1}$ to $\rm 1~km~s^{-1}$. 
Many partial-shell and/or cavity structures can be seen in the channel maps. For example, a large cavity-like structure with a long axis of $\sim~0\fdg8$ centered at $(l\sim202\fdg0,~b\sim1\fdg1)$ is revealed from $\sim\rm 1~km~s^{-1}$ to $\sim\rm 6~km~s^{-1}$ in the channel maps (see the red dashed oval in [2,3] panel of Figure \ref{fig:3}). 
Cavity structures with a smaller scale are also found in the east cloud, for example, the red dashed circles in the [6,7] panel of Figure \ref{fig:4}. We analyze one of the cavity structures ($l\sim203\fdg65,~b\sim1\fdg90$) in Section \ref{subsec:4.2} (see also Figure \ref{fig:15}).

A horseshoe-shaped cloud located at ($l\sim202\fdg0,~b\sim2\fdg8$) is found at the northwest of NGC 2264 in [5,6] $\rm~km~s^{-1}$, which is consistent with the result of \citet{2004hreipurthalpha}. As shown in Figure \ref{fig:4}, we also find that another horseshoe structure is located at $(l\sim203\fdg3,~b\sim1\fdg7)$ in [6,7] $\rm~km~s^{-1}$. Both structures are indicated in the [5,6] and [6,7] panels of Figure \ref{fig:4}.
A cavity-like structure at ($l\sim202\fdg7,~b\sim2\fdg1$) in [5,7] $\rm~km~s^{-1}$ is roughly located between the two horseshoe structures. 

A concave structure located at ($l\sim203\fdg0, b\sim2\fdg1$) with high integrated intensity is clearly exhibited from $\rm 6~km~s^{-1}$ to $\rm 11~km~s^{-1}$ (red dashed rectangle in [9,10] panel of Figure \ref{fig:4}). We make a further study toward the structure in Section \ref{subsec:4.1} (also see Figure \ref{fig:10}). 
Additionally, filamentary structures and diverse morphology of CO emission are prevalent in channel maps, showing the complicated distribution of molecular gas toward the Mon OB1 region. 

\subsection{Physical Properties of Molecular Gas} \label{subsec: physical_properties}
Many studies have been done to study the physical properties of molecular gas based on CO observations, for example, excitation temperature, optical depth, and column density (e.g., \citealt{onishi1996ac, Bourke_1997, kawamura1998a13co, pineda2010relation}). 
We follow these studies and assume $\rm ^{12}CO$ emission is optically thick. The excitation temperature ($T_{\mathrm{ex}}$) of $\rm ^{12}CO$ can be derived as

\begin{equation} \label{eq:Tex}
  T_{\mathrm{ex}} = \frac{h\nu_{12}/k}{\mathrm{ln}\left [1+\frac{h\nu_{12}/k}{T_{\mathrm{MB,12_{peak}}+\frac{h\nu_{12}/k}{\mathrm{exp}(h\nu_{12}/kT_{\mathrm{bg}})-1}}}\right ]},
\end{equation}

\noindent where $\nu_{12}=\mathrm{115.27~GHz}$ is the frequency of $\rm ^{12}CO~(J=1\rightarrow 0$) emission. $T_\mathrm{{MB,12_{peak}}}$ is the peak main-beam temperature of $\rm ^{12}CO$. $T_{\mathrm{bg}}\approx2.7$$\rm ~K$ is the temperature of cosmic microwave background. Replacing the physical constants by specific values, equation (\ref{eq:Tex}) can be simplified as

\begin{equation}
  T_{\mathrm{ex}} = \frac{5.53}{\mathrm{ln}(1+\frac{5.53}{T_{\mathrm{MB,12_{peak}}}+0.819})}.
\end{equation}

Assuming that $T_{\mathrm{ex}}$ is the same for $\rm ^{12}CO, ~^{13}CO$, and $\rm C^{18}O$ in MCs, the optical depth of $\rm ^{13}CO$ and $\rm C^{18}O$ can be derived by

\begin{eqnarray}
  \tau(\mathrm{^{13}CO}) = -\mathrm{ln}\left\{1-\frac{T_{\mathrm{MB,13_{peak}}}}{5.29}\left [(e^{5.29/T_{\mathrm{ex}}}-1)^{-1} -0.164\right ]^{-1}\right\},
\end{eqnarray}

\begin{eqnarray}
  \tau(\mathrm{^{18}CO}) = -\mathrm{ln}\left\{1-\frac{T_{\mathrm{MB,18_{peak}}}}{5.27}\left [(e^{5.27/T_{\mathrm{ex}}}-1)^{-1} -0.166\right ]^{-1}\right\},
\end{eqnarray}

\noindent where $T_{\mathrm{MB,13_{peak}}}$ and $T_{\mathrm{MB,18_{peak}}}$ are the peak main-beam temperatures of $\rm ^{13}CO$ and $\rm C^{18}O$. The column density of $\rm ^{13}CO$ and $\rm C^{18}O$ emission can then be estimated by 

\begin{equation}
  N_{\mathrm{^{13}CO}} = 2.42\times10^{14}\times\frac{\tau(\mathrm{^{13}CO})}{1-e^{-\tau(\mathrm{^{13}CO})}}\times\frac{1+0.88/T_{\mathrm{ex}}}{1-e^{-5.29/T_{\mathrm{ex}}}}\int T_{\mathrm{MB}}(\mathrm{^{13}CO})\mathrm{d}v,
\end{equation}

\begin{equation}
  N_{\mathrm{C^{18}O}} = 2.54\times10^{14}\times\frac{\tau(\mathrm{^{13}CO})}{1-e^{-\tau(\mathrm{^{13}CO})}}\times\frac{1+0.88/T_{\mathrm{ex}}}{1-e^{-5.27/T_{\mathrm{ex}}}}\int T_{\mathrm{MB}}(\mathrm{C^{18}O})\mathrm{d}v.
\end{equation}

The column density of molecular hydrogen traced by $\mathrm{^{13}CO}$ ($N_{\mathrm{H_2}}(\mathrm{LTE},13)$) and $\rm C^{18}O$ ($N_{\mathrm{H_2}}(\mathrm{LTE},18)$) can therefore be calculated by

\begin{equation} \label{eq:N_LTE}
  N_{\mathrm{H_2}}(\mathrm{LTE},13) = f(\mathrm{H_2/^{12}CO})\times f(\mathrm{^{12}C/^{13}C})\times N_{\mathrm{^{13}CO}} \simeq 8.5\times10^5\times N_{\mathrm{^{13}CO}},
\end{equation}

\begin{equation} \label{eq:N_LTE18}
  N_{\mathrm{H_2}}(\mathrm{LTE},18) = f(\mathrm{^{16}O/^{18}O})\times N_{\mathrm{^{18}CO}} \simeq 6.2\times10^6\times N_{\mathrm{C^{18}O}},
\end{equation}

\noindent where $f(\mathrm{^{12}C/^{13}C})$ (77; \citealt{wilson1994abundances}), $f(\mathrm{H_2/^{12}CO})$ ($1.1\times10^4$; \citealt{frerking1982relationship}) and $f(\mathrm{^{16}O/^{18}O})$ (560; \citealt{wilson1994abundances}) are abundance ratios.

From Equations (\ref{eq:N_LTE}) and (\ref{eq:N_LTE18}) we obtain the column density of each pixel based on $\rm ^{13}CO$ and $\rm C^{18}O$ emission, respectively. And then the mass of the MCs can be estimated by

\begin{equation} \label{eq:mass}
  M = N_{\mathrm{tot}}d^2\Omega\mu m_{\mathrm{H}},
\end{equation}
where $N_{\mathrm{tot}} = \sum N_{\mathrm{H_2}}$ denotes the sum of the $N_{\mathrm{H_2}}$ over the cloud area, $d$ is the distance of clouds, $\Omega$ is the solid angle corresponding to each pixel, $\mu = 2.8$ is the mean atomic weight \citep{2008A&A...487..993K}, and $m_\mathrm{H}$ represents the mass of atomic hydrogen.

Additionally, we can also estimate the column density of molecular gas by $\rm ^{12}CO$ emission (X-factor method) directly by 

\begin{equation} \label{eq:Xco}
  N_{\mathrm{H_2}}(X_{\mathrm{CO}}) = X_{\mathrm{CO}}\int T_{\mathrm{MB}}(\mathrm{^{12}CO})\mathrm{d}v,
\end{equation}

\noindent where $X_{\mathrm{CO}} = 2.0\times10^{20}~(\rm K~km~s^{-1})^{-1}$ is the conversion factor (e.g., \citealt{dame2001milky, bolatto2013co}).

\begin{figure}[ht!]
  \gridline{\fig{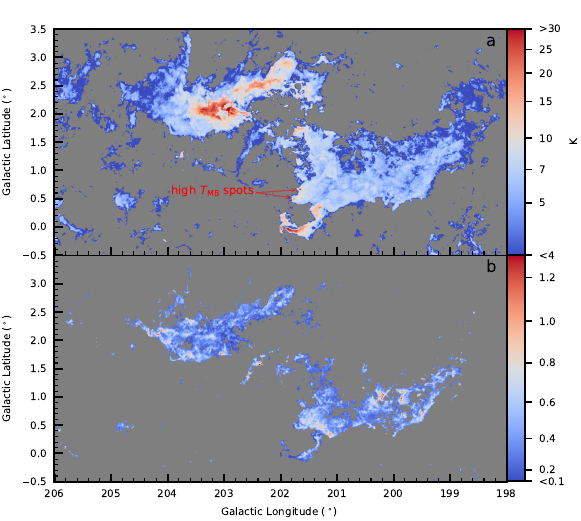}{0.49\textwidth}{}
            \fig{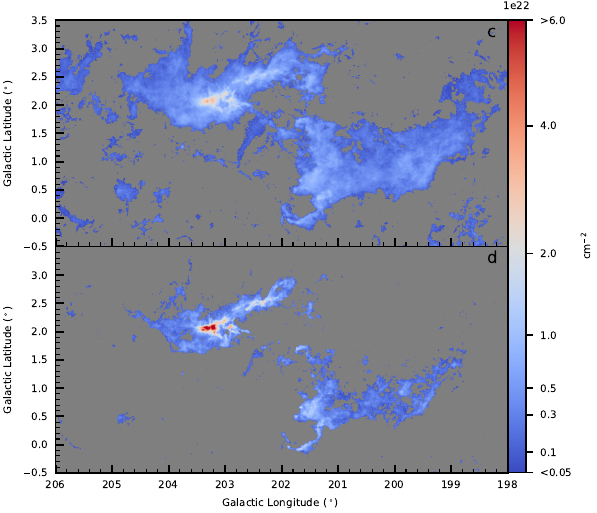}{0.50\textwidth}{}
            }
  \caption{Physical properties of molecular gas toward the Mon OB1 region.
  Panels (a)--(d) are distributions of $T_{\mathrm{MB,12_{peak}}}$, $\tau(\mathrm{^{13}CO})$, $N_{\mathrm{H_2}}(X_{\mathrm{CO}})$, and $N_{\mathrm{H_2}}(\mathrm{LTE},13)$, respectively. the moment masking criterion (all pixels have three continuous channels larger than three times the rms noise) is applied to extract the $\rm ^{12}CO$ and $\rm ^{13}CO$ signal.
  \label{fig:5}}
\end{figure}

Figure \ref{fig:5} exhibits the distribution of peak temperature of $\rm ^{12}CO$ ($T_{\mathrm{MB,12_{peak}}}$), optical depth of $\rm ^{13}CO$ ($\tau(\mathrm{^{13}CO})$), and column density of molecular gas ($N_{\mathrm{H_2}}(X_{\mathrm{CO}})$, $N_{\mathrm{H_2}}(\mathrm{LTE},13)$) toward the Mon OB1 region. 
In panel (a), the peak value of $T_{\mathrm{MB,12_{peak}}}$, 37.5 K, occurs near a B-type star (V* V641 Mon) and HII region Sh-273.
Molecular gas near the Mon R1 loop and the HII region G201.535+01.597 also shows a high peak temperature of 20 -- 30 K.
We also find two localized high-temperature spots (indicated in Figure \ref{fig:5} (a)) at ($l=201\fdg76,~b=0\fdg50$) and ($l=201\fdg67,~b=0\fdg67$), which correspond to two B-type stars, V699 Mon (703 pc) and HD 259431 (712 pc) \citep{arun2019mass}, respectively. The distances of these two stars are well matched with the surrounding gas (Section \ref{subsubsec:distances}), indicating that the gas is illuminated by the early-type stars. 

The mean value of $\tau(\mathrm{^{13}CO})$ based on Figure \ref{fig:5} (b) is 0.38. Over 90\% of pixels have an optical depth of less than 0.6, indicating an optical thinning of $\rm ^{13}CO$ toward the entire Mon OB1 region. Generally, the $\tau(\mathrm{^{13}CO})$ map outlines the main body of cloud.
The $\rm ^{13}CO$ emission covers 37.2\% of the $\rm ^{12}CO$ emission area.
In some regions, $\rm ^{13}CO$ emission reveals some localized dense gas structures with $\tau(\mathrm{^{13}CO}) \geq 0.8$. 
Note that the 2D map only shows limited information of the distribution of $\tau(\mathrm{^{13}CO})$, since we know that in some regions there are several velocity components of gas superposed along the line of sight (LOS). We further discuss the distribution of $\tau(\mathrm{^{13}CO})$ according to the identified $\rm ^{13}CO$ structures in position-position-velocity (PPV) space (Section \ref{subsubsec:3.3.3}). 

Nearly 60\% of pixels have a low temperature of $T_{\mathrm{MB,12_{peak}}}<5~\mathrm{K}$, while only $\sim$ 5\% of pixels have a $T_{\mathrm{MB,12_{peak}}}>10~\mathrm{K}$ based on Figure \ref{fig:5} a. We also find that over 50\% of pixels have a low integrated intensity of $<$ 10 $\mathrm{K~km~s^{-1}}$. These statistics indicate that $\rm ^{12}CO$ emission can well trace diffuse and extended molecular gas of the GMC complex. 

$ N_{\mathrm{H_2}}(X_{\mathrm{CO}})$ in panel (c) is more extended than $N_{\mathrm{H_2}}(\mathrm{LTE},13)$ in panel d. $ N_{\mathrm{H_2}}(X_{\mathrm{CO}})$ ranges from $\rm \sim 1\times10^{20}~cm^{-2}$ to $\rm \sim 4\times10^{22}~cm^{-2}$. The peak value occurs in the NGC 2264 region. On the other hand, $N_{\mathrm{H_2}}(\mathrm{LTE},13)$ exhibits a larger range of values from $\rm \sim 2\times10^{20}~cm^{-2}$ to $\rm \sim 1\times10^{23}~cm^{-2}$. The peak value of $N_{\mathrm{H_2}}(\mathrm{LTE},13)$ is also located in the NGC 2264 region, but is 2.5 times larger than that of $N_{\mathrm{H_2}}(X_{\mathrm{CO}})$.
The discrepancy indicates that $\rm ^{13}CO$ can trace larger dynamic range of column density distribution compared to $\rm ^{12}CO$.
Regions with high column density traced by $\rm ^{13}CO$ ($\rm \geq 3\sim4\times10^{21}~cm^{-2}$) correspond well to the $\rm C^{18}O$ emission seen in Figure \ref{fig:4}.

From Figure \ref{fig:5}, molecular gas in NGC 2264 and the Mon R1 loop exhibits higher temperature and density. On the other hand, NGC 2264 and Mon R1 loop both contain active star-forming activities (e.g., \citealt{2008hsf1.book..966D, movsessian2021new}). This suggests a connection between high-temperature, high-density gas and star-forming activities in the Mon OB1 region. Such connection was also shown in previous studies that MCs with high $T_{\mathrm{peak}}$ ($>$10 K, corresponding to $T_{\mathrm{ex}}$ $>$13.4 K, \citealt{wang2019molecular}, \citeyear{2023AJ....166..121W}) or high column density (e.g., $N_{\mathrm{H_2}}(\mathrm{LTE},13)$ $\rm \gtrsim 3 \times 10^{21}~cm^{-2}$, \citealt{ma2022gas, 2023AJ....166..121W}) may be related to possible star-forming activities. 

The total mass of molecular gas traced by $\mathrm{^{12}CO}$ in the Mon OB1 region is $1.1\times10^5~M_\odot$, which is about 30\% larger than the result of \citet{oliver1996new} after scaling to the same distance of 729 pc (see section \ref{subsubsec:3.3.3}). Obviously, the MWISP survey reveals more molecular gas compared to the CfA 1.2 m survey \citep[see detailed comparisons in][]{2021ApJS..256...32S}.
On the other hand, the total mass traced by $\mathrm{^{13}CO}$ is $5.4\times10^4~M_\odot$.
The mass ratio of $M\mathrm{(^{13}CO)}/M\mathrm{(^{12}CO)}$ is 0.48, which is slightly larger than the ratio of emission area obtained above (0.37).

We also try to estimate the mass of dense gas traced by $\rm C^{18}O$. Here we use the stacking bump algorithm (\citealt{wang2023molecular}) to search for the $\rm C^{18}O$ signal within the $\rm ^{13}CO$ boundary. Briefly, the identified $\rm C^{18}O$ signal satisfies the following conditions: (1) For each identified pixel, at least three continuous channels of the spectrum are larger than two times of the rms noise ($\rm \sim 0.5~K$). (2) At least one of the surrounding pixels identically fulfills condition 1. (3) The average spectrum of identified $\rm C^{18}O$ structures has three continuous channels above three times the noise of the average spectrum. According to \citet{wang2023molecular}, the stacking bump algorithm can effectively avoid false detections generated by noise and can find weak but reliable $\rm C^{18}O$ signal.
In total, the $\rm C^{18}O$ emission area makes up 3.2\% of the $\rm ^{12}CO$ emission area. The mass of dense molecular gas traced by $\rm C^{18}O$ is $M\mathrm{(C^{18}O)}=9.1\times10^3~M_\odot$.

\subsection{$^{13}CO$ MCs Toward the Mon OB1 region} \label{subsec:cloud-identification}

\subsubsection{Cloud Identification} \label{subsubsec:identification}
MCs present hierarchical structures. As shown in Figures \ref{fig:3} and \ref{fig:4}, molecular gas displays extended and complicated morphology, for example, cavity-like, filamentary, and other irregular structures. On the other hand, Figure \ref{fig:2} indicates that multiple velocity components are superposed in the Mon OB1 region. Therefore, the 2D distribution of physical properties in Figure \ref{fig:5} cannot well reveal the details of molecular gas in velocity-crowding regions (e.g., the NGC 2264 region).

Following the work of \citet{zhang2024multilayer}, we try to identify MC structures to study the distribution and properties of molecular gas based on MWISP $\mathrm{^{13}CO}$ data and Gaussian decomposition + hierarchical clustering methods (e.g., \citealt{miville2017physical, henshaw2019brick}). 
The reason we utilize $\rm{^{13}}CO$ data for cloud identification is that $\rm{^{13}}CO$ is usually optically thin (Figure \ref{fig:5} (b), see also \citealt{wang2023molecular}). In addition, statistics of large MC samples show that $\rm{^{13}}CO$ can trace the main structures of MCs (\citealt{su2020local, yuan2023spatial, wang2023molecular}). Moreover, the $\rm{^{13}}CO$ spectral line profile is velocity separated compared to $\rm{^{12}}CO$ spectral line profile for MCs with multivelocity components. All these features can largely avoid overfitting problems in velocity-crowding regions.  
 
Here we briefly summarize the steps to identify $\mathrm{^{13}CO}$ structures. We first perform a Gaussian decomposition of the velocity component of each pixel of the data from $\mathrm{^{13}CO}$ using the GaussPy+ algorithm (\citealt{riener2019gausspy+}, \citeyear{riener2020autonomous}), and then we use the Agglomerative Clustering for ORganizing Nested Structures (ACORNS) algorithm (\citealt{henshaw2019brick}) to cluster the results of the GaussPy+ fit to construct $\mathrm{^{13}CO}$ structures with similar properties.

GaussPy+ is a fully automated machine-learning-based Gaussian decomposition algorithm that allows for fast fitting of spectral lines with multivelocity components, making it ideal for application to a large-scale 3D data set, for example, the MWISP CO survey \citep{riener2020autonomous}. Using this algorithm, multiple parameters of each pixel can be obtained, such as center velocity, peak intensity, and FWHM. These parameters are the basis for subsequent calculations of the physical properties of the identified $\mathrm{^{13}CO}$ structures. Then, it is possible to distinguish the different components on each pixel and to prepare for the next step of clustering using the ACORNS algorithm. 

In this paper, we adopt the following parameters:
\begin{itemize}
  \item[1.] Signal-to-noise ratio. We set it to 5 for skipping the weak emission. The signal of $\mathrm{^{13}CO}$ emission with level $\gtrsim$ 1 K remains.
  \item[2.] Minimum FWHM. We set it to be at least two channels ($\sim 0.3~\rm~km~s^{-1}$) for the MWISP data.
  \item[3.] Smoothing parameter. We take the optimal smoothing parameters (i.e., $\alpha_1$=2.18, $\alpha_2$=4.94; \citealt {riener2020autonomous}) for the MWISP data.
\end{itemize}

For other parameters, we adopt the defaults. In the whole map, the total recovered flux from Gaussian decomposition makes up about 88.0\% of the raw data flux.

\begin{figure}[ht!]
  \centering
  \includegraphics[scale=0.7]{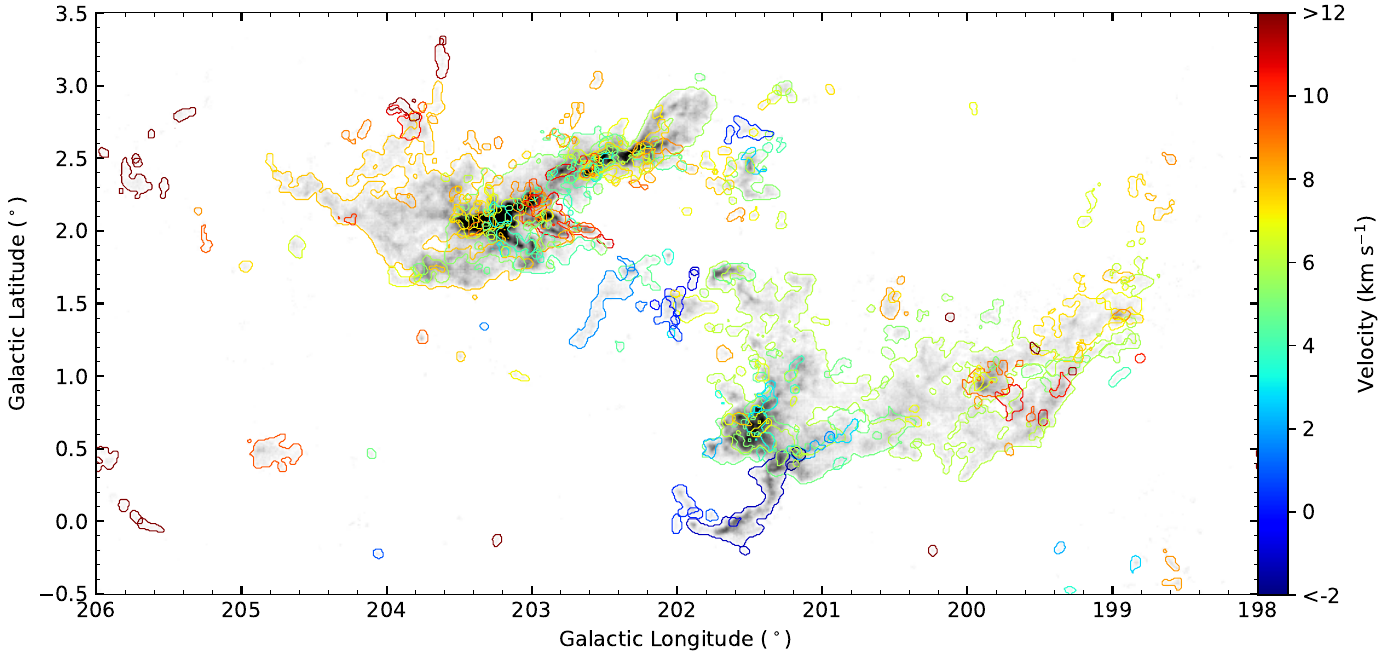}
  \caption{Velocity distribution of identified $\mathrm{^{13}CO}$ structures toward the Mon OB1 region. Different colors of the contours represent the velocities of structures. The background shows the integrated intensity of $\mathrm{^{13}CO}$.
  \label{fig:6}}
\end{figure}

ACORNS is a dendrogram-based (\citealt{rosolowsky2008structural}) algorithm that can construct spatially coherent pixels with similar properties (e.g., center velocity, intensity, and FWHM). In this paper, we consider a $\rm^{13}CO$ structure to be larger than 9 pixels ($\rm\sim2.25~arcmin^2$); other parameters in the algorithm are set to their default values \citep{henshaw2019brick}. As a result, we identify 263 $\rm^{13}CO$ structures in the Mon OB1 region (see Figure \ref{fig:6}).
The total flux of the identified $\rm^{13}CO$ structures makes up about 99.2\% of the recovered flux from Gaussian decomposition.

\begin{figure}[ht!]
  \centering
  \includegraphics[scale=0.5]{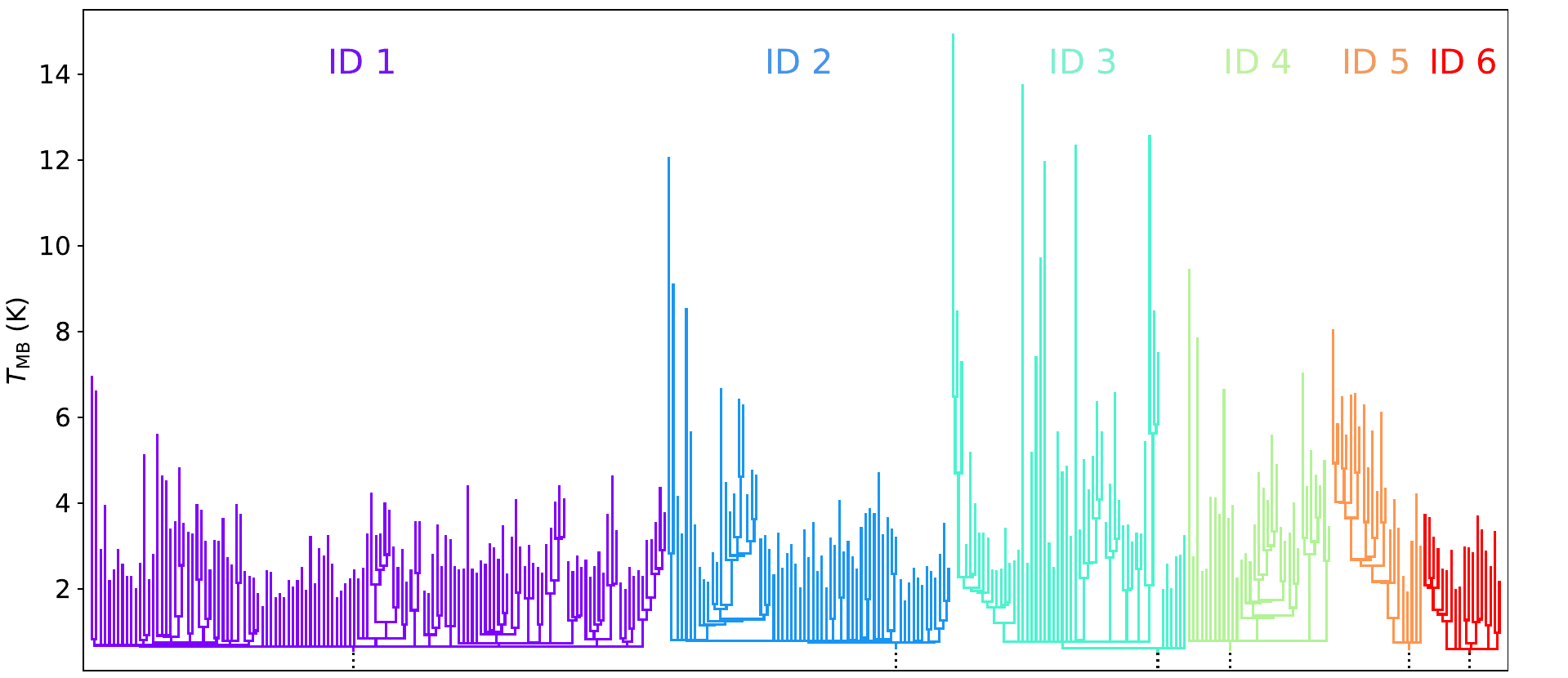}
  \caption{ACORNS clustering displayed as a dendrogram. Different colors represent different trees. Each tree can be further divided into branches and leaves. Here we display the forest of the six largest $\rm^{13}CO$ structures (from left to right for IDs 1--6 in Table \ref{table:phy-para}).
  \label{fig:6.1}}
\end{figure}

The $\rm^{13}CO$ structures can also be represented as hierarchical structures. We show the six largest $\mathrm{^{13}CO}$ structures (containing 68.7\% of the total flux) as dendrogram in Figure \ref{fig:6.1}. The difference between these $\rm^{13}CO$ structures can be revealed from the dendrogram. For example, leaves and branches of IDs 1 and 6 have low $T_{\mathrm{MB}}$ compared to IDs 2, 3, and 5. 
These features also reveal the difference in $T_{\mathrm{MB}}$ distribution between the east cloud (IDs 2, 3, and 5) and west cloud (IDs 1 and 6). 
On the other hand, structure ID 4 near the eastern edge of the west cloud also shows somewhat bright emission, indicating an association between the gas and the surrounding star-forming activities (also see the high-temperature spots in Figure \ref{fig:5} (a)).
We propose that MCs closely associated with active star-forming regions have a large contrast between the peak and average temperature (as well as the column density) of $\rm^{13}CO$ emission. These MCs also exhibit enhanced $\rm C^{18}O$ emission compared to other MCs.  


\subsubsection{Distance Measurements} \label{subsubsec:distances}
MCs with different velocity components along a certain LOS may not be located at the same distance (e.g., the velocity-crowding regions toward the Cygnus region; \citealt{zhang2024multilayer}). On the other hand, MCs with similar velocities may be located at different distances (\citealt{peek2022burton}). Whether MCs in Mon OB1 region with different velocities belong to a physically extended GMC complex needs to be further studied.

\begin{figure}[ht!]
  \centering
  \includegraphics[scale=0.45]{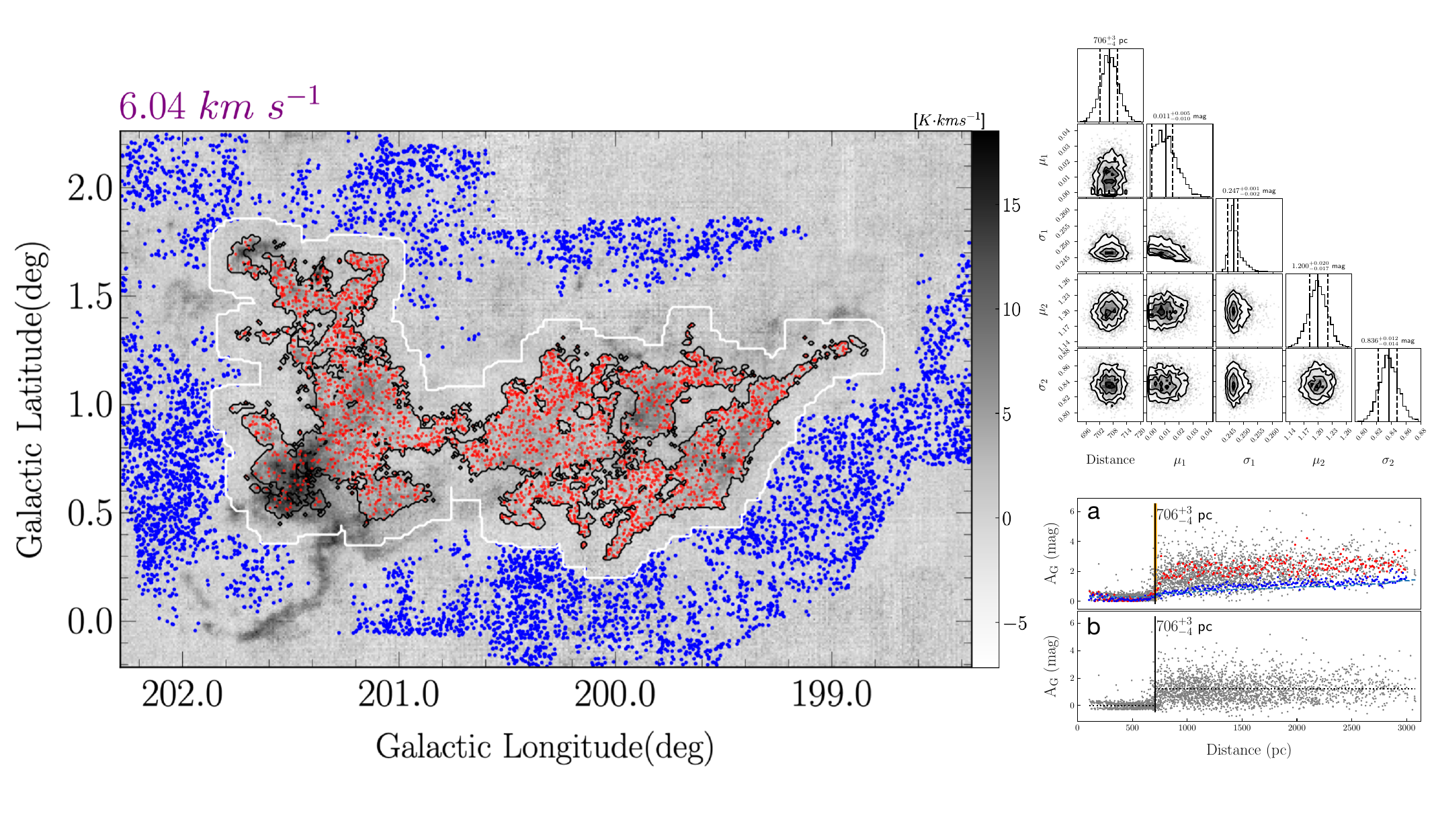}
  \caption{Example of the distance measurement for G200.663+0.897 (ID 1 in Table \ref{table:phy-para}). 
  In the left panel, the background shows the integrated intensity of $\rm ^{13}CO$ emission. The black contour is the boundary of $\rm{^{13}}CO$ structure identified by ACORNS (see section \ref{subsubsec:identification}), while the white contour expands the black contour by $5'$. 
  Blue and red circles represent off-cloud and on-cloud stars, respectively. The corner plot in the upper right corner shows the result of MCMC sampling. Parameters in the panel are extinction $A_G$ ($\mu_1$) and its dispersion ($\sigma_1$) of foreground stars and extinction $A_G$ ($\mu_2$) and its dispersion ($\sigma_2$) of background stars. The median value and 1 $\sigma$ confidence interval are shown with solid and dashed lines. Panels (a) and (b) in the lower right corner show the $A_G$-distance distribution for on-cloud stars. The gray circles represent raw star data. Blue and red stars are bin values of off-cloud and on-cloud stars for a 10 pc interval. Panel (a) gives the $A_G$ values before baseline subtraction. The blue dashed line represents the fit baseline of off-cloud stars. Panel (b) gives the result after baseline subtraction. The black dotted lines to the left and right of the jump point (black solid line) indicate the values of $\mu_1$ and $\mu_2$. The yellow vertical lines in panel (a) are the distance sampling of MCMC.
  \\
  (The complete figure set (32 images) is available.)
  \label{fig:7}}
\end{figure}

In this subsection, we measure the distances of the identified $\rm^{13}CO$ structures in combination with Gaia DR3 data following the BEEP method in \citet{yan2019molecular} and \citet{zhang2024multilayer}.
The method is divided into three steps. First, we select on-cloud stars and off-cloud stars to represent the extinction environment toward the studied cloud and surrounding background, respectively. 
For on-cloud stars, we select those within the boundary of the identified $\rm{^{13}}CO$ structures (represented by red circles in Figure \ref{fig:7}). 
For off-cloud stars (represented by blue circles in Figure \ref{fig:7}), we select those in $\rm{^{12}}CO$ emission-free regions and close to the $\rm{^{13}}CO$ structures. That is, off-cloud stars are located between $5'$ (white contour in Figure \ref{fig:7}) and $30'$ away from the boundary of the studied $\rm{^{13}}CO$ structure.
These stars can well represent the background extinction features around the studied $\rm{^{13}}CO$ structure.
The second step is to estimate the baseline of $A_G$ by utilizing the off-cloud stars. 
The third step is to subtract the $A_G$ of on-cloud stars with $A_G$ baseline from above off-cloud stars. Then, we pick up the jump feature in the $A_G$-distance map (see Figure \ref{fig:7} (a) and (b)) by Bayesian analysis and Markov Chain Monte Carlo (MCMC) methods. The model contains five parameters: the cloud distance ($D$), extinction $A_G$ ($\mu_1$) and its dispersion ($\sigma_1$) of foreground stars, and extinction $A_G$ ($\mu_2$) and its dispersion ($\sigma_2$) of background stars. The likelihoods and priors of the model are similar to \citet{yan2019molecular}. The MCMC algorithm used in this work is emcee \citep{Foreman2013emcee}.

Since the error of the Gaia $G$-band extinction is 0.07 mag \citep{2023A&A...674A..27A}, we considered $\Delta A_G \geq$ 3 sigma (0.2 mag) as jump features caused by extinction from the studied $\rm{^{13}}CO$ structures (e.g., see the vertical line in the lower right corner of Figure \ref{fig:7}). Additionally, the uncertainties of MCMC fit results should be less than 20\%, and there should be a sufficient number of stars between jump features to ensure the quality of fitting.

Due to lacking the on-cloud stars, we are unable to obtain the distances of $\rm{^{13}}CO$ structures with small angular areas (i.e., $\rm\leq30~arcmin^2$). Finally, we obtain distances of 32 $\rm{^{13}}CO$ structures (see Table \ref{table:phy-para}). An example of distance measurements is shown in Figure \ref{fig:7}. The flux of $\rm{^{13}}CO$ structures with distance contributes to 90.9\% of total flux of identified $\rm{^{13}}CO$ structures. 

\clearpage

\begin{figure}[ht!]
  \centering
  \includegraphics[scale=0.7]{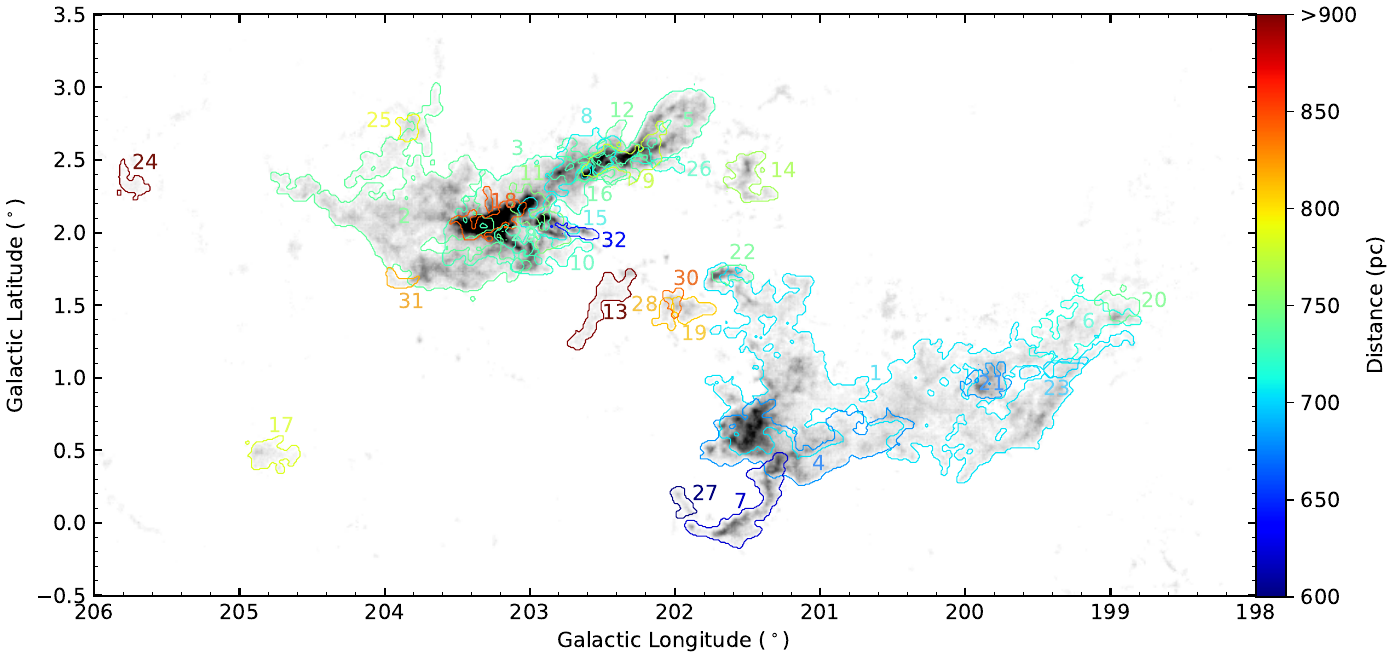}
  \caption{Identified $\mathrm{^{13}CO}$ structures with distance measurements toward the Mon OB1 region. Different colors of the contours represent the distances of identified $\mathrm{^{13}CO}$ structures, while the numbers indicate the structure ID from Table \ref{table:phy-para}. The background is the same as Figure \ref{fig:6}.
  \label{fig:8}}
\end{figure}

\figsetstart
\figsetnum{8}
\figsettitle{Distance measurements for 32 large $\rm{^{13}}CO$ structures}

\figsetend

Figure \ref{fig:8} displays the distribution of $\rm{^{13}}CO$ structures for which distances are measured. 
The emission from these $\rm{^{13}}CO$ structures well represents the outline of the whole $\rm{^{13}}CO$ emission in the Mon OB1 region.
The regions with high integrated intensity (or column density) correspond to the area where multivelocity components of $\rm ^{13}CO$ structures are superimposed. A similar scenario was also discussed for MCs toward the Aquila Rift region \citep{su2020local}.

\begin{figure}[ht!]
  \centering
  \includegraphics[scale=0.65]{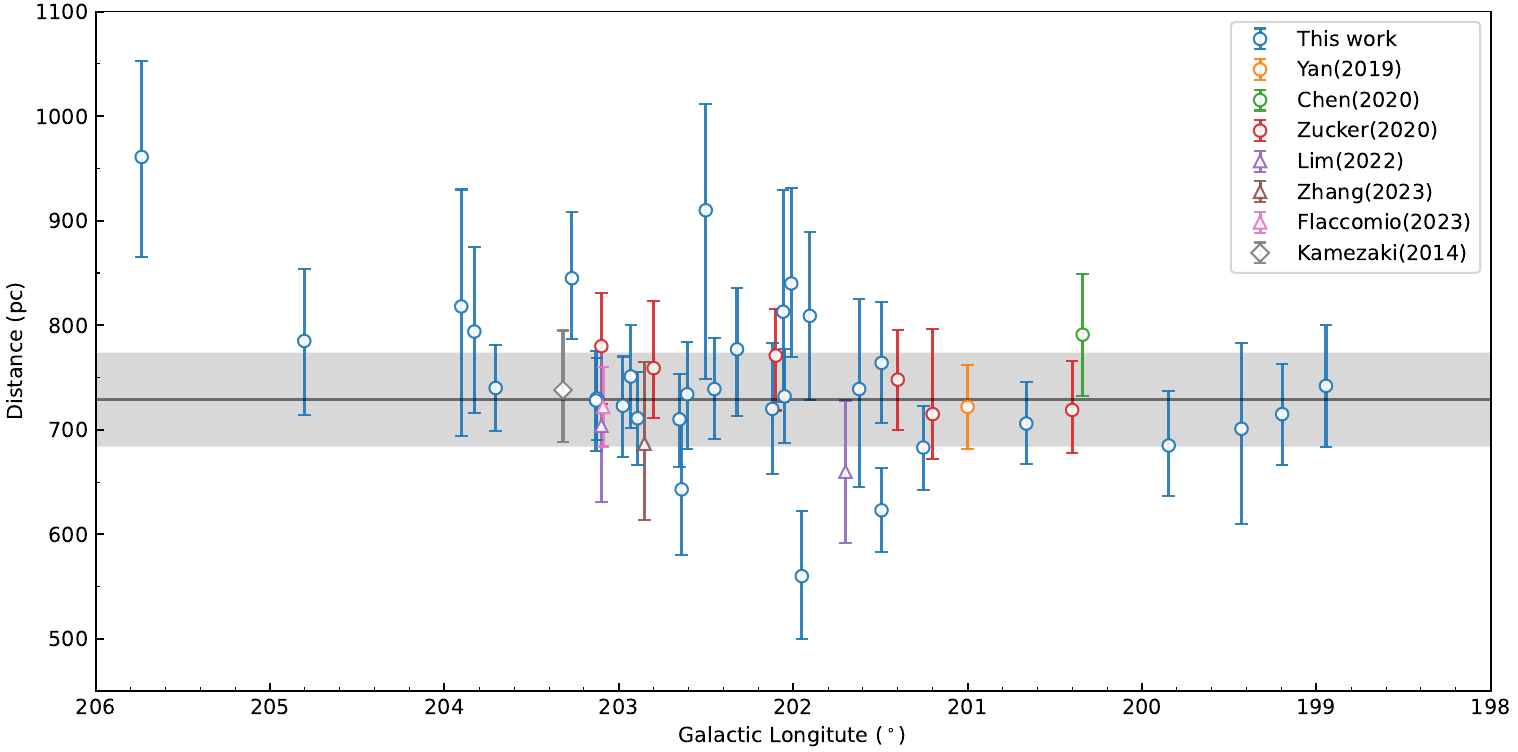}
  \caption{Distribution of distances of $\rm ^{13}CO$ structures (blue circles) in the Mon OB1 region. The horizontal black line represents the mass-weighted average distance of 729 pc for the 32 $\rm ^{13}CO$ structures, and the gray shaded area represents the uncertainties of the average distance.
  For comparison, we also show the previous distance estimations toward the region. 
  Circles indicate distances obtained by extinction measurements of clouds (\citealt{yan2019molecular, chen2020large, 2020A&A...633A..51Z}). Triangles indicate distances obtained using star parallax (\citealt{lim2022gaia, zhang2023distances, flaccomio2023spatial}). Diamond indicates distance obtained from maser measurements (\citealt{kamezaki2014annual}). For studies involving Gaia data, the uncertainties shown in the figure include the 5\% systematic error of Gaia parallax. 
  \label{fig:9}}
\end{figure}

We exhibit the distribution of distances of $\mathrm{^{13}CO}$ structures along the Galactic longitude in Figure \ref{fig:9}. 
The mass-weighted average distance of these $\mathrm{^{13}CO}$ structures is $D_{\mathrm{avg}} = \mathrm{729^{+45}_{-45}~pc}$ (including the 5\% systematic error of Gaia parallax). 
The $\mathrm{^{13}CO}$ structures in the west cloud have distances of roughly 700 pc. The largest $\mathrm{^{13}CO}$ structure of the west cloud (ID 1 in Table \ref{table:phy-para}) has a distance of $\mathrm{706^{+39}_{-38}~pc}$. 
The distance of Mon R1 loop (e.g., ID 7 in Table \ref{table:phy-para}) is $\mathrm{623^{+40}_{-40}~pc}$, which is consistent with the distance of $\mathrm{660^{+68}_{-68}~pc}$ from \citet{lim2022gaia}. 
It seems that the Mon R1 loop is somewhat located in front of other $\rm ^{13}CO$ structures in the west cloud.
The CO gas with negative velocities of [-15, -5] $\rm~km~s^{-1}$ is located at $\sim~$0.7 kpc based on our $\mathrm{^{12}CO}$ and Gaia DR3 data.

The $\mathrm{^{13}CO}$ structures in the east cloud are mainly distributed around 700-750 pc. The two largest $\mathrm{^{13}CO}$ structures (IDs 2 and 3 in Table \ref{table:phy-para}) are at distances of $\mathrm{740^{+41}_{-41}~pc}$ and $\mathrm{730^{+38}_{-39}~pc}$, respectively. Some small structures (i.e., structures 18, 25, and 31 in Table \ref{table:phy-para}) in the east cloud are probably at a slightly larger distance of $\geq$ 800 pc. 
There are also some $\rm ^{13}CO$ structures with a comparatively larger distance of $>800$ pc toward the center of the Mon OB1 region (e.g., IDs 13, 19, 28, and 30). 

We compared the MC distances with recent 3D dust map of \citet{Edenhofer2023}. The dust map shows high extinction in 650 -- 850 pc, which is agreed with the distance of main CO structure in Mon OB1 region.
That is the $\sim$ 720 pc slice and $\sim$ 760 pc slice are roughly agreed with the west cloud ($\sim$ 700 pc) and east cloud ($\sim$ 750 pc) that are observed in this work. In addition, the dust map also shows high extinction in $\sim~900$ pc, which are roughly matched with some $\rm ^{13}CO$ structures with slightly larger distance (i.e., IDs 18 and 24). The agreement in distance measurements between the two works enhances the reliability of distance results for the $\rm ^{13}CO$ structures (see Figure \ref{fig:8}). 

Considering the distance uncertainties of identified $\rm ^{13}CO$ structures (see Table \ref{table:phy-para} and Figure \ref{fig:9}), we do not find a significant discrepancy in measured distances between the east cloud and the west cloud. Therefore, we propose that these $\rm ^{13}CO$ structures constitute a large GMC complex at a distance of 680$\sim$770 pc. This result is in good agreement with previous studies toward the Mon OB1 region (\citealt{kamezaki2014annual,yan2019molecular, chen2020large, 2020A&A...633A..51Z,lim2022gaia,zhang2023distances,flaccomio2023spatial}). 

The entire GMC complex may span a large depth of $\rm \sim 150$ pc or more along the LOS, which is about twice of its projection size of $\sim$80 pc (i.e., $\sim6^{\circ}$ at 729 pc). This scenario is comparable to the case of the Orion GMC (\citealt{kounkel2017gould, grossschedl20183d}).

In addition, the identified $\rm ^{13}CO$ structures near the edge of the east and west clouds, as well as Mon R1 loop, seem to consist of a large cavity-like structure centered at $(l\sim202\fdg1,~b\sim1\fdg2)$ with a major axis of $\sim2\fdg6$ (see Figure \ref{fig:6}).
The structure can also be discerned in the channel maps (Figures \ref{fig:3} and \ref{fig:4}) in the velocity range of [-2,6] $\rm~km~s^{-1}$. 
The parallelogram-like velocity structure seen in Figure \ref{fig:2} likely corresponds to the large cavity-like structure.
The origin of this structure is unknown. Interestingly, we find that the Monoceros OB4 discovered by \citet{teixeira2021monoceros} is located near the projected center of the cavity-like structure. By considering an LOS velocity of about $\rm 30~km~s^{-1}$ for the cluster (\citealt{naik2023missing}), the velocity of the cluster relative to MCs in the Mon OB1 region can be estimated as $\rm \sim20~km~s^{-1}$. 
Therefore, it is likely that the Mon OB4 association at $\sim$ 1 kpc was probably located near the center of the cavity-like structure  $\sim$ 20 Myr ago, which is roughly consistent with the estimated age of the cluster \citep{teixeira2021monoceros}.

Some clouds with high velocity ($\sim \rm 20~km~s^{-1}$ -- $\rm 35~km~s^{-1}$) are located at distances of $\sim$ 2.1 -- 2.7 kpc based on our $\mathrm{^{12}CO}$ data and Gaia DR3 data. The estimated distances of the clouds agree with their kinematic distances of $\sim$ 2.3 -- 3.8 kpc \citep{reid2019trigonometric}. These results place the clouds in the Perseus arm, which is not related to the Mon OB1 GMC complex.

\subsubsection{Physical Properties Based on Identified $^{13}CO$ Structures}\label{subsubsec:3.3.3}
The physical properties of each $\rm ^{13}CO$ structure are shown in Table \ref{table:phy-para}. We compare the properties of the top 10 largest $\rm ^{13}CO$ structures. 
Compared to the 2D distribution in Figure \ref{fig:5}, more details of $T_{\mathrm{MB,12_{peak}}}$, $\tau(\mathrm{^{13}CO})$, and $N_{\mathrm{H_2}}(\mathrm{LTE},13)$ are revealed by the identified $\rm ^{13}CO$ structures.
Among these 10 $\rm ^{13}CO$ structures, IDs 1, 4, 6, and 7 belong to the west cloud, while the others belong to the east cloud. 
We analyze the distribution of $T_{\mathrm{MB,12_{peak}}}$, $\tau(\mathrm{^{13}CO})$, and $N_{\mathrm{H_2}}(\mathrm{LTE},13)$ pixel by pixel. 

The mean value of $T_{\mathrm{MB,12_{peak}}}$ mainly ranges from $\sim$ 6 -- 10 K, excluding IDs 7 (12.2 K) and 9 (14.0 K).
IDs 1, 2, 4, and 6 show relatively lower mean $T_{\mathrm{MB,12_{peak}}}$ of $\rm \lesssim7~K$. 
Interestingly, some structures (i.e., IDs 2, 3, 7, and 9) exhibit another high-temperature component ranging from $\sim$ 10 -- 30 K, indicating the association between molecular gas and the surrounding star-forming activities (e.g., \citealt{2008hsf1.book..966D, movsessian2021new}).

The average values of $\tau(\mathrm{^{13}CO})$ for most structures range from 0.35 to 0.43, which is consistent with the average value of 0.37 in Figure \ref{fig:5} b. As an exception, we find that ID 4 shows a relatively high mean $\tau(\mathrm{^{13}CO})$ of  $\sim$0.5.
The relatively high optical depth environment where ID 4 is located also can be directly seen in Figure \ref{fig:5} b.
Additionally, the $\tau(\mathrm{^{13}CO})$ probability distribution of all 10 structures shows a high optical depth tail beginning from $\gtrsim$ 0.5 to 0.6.

The average $N_{\mathrm{H_2}}(\mathrm{LTE},13)$ of structures ranges from $\rm \sim 1.3\times10^{21}~cm^{-2}$ to $\rm \sim 5.8\times10^{21}~cm^{-2}$ (see Table \ref{table:phy-para}). The two large $\rm ^{13}CO$ structures in the west cloud (IDs 1 and 6) exhibit relatively low $N_{\mathrm{H_2}}(\mathrm{LTE},13)$.
Structures near star-forming regions (e.g., IDs 3 and 9) have a relatively high average $N_{\mathrm{H_2}}(\mathrm{LTE},13)$ of $\rm \sim 6\times10^{21}~cm^{-2}$. The distribution of $N_{\mathrm{H_2}}(\mathrm{LTE},13)$ for most structures has the shape of a lognormal (LN) distribution in low to medium density range ($\rm \lesssim3-4\times10^{21}~cm^{-2}$), while the distribution of some $\rm ^{13}CO$ structures also shows power-law (PL) high-density tails, which are likely related to star-forming activities \citep{ma2022gas}.

We estimate the state of gravitational stability of $\rm ^{13}CO$ structures by virial parameters \citep{kauffmann2013low}

\begin{equation}\label{eq:alpha}
  \alpha_{\mathrm{vir}} = \frac{M(\mathrm{vir})}{M(\mathrm{^{12}CO})},
\end{equation}
where $M(\mathrm{vir}) = \frac{5\delta_v^2R}{G}\approx209(\frac{R}{\mathrm{pc}})(\frac{\Delta v}{\mathrm{km~s^{-1}}})^2~M_\odot$. Here $\delta_v$ represents velocity dispersion and $\Delta v$ is the FWHM of the spectrum. An $\alpha_{\mathrm{vir}}$ smaller than the critical value of $\sim$1 -- 2 means that the cloud is in a gravitationally bound state. 
We tentatively use $\rm ^{12}CO$ emission to trace the total mass of molecular gas, while we estimated $M(\mathrm{vir})$ by $\delta_v$ and effective radius $R=\frac{{Distance}}{2}\times\sqrt{\frac{Area}{\pi}-\theta_{\mathrm{beam}}^2}$ for the identified $\rm ^{13}CO$ structures in Table \ref{table:phy-para}, where $\theta_{\mathrm{beam}}$ is the HPBW of the telescope.


The median value of the virial parameters is 2.2, indicating that most of the structures are in an approximate virial equilibrium state. Some $\rm ^{13}CO$ structures with $\rm \alpha_{vir}<1$ are gravitationally dominated (e.g., IDs 3 and 5), while some $\rm ^{13}CO$ structures show very high $\rm \alpha_{vir}$, for example, ID 25 with $\rm \alpha_{vir}=9.3$ is likely in a nongravitationally bound state. 

$M\mathrm{(^{12}CO)}$ and $M\mathrm{(C^{18}O)}$ are measured within the boundary of each identified $\mathrm{^{13}CO}$ structure. 
The measurements of $M\mathrm{(^{12}CO)}$ are directly from the integrated $\mathrm{^{12}CO}$ emission within the 3D $\mathrm{^{13}CO}$ structures.
We use the same method introduced in section \ref{subsec: physical_properties} (the stacking bump algorithm) to search for weak $\rm C^{18}O$ signals in the boundary of the $\rm ^{13}CO$ structures, and then we obtain $M\mathrm{(C^{18}O)}$.

The total $M\mathrm{(^{13}CO)}$ and $M\mathrm{(C^{18}O)}$ of the identified $\rm ^{13}CO$ structures is $4.3~\times10^4~M_\odot$ and $8.4\times10^3~M_\odot$, respectively. The $M\mathrm{(^{13}CO)}$ of $4.3~\times10^4~M_\odot$ here is somewhat less than that estimated from the raw data ($5.4\times10^4~M_\odot$ in Section \ref{subsec: physical_properties}). We find that over 90\% of the mass of $\rm ^{13}CO$ structures traced by $\rm ^{13}CO$ and $\rm C^{18}O$ is distributed in $\rm ^{13}CO$ structures with large angular sizes (see Table \ref{table:phy-para}). The total $M\mathrm{(^{12}CO)}$ of the identified $\rm ^{13}CO$ structures is $6.1~\times10^4~M_\odot$, meaning that nearly half of the mass of molecular gas ($4.9\times10^4~M_\odot$) is in the region with $\rm ^{12}CO$ but no $\rm ^{13}CO$. This result generally agrees with the statistics of molecular gas in the Taurus region based on FCRAO large-scale $\rm ^{12}CO$ and $\rm ^{13}CO$ observations \citep{2008ApJ...680..428G}.

We compare the dense gas fraction ($\frac{M\mathrm{(C^{18}O)}}{M\mathrm{(^{12}CO)}}$) for the east and west clouds. Emission from the region where $l<202^\circ$ and $b<2^\circ$ is assigned to the west cloud.
Emission from the region where $201^\circ<l<205^\circ$, $b>1^\circ$, and not overlapping with the west cloud region is assigned to the east cloud.
The total mass of the east and west clouds traced by $\rm ^{12}CO$ is $5.6\times10^4~M_\odot$ and $4.6\times10^4~M_\odot$, respectively.
In the east cloud there are a total of 35 $\rm ^{13}CO$ structures containing $\rm C^{18}O$ emission, while in the west cloud there are a total of 17 $\rm ^{13}CO$ structures with $\rm C^{18}O$ emission. For those $\rm ^{13}CO$ structures with distance, we apply this distance to the corresponding $\rm C^{18}O$ emission. If there is no distance, we adopt the average distance of 729 pc to calculate the mass. The total mass of dense gas traced by $\rm C^{18}O$ in the east cloud and the west cloud is $6.9\times10^3~M_\odot$ and $1.5\times10^3~M_\odot$, corresponding to dense gas mass fraction of 12.4\%, and 3.3\%, respectively. 
The dense gas fraction of the east cloud is $\sim$ 3.5 times larger than that of the gas in $l=[10^{\circ},50^{\circ}]$ (3.7\%; \citealt{torii2019forest}). Actually, many star-forming activities (e.g., infalls, outflows, and $\rm H_2~jets$) are ongoing therein.

\begin{longrotatetable}
\begin{deluxetable*}{cccccccccccccc}
  \tablecaption{Physical parameters of $\rm ^{13}CO$ structures with distance measurements\label{table:phy-para}}
  \tablewidth{700pt}
  \tabletypesize{\scriptsize}
  \renewcommand{\arraystretch}{1.5}
  \tablehead{
    \colhead{ID} & 
    \colhead{$l$} & 
    \colhead{$b$} & 
    \colhead{$v_{\mathrm{{LSR}}}$} & 
    \colhead{$\delta_v$} & 
    \colhead{$Distance$} & 
    \colhead{$Area$} & 
    \colhead{$N_{\mathrm{xfactor}}$} & 
    \colhead{$N_{\mathrm{LTE}}$} & 
    \colhead{$M\mathrm{(^{12}CO)}$} &  
    \colhead{$M\mathrm{(^{13}CO)}$} &
    \colhead{$M\mathrm{(C^{18}O)}$} &
    \colhead{$\frac{M(\mathrm{C^{18}O})}{M(\mathrm{^{13}CO})}$} & 
    \colhead{$\alpha_{\mathrm{vir}}$} \\  
    \colhead{} & 
    \colhead{(deg)} & 
    \colhead{(deg)} & 
    \colhead{($\mathrm{km~s^{-1}}$)} & 
    \colhead{($\mathrm{km~s^{-1}}$)} &
    \colhead{(pc)} & 
    \colhead{($\mathrm{arcmin^2}$)} & 
    \colhead{($\mathrm{10^{21}cm^{-2}}$)} &
    \colhead{($\mathrm{10^{21}cm^{-2}}$)} & 
    \colhead{($M_\odot$)} & 
    \colhead{($M_\odot$)} &
    \colhead{($M_\odot$)} & 
    \colhead{} & 
    \colhead{}
    } 
\startdata
1 & 200.663 & 0.897 &  6.04 &   1.64 &    $706^{+4}_{-3}\pm35$ &   4073.25 &        3.3 &     2.0 &   12770 &    7800 &     480 &    0.062 &    1.9 \\
2 & 203.706 & 2.062 &  7.82 &   1.39 &    $740^{+4}_{-4}\pm37$ &   2208.00 &        3.8 &     2.7 &    8810 &    6370 &    1310 &    0.206 &    1.5 \\
3 & 203.126 & 2.129 &  5.34 &   1.16 &    $730^{+2}_{-3}\pm36$ &   1343.75 &        6.5 &     5.6 &    8860 &    7630 &    2200 &    0.288 &    0.8 \\
4 & 201.254 & 0.539 &  4.86 &   1.25 &    $683^{+6}_{-6}\pm34$ &   1137.75 &        3.9 &     3.2 &    3990 &    3290 &     730 &    0.222 &    1.8 \\
5 & 202.049 & 2.723 &  5.66 &   0.74 &    $732^{+9}_{-8}\pm37$ &    540.75 &        3.8 &     3.4 &    2090 &    1860 &     400 &    0.215 &    0.9 \\
6 & 199.196 & 1.341 &  7.45 &   0.87 &  $715^{+12}_{-13}\pm36$ &    402.50 &        2.2 &     1.3 &     850 &     520 &      10 &    0.019 &    2.5 \\
7 & 201.493 & 0.100 & -1.28 &   0.87 &    $623^{+9}_{-9}\pm31$ &    389.00 &        4.4 &     3.9 &    1290 &    1140 &     130 &    0.114 &    1.4 \\
8 & 202.652 & 2.501 &  4.62 &   0.98 &   $710^{+8}_{-10}\pm36$ &    268.50 &        4.1 &     2.8 &    1060 &     730 &      40 &    0.055 &    2.1 \\
9 & 202.322 & 2.513 &  7.57 &   1.24 &  $777^{+20}_{-25}\pm39$ &    225.25 &        6.5 &     5.8 &    1710 &    1520 &     300 &    0.197 &    2.1 \\
10 & 202.977 & 1.838 &  4.25 &   1.07 &  $723^{+11}_{-13}\pm36$ &    219.50 &        5.2 &     3.8 &    1150 &     830 &     150 &    0.181 &    2.1 \\
11 & 202.931 & 2.099 &  8.59 &   0.91 &  $751^{+12}_{-12}\pm38$ &    214.25 &        7.1 &     4.6 &    1640 &    1070 &      70 &    0.065 &    1.1 \\
12 & 202.452 & 2.503 &  6.98 &   1.16 &  $739^{+12}_{-11}\pm37$ &    202.25 &        6.2 &     3.8 &    1310 &     800 &     260 &    0.325 &    2.1 \\
13 & 202.501 & 1.501 &  1.74 &   0.43 & $910^{+56}_{-116}\pm46$ &    177.25 &        1.1 &     0.9 &     310 &     250 & \nodata &  \nodata &    1.4 \\
14 & 201.493 & 2.379 &  5.25 &   1.21 &  $764^{+20}_{-19}\pm38$ &    169.00 &        4.8 &     2.1 &     900 &     400 & \nodata &  \nodata &    3.2 \\
15 & 202.892 & 2.086 & 10.59 &   0.81 &    $711^{+9}_{-9}\pm36$ &    151.75 &        7.6 &     6.0 &    1120 &     890 &      80 &     0.09 &    1.0 \\
16 & 202.608 & 2.436 &  6.63 &   1.22 &  $734^{+13}_{-16}\pm37$ &    126.00 &        5.2 &     2.4 &     680 &     320 &      20 &    0.062 &    3.5 \\
17 & 204.804 & 0.474 &  9.53 &   0.70 &  $785^{+30}_{-32}\pm39$ &    120.50 &        2.0 &     1.1 &     280 &     150 & \nodata &  \nodata &    2.9 \\
18 & 203.269 & 2.087 &  6.58 &   1.25 &  $845^{+21}_{-16}\pm42$ &    109.75 &       12.9 &    10.8 &    1940 &    1620 &    1280 &     0.79 &    1.4 \\
19 & 201.905 & 1.474 &  6.72 &   0.75 &  $809^{+40}_{-40}\pm40$ &     95.00 &        2.8 &     1.7 &     330 &     200 & \nodata &  \nodata &    2.6 \\
20 & 198.945 & 1.487 &  7.01 &   0.65 &  $742^{+21}_{-21}\pm37$ &     94.50 &        2.2 &     1.2 &     220 &     120 & \nodata &  \nodata &    2.7 \\
21 & 199.846 & 0.949 &  9.16 &   0.93 &  $685^{+18}_{-14}\pm34$ &     90.50 &        2.7 &     1.2 &     220 &     100 &      10 &      0.1 &    4.9 \\
22 & 201.621 & 1.716 &  4.72 &   0.82 &  $739^{+49}_{-57}\pm37$ &     64.50 &        5.0 &     4.1 &     340 &     280 &      10 &    0.036 &    2.3 \\
23 & 199.429 & 1.057 &  5.07 &   0.43 &  $701^{+47}_{-56}\pm35$ &     64.00 &        1.3 &     0.6 &      80 &      40 & \nodata &  \nodata &    2.5 \\
24 & 205.736 & 2.334 & 12.07 &   0.49 &  $961^{+44}_{-48}\pm48$ &     52.25 &        1.4 &     0.8 &     130 &      70 & \nodata &  \nodata &    2.5 \\
25 & 203.828 & 2.725 & 10.47 &   1.28 &  $794^{+41}_{-38}\pm40$ &     50.25 &        3.2 &     1.6 &     200 &     100 & \nodata &  \nodata &    9.3 \\
26 & 202.120 & 2.488 &  4.44 &   0.63 &  $720^{+27}_{-26}\pm36$ &     48.75 &        2.6 &     1.4 &     120 &      70 & \nodata &  \nodata &    3.1 \\
27 & 201.950 & 0.151 &  0.32 &   0.45 &  $560^{+34}_{-32}\pm28$ &     47.00 &        3.2 &     2.1 &      90 &      60 & \nodata &  \nodata &    1.7 \\
28 & 202.058 & 1.455 &  0.78 &   0.65 &  $813^{+76}_{-46}\pm41$ &     44.25 &        2.1 &     1.3 &     120 &      70 & \nodata &  \nodata &    3.8 \\
29 & 203.131 & 1.938 &  3.95 &   0.87 &  $728^{+11}_{-12}\pm36$ &     43.25 &        8.1 &     7.4 &     360 &     330 &      90 &    0.273 &    2.0 \\
30 & 202.011 & 1.534 & -0.14 &   0.56 &  $840^{+49}_{-28}\pm42$ &     36.50 &        1.7 &     1.1 &      80 &      60 & \nodata &  \nodata &    3.7 \\
31 & 203.903 & 1.669 &  6.86 &   0.61 &  $818^{+71}_{-83}\pm41$ &     32.50 &        1.9 &     1.3 &      80 &      50 & \nodata &  \nodata &    4.2 \\
32 & 202.640 & 2.000 & 10.07 &   0.85 &  $643^{+33}_{-31}\pm32$ &     29.50 &        5.7 &     3.5 &     130 &      80 & \nodata &  \nodata &    3.7 \\
\enddata
\tablecomments{Column (1): ID of $\rm ^{13}CO$ structures. \\
Column (2)-(3): Center coordinate of $\rm ^{13}CO$ structures. \\
Column (4): LSR velocity of $\rm ^{13}CO$ structures. \\
Column (5): Velocity dispersion of $\rm ^{13}CO$ structures. \\
Column (6): Distance of $\rm ^{13}CO$ structures. The first uncertainties are given by MCMC method and the second uncertainties are the 5\% systematic error due to Gaia parallax data. \\
Column (7): Area of $\rm ^{13}CO$ structures. \\
Column (8): Column density of $\rm H_2$ traced by $\mathrm{^{12}CO}$(x-factor method). \\
Column (9): Column density of $\rm H_2$ traced by $\mathrm{^{13}CO}$(LTE assumption). \\
Column (10)-(12): Molecular gas ($\rm H_2$) mass of $\rm ^{13}CO$ structures traced by $\mathrm{^{12}CO}$, $\mathrm{^{13}CO}$ and $\mathrm{C^{18}O}$ emission, the results are rounded to the nearest ten. $M\mathrm{(C^{18}O)}<10M_\odot$ are ignored, due to the uncertainties of measuring $\mathrm{C^{18}O}$ weak signal.\\
Column (13): Fraction of $M\mathrm{(C^{18}O)}$ and $M\mathrm{(^{13}CO)}$. \\
Column (14): virial parameters.}
\end{deluxetable*}
\end{longrotatetable}

\section{Discussion}\label{sec:discuss}
\subsection{$^{13}CO$ Structures Associated with NGC 2264}\label{subsec:4.1}
Star activities affect the distribution of surrounding molecular gas, leading to various gas structures in the interstellar medium. Our high-sensitivity CO observation reveals that many partial-shell and arc structures are widely presented in the whole Mon OB1 region. As a young open cluster, NGC 2264 contains an HII region and the massive O star S Mon. Both have an impact on the surrounding gas. Here we investigate how the distribution and properties of molecular gas are affected by nearby star activities in the NGC 2264 region.

\begin{figure}[ht!]
  \centering
  \includegraphics[scale=1.3]{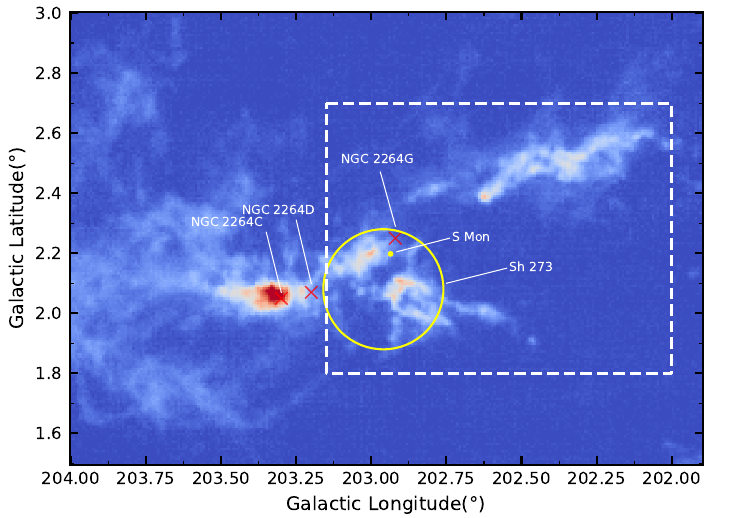}
  \caption{Integrated intensity map of $\mathrm{^{13}CO}$ emission in the velocity range of [6.5,13] $\rm km~s^{-1}$. The yellow circle indicates the HII region Sh 273, while the yellow dot indicates the O star S Mon. Crosses mark protostar NGC 2264G and star clusters, NGC 2264C and NGC 2264D. The concave structure (see text in Section \ref{subsec:4.1}) is framed out by the white dashed rectangle.
  \label{fig:10}}
\end{figure}

Based on CO channel maps, we integrate $\mathrm{^{13}CO}$ emission in the velocity range of [6.5,13]$\rm ~km~s^{-1}$ to show many partial-shell structures near NGC 2264 (Figure \ref{fig:10}). 
As shown in the white dashed rectangle of the figure, the partial-shell gas with velocity larger than $\rm 6.5~km~s^{-1}$ forms a concave structure at the west of NGC 2264. 
The HII region, Sh 273 (see the yellow circle in Figure \ref{fig:10}), is located toward the bright part of the concave structure. 

\begin{figure}[ht!]
  \centering
  \includegraphics[scale=0.7]{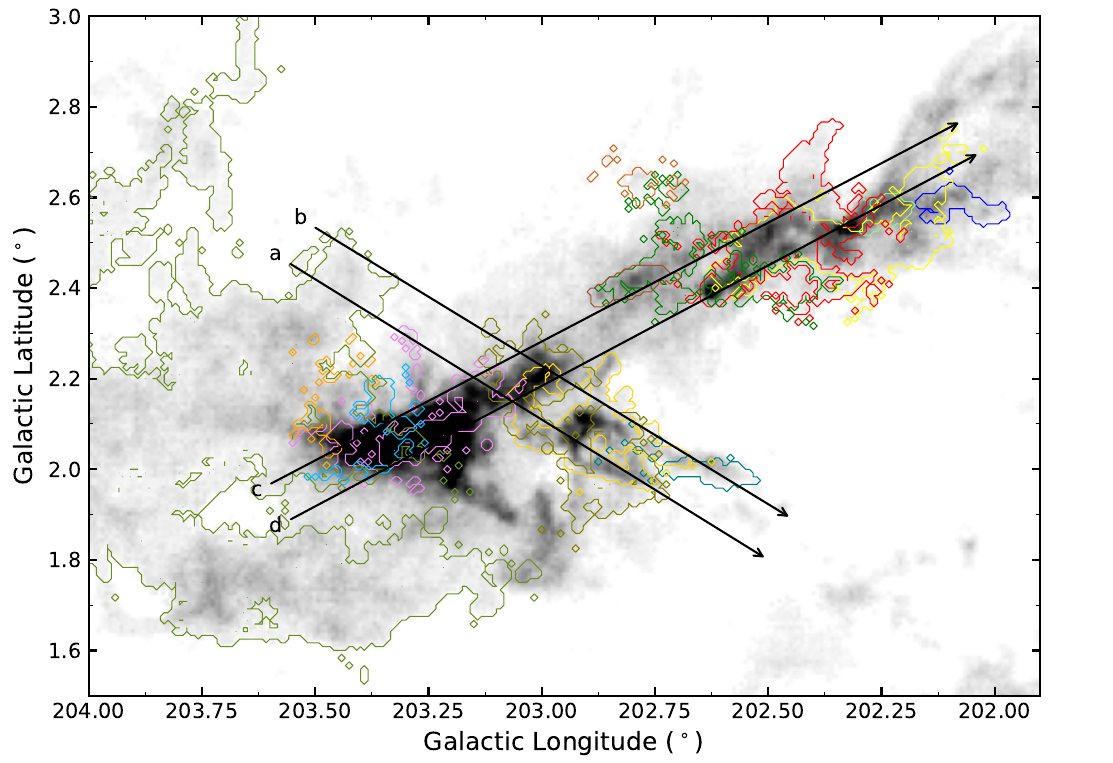}
  \vspace{-5mm}
  \gridline{\fig{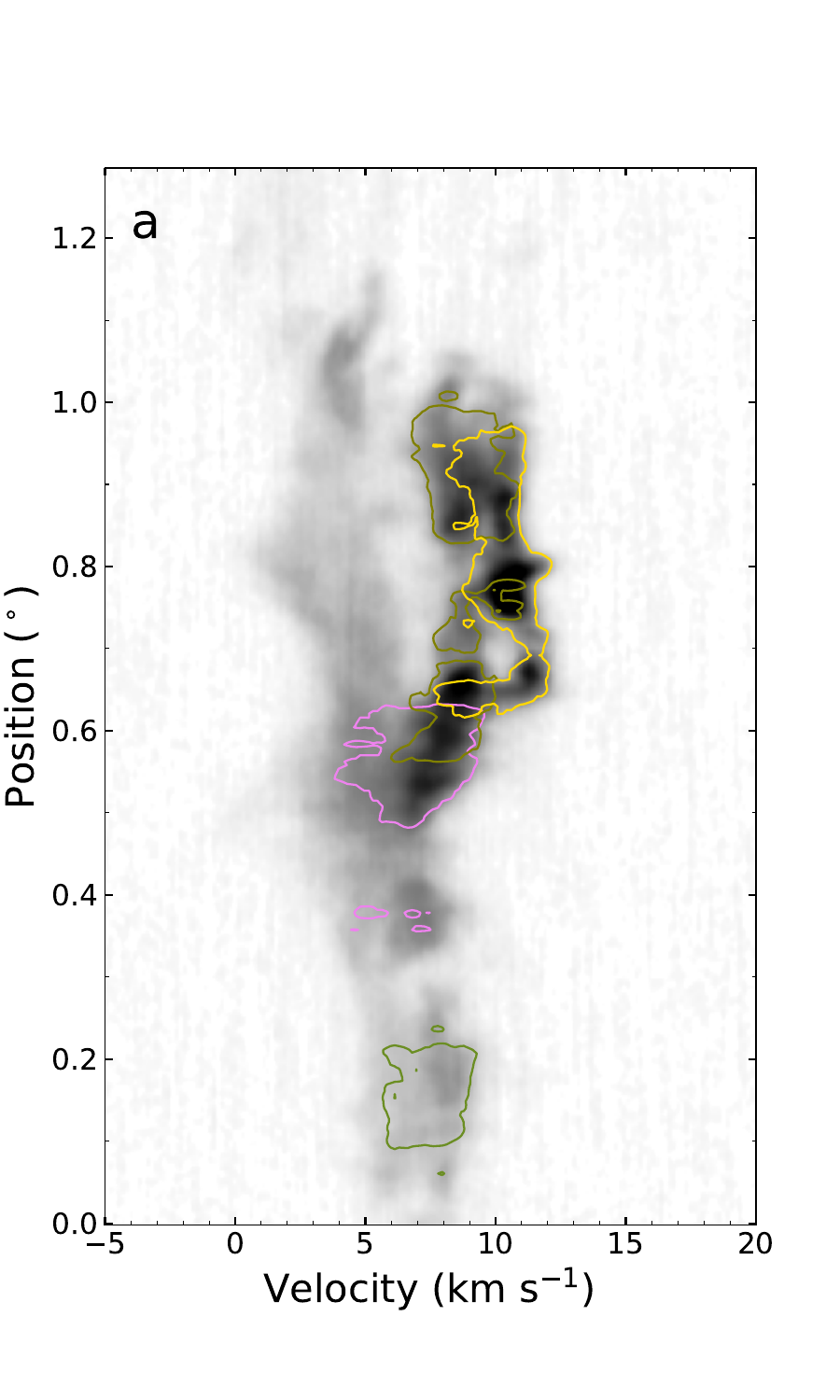}{0.24\textwidth}{}
            \fig{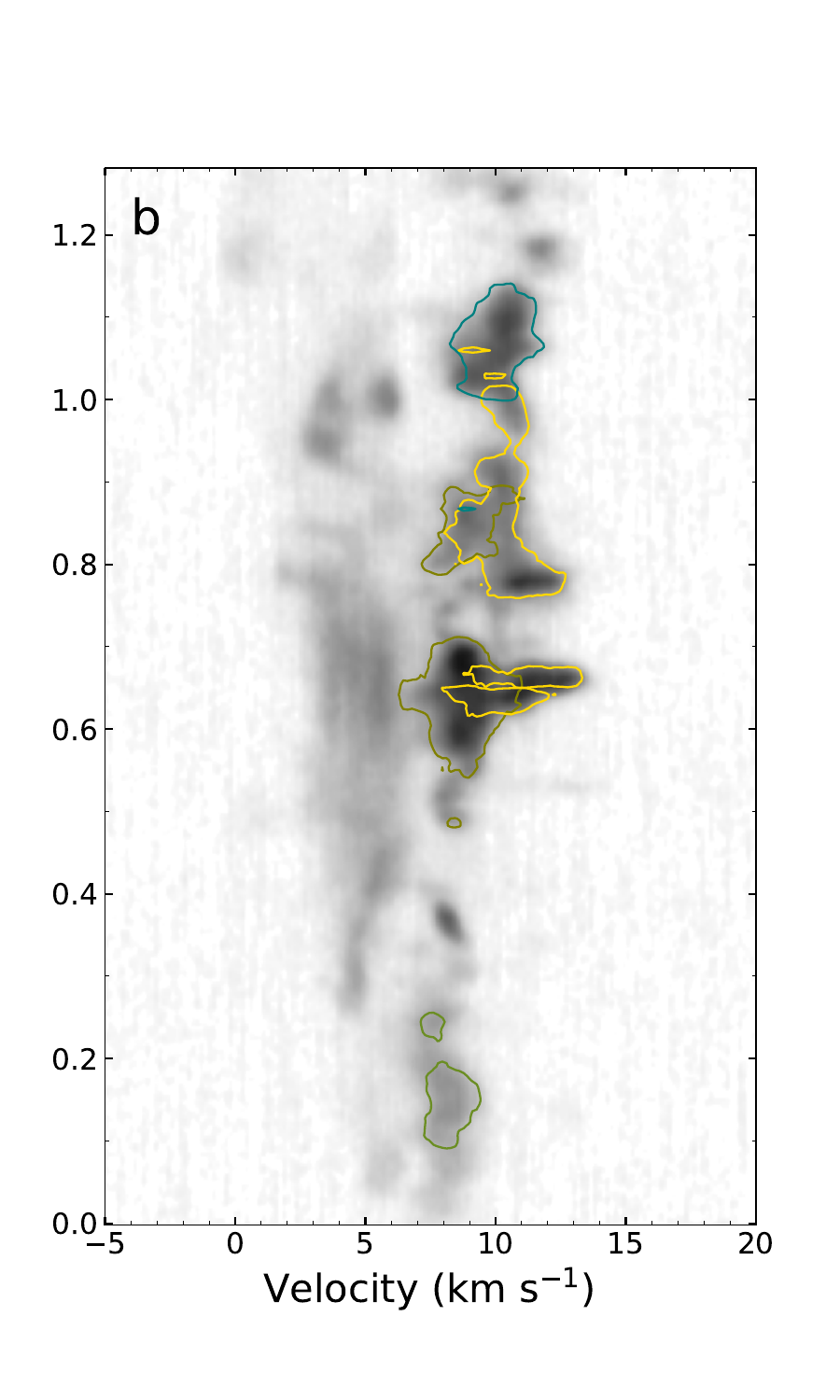}{0.24\textwidth}{}
            \fig{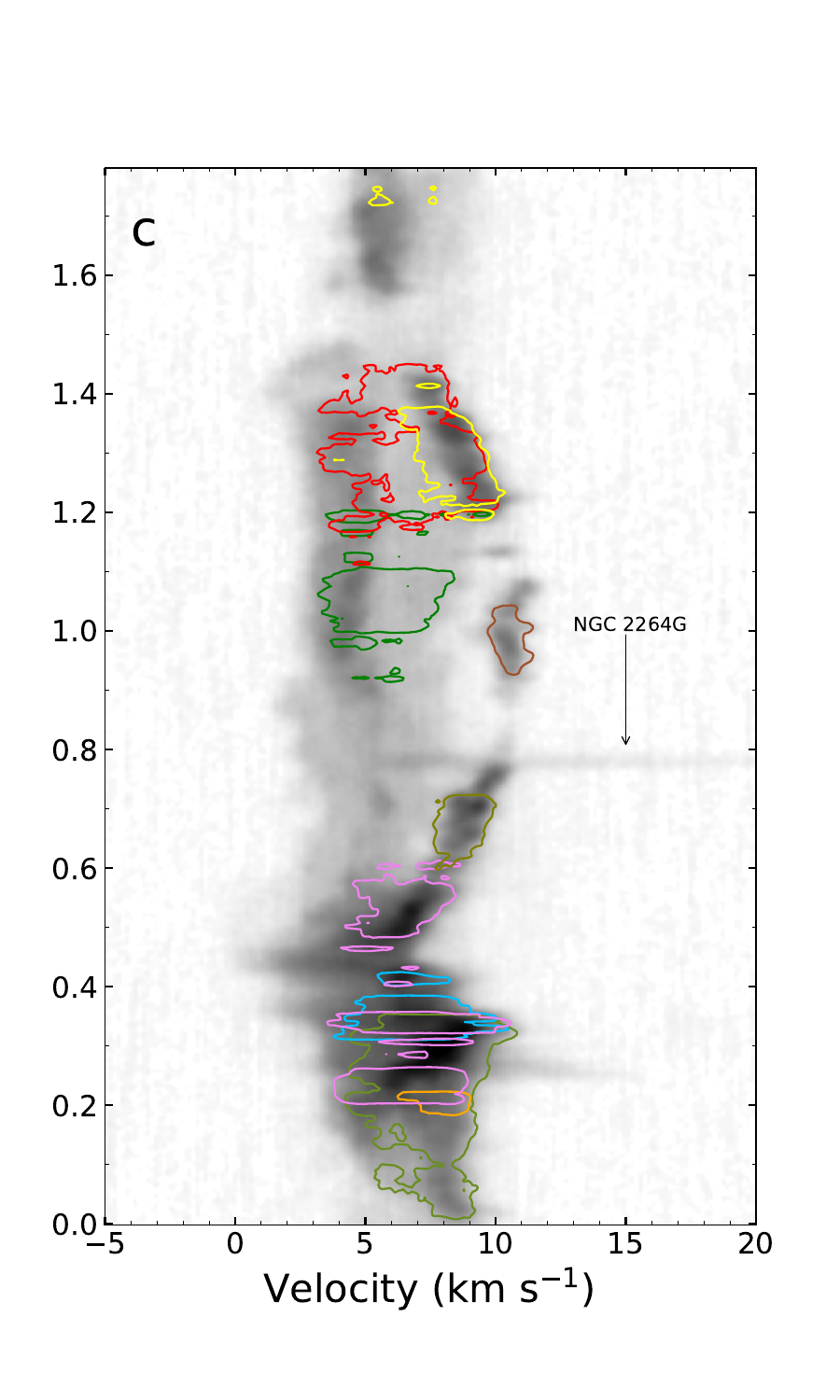}{0.24\textwidth}{}
            \fig{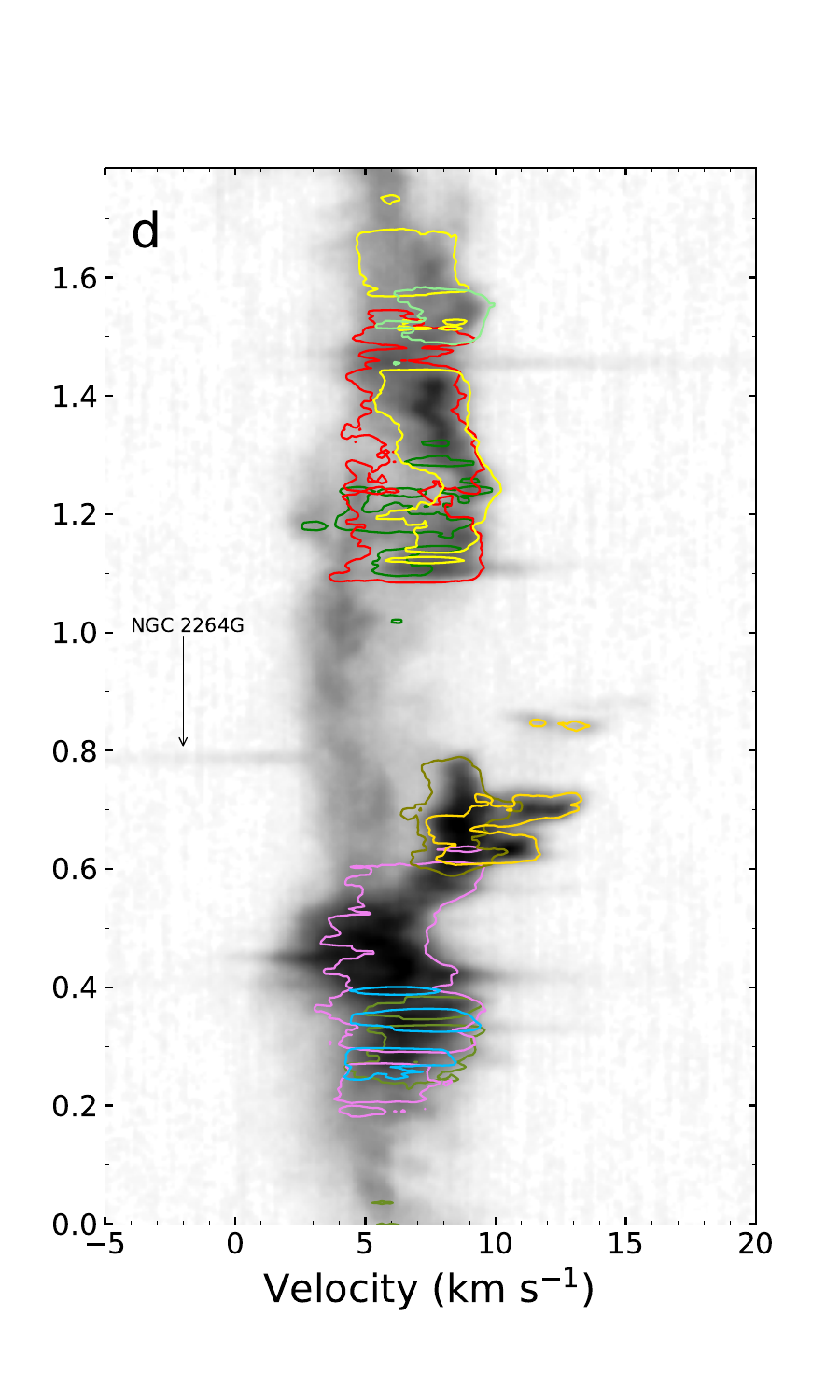}{0.24\textwidth}{}
            }
  \caption{Top panel: integrated intensity map of $\mathrm{^{13}CO}$ emission, with arrows indicating the directions of PV maps. The contours show the different $\rm ^{13}CO$ structures with velocity larger than $\rm 6.5~km~s^{-1}$. Bottom panels: PV maps along the arrows in the top panel with a width of $2\farcm5$. The background displays $\mathrm{^{12}CO}$ emission along the arrows, while contours indicate the $\mathrm{^{13}CO}$ emission. Different colors of contours represent the emission from the $\rm ^{13}CO$ structure of the same color in the top panel. The threshold of contours is $\rm 0.5~deg~K$.
  \label{fig:11}}
\end{figure}

In order to investigate the kinematic features of the molecular gas, we make position-velocity (PV) maps across the main emission of the concave structure (see Figure \ref{fig:11}). The contours with different colors in the map represent the identified $\mathrm{^{13}CO}$ structures with velocities larger than $\rm 6.5~km~s^{-1}$. The PV maps along the arrows (bottom panels of Figure \ref{fig:11}) exhibit multiple velocity components. We overlay the high-velocity components (color contours) on the background $\mathrm{^{12}CO}$ emission. Compared to the low-velocity structures, the identified high-velocity $\mathrm{^{13}CO}$ gas merges with the low-velocity components at $\rm 6.5~km~s^{-1}$. The high-velocity components in the PV maps (especially see Figure \ref{fig:11}(c)) display expanding features of the molecular gas. The expanding features in the PV maps roughly correspond to the concave structure in Figure \ref{fig:10}. We suggest that the molecular gas has been influenced by nearby massive stars, which is consistent with a previous study \citep{2004hreipurthalpha}.

Excluding the kinematic features of the large-scale concave structure ($\rm \sim15\times9~pc$), there are many narrow HVFs in these PV maps (Figure \ref{fig:11}). Notably, the most prominent HVF is exactly associated with the $\rm H_2$ jet of NGC 2264G (\citealt{o2006h2}). These features indicate localized disturbances of outflows embedded in the cloud. We propose that the clouds of the concave structure and gas associated with HVF have been perturbed by nearby star-forming activities. 

\begin{figure}[ht!]
  \centering
  \includegraphics[scale=0.7]{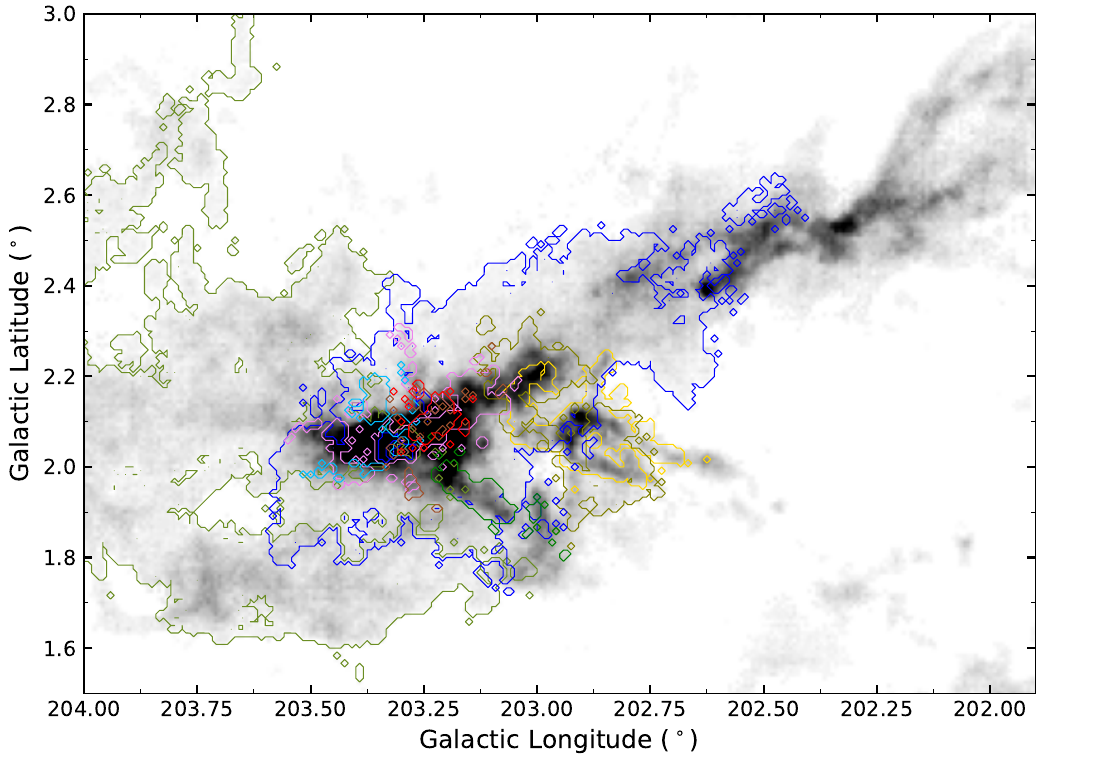}
  \caption{Same as the top panel of Figure \ref{fig:11}, but the contours show the $\rm ^{13}CO$ structures with high peak temperature or column density.
  \label{fig:12}}
\end{figure}

Additionally, molecular gas with high temperature and density is probably related to star-forming regions therein (e.g., \citealt{wang2019molecular, ma2022gas}). Our results also show that molecular gas with high temperature and column density is mainly concentrated toward the NGC 2264 region (see Figure \ref{fig:5} (a) and (d)). The identified $\mathrm{^{13}CO}$ structures with high temperature and column density are presented in Figure \ref{fig:12}. We consider that the MCs, together with high-velocity clouds in Figure \ref{fig:11}, are related to star-forming activities in NGC 2264. 

\begin{figure}[ht!]
  \centering
  \includegraphics[scale=0.7]{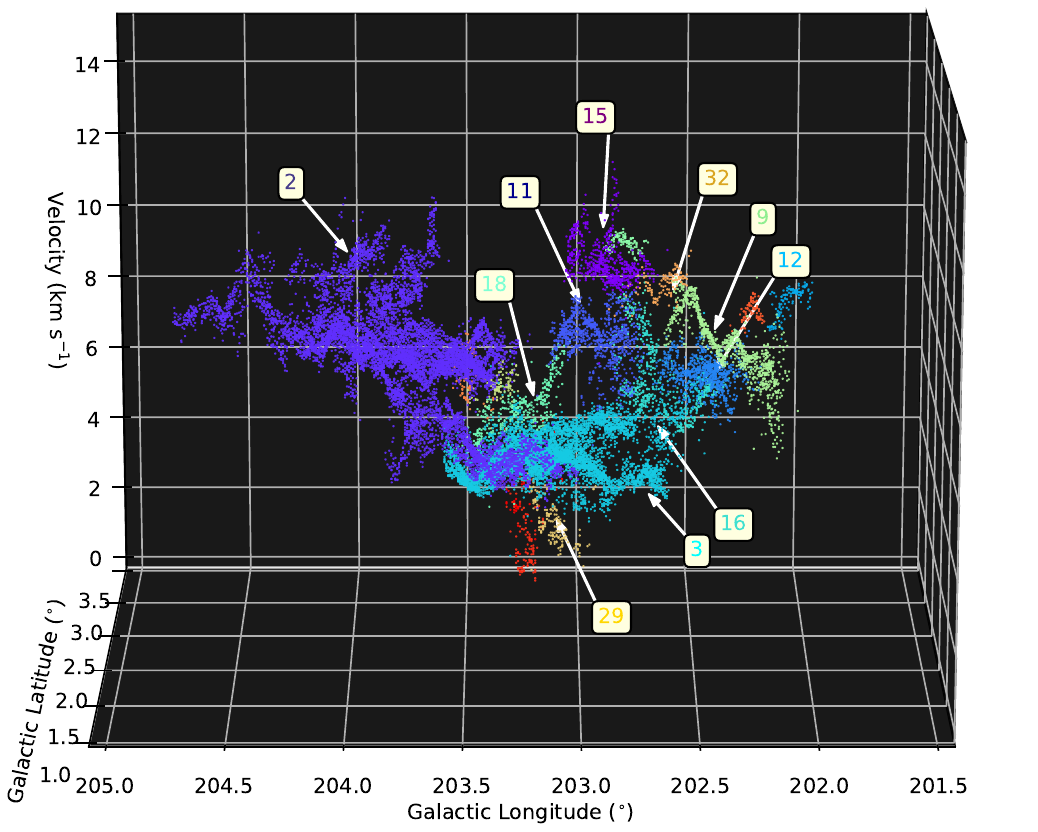}
  \caption{PPV map of 18 identified $\rm ^{13}CO$ structures related to star-forming activities in the NGC 2264 region. The $\rm ^{13}CO$ structures listed in Table \ref{table:phy-para} are labeled with corresponding structure ID. The colors of text are selected similar to those of corresponding structures. 
  \label{fig:13}}
\end{figure}

The PPV distribution of these $\rm ^{13}CO$ structures is shown in Figure \ref{fig:13}.
We then compare the differences between the $\rm ^{13}CO$ structures associated with NGC 2264 (Sample A, 18 in total) and other $\rm ^{13}CO$ structures with angular areas larger than 15 $\rm arcmin^2$ in the whole Mon OB1 region (Sample B, 60 in total).

\begin{figure*}[ht!]
  \gridline{\fig{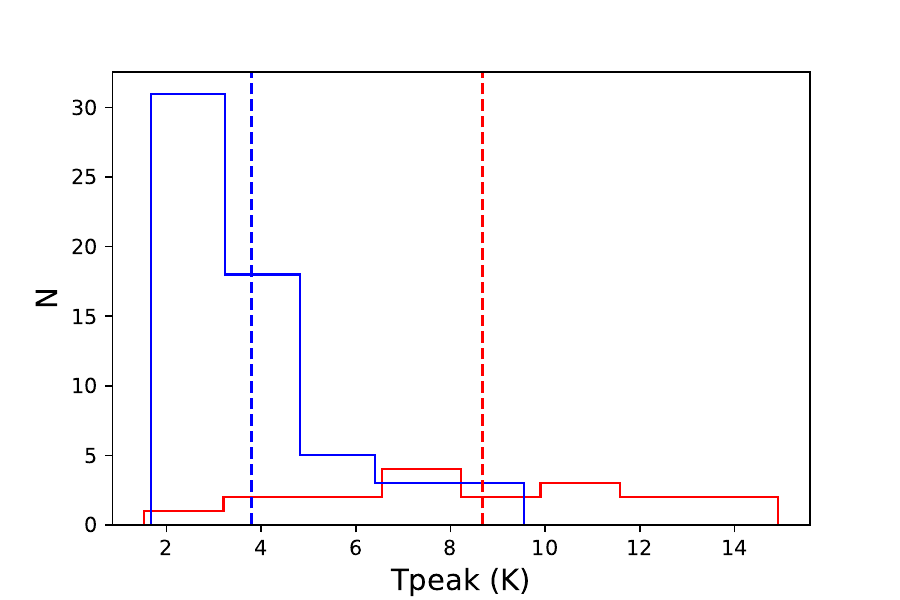}{0.4\textwidth}{}
            \fig{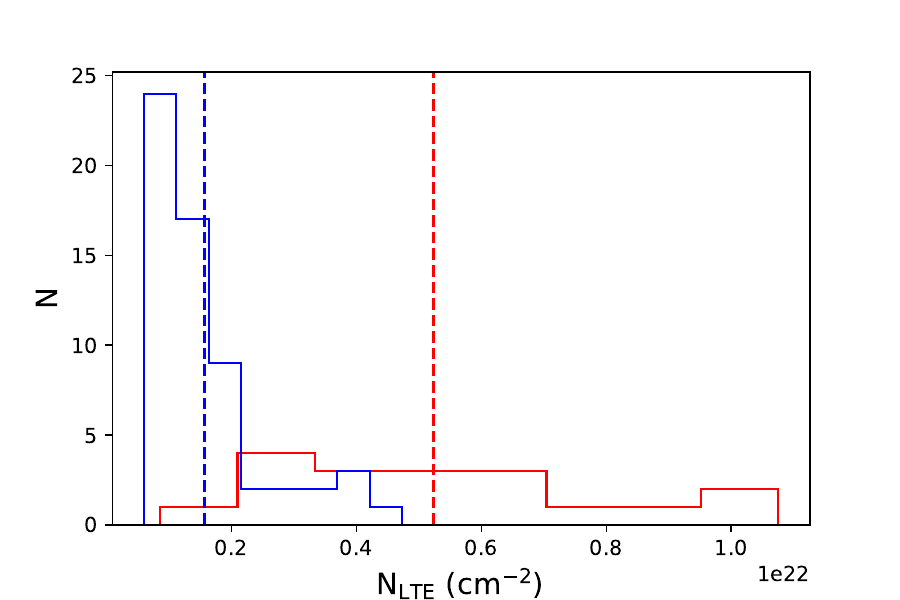}{0.4\textwidth}{}
            }
            \vspace{-15mm}
  \gridline{\fig{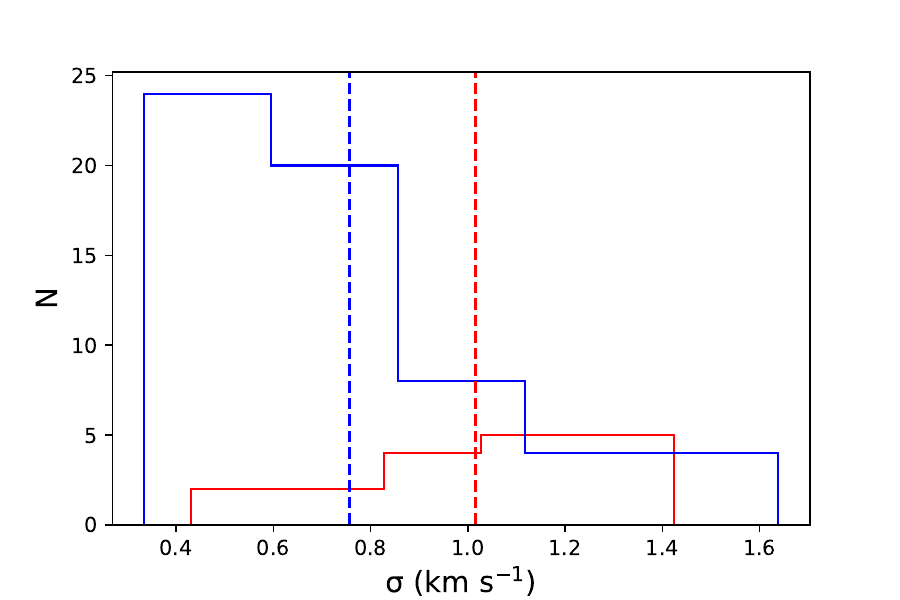}{0.4\textwidth}{}
            \fig{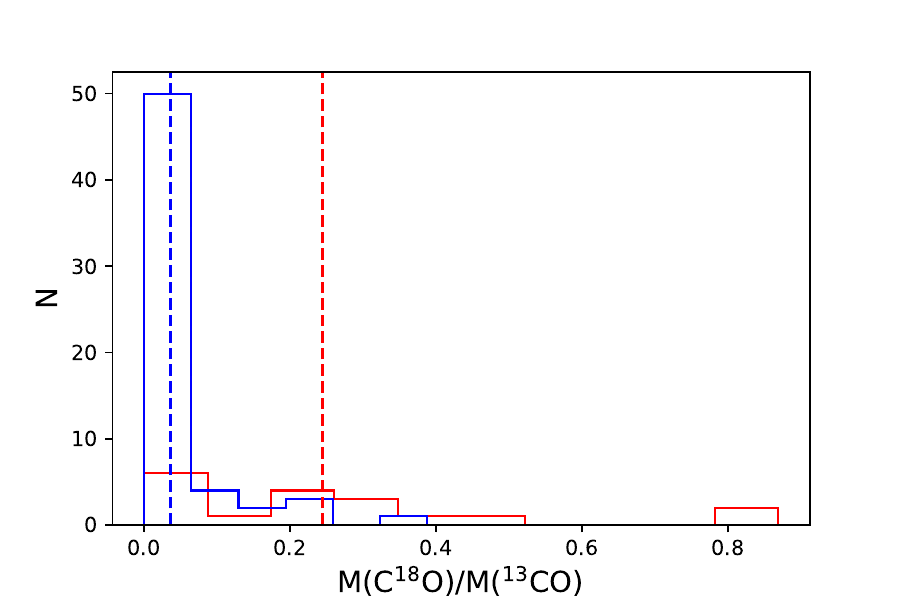}{0.4\textwidth}{}
            }
  \caption{Distribution of physical properties of $\rm ^{13}CO$ structures in sample A (red bars) and sample B (blue bars; see text in Section \ref{subsec:4.1}). The y-axis of each panel represents the number of $\rm ^{13}CO$ structures. The vertical dashed lines in each panel indicate the mean values of different samples.
  \label{fig:14}}
  \end{figure*}

Based on the statistical results of Figure \ref{fig:14}, sample A shows distinct differences from sample B, with low peak temperature $T_{\mathrm{peak}}$ and column density $N_\mathrm{LTE}$. Additionally, sample A has a larger velocity dispersion $\sigma(v)$ and a higher mass fraction of $M\mathrm{(C^{18}O)}/M\mathrm{(^{13}CO)}$ (see details in Table \ref{table:phy-para}). Some $\mathrm{^{13}CO}$ structures in sample A have very high $M\mathrm{(C^{18}O)}/M\mathrm{(^{13}CO)}$ (i.e., ID 18 in Table \ref{table:phy-para} and Figure \ref{fig:13}), which indicates a denser gas environment. The dense gas environment indicates ongoing and forthcoming star-forming activities.

The discussions show that the perturbed kinematic features of the molecular gas (expanding shells and outflows), together with gas properties (i.e., high temperature and large velocity dispersion), are related to nearby star-forming activities.

\clearpage
\subsection{Gas Cavity Associated with HD 262042}\label{subsec:4.2}
Figures \ref{fig:3} and \ref{fig:4} reveal that there are many cavity structures in the Mon OB1 region. The origin of these cavity structures is not clear. One possible explanation is that cavity structures are carved by wind bubbles of massive stars (e.g., see B0 star HD 289291 and surrounding arc-like molecular gas in Figure 19 of \citealt{su2017molecular}).

\begin{figure}[ht!]
  \centering
  \includegraphics[scale=1]{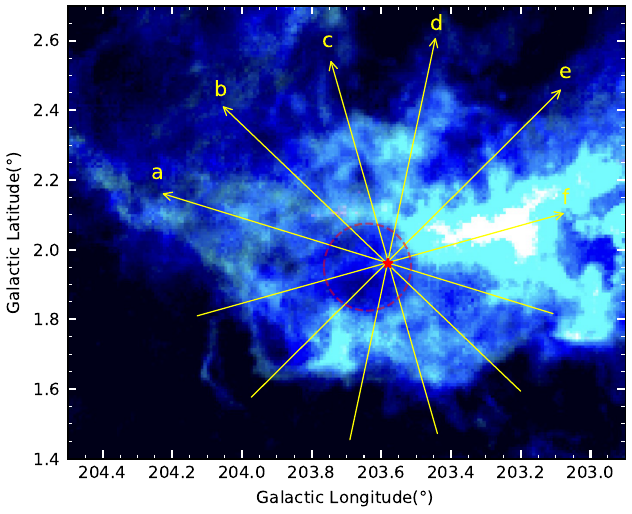}

  \caption{Integrated intensity map of $\rm ^{12}CO$ (blue), $\rm ^{13}CO$ (green), and $\rm C^{18}O$ (red) toward the cavity structure near NGC 2264. 
  The red pentagram indicates the position of B1.8 V star HD 262042 ($l=203\fdg581,~b=1\fdg960$).
  Yellow arrows indicate the PV maps across the gas cavity and the star.
  The dashed red circle indicates the manual fitting of the gas cavity.
  \label{fig:15}}
\end{figure}

We find a gas cavity centered at ($l\sim203\fdg65,~b\sim1\fdg90$) and located east of NGC 2264 (see the red dashed circle in Figure \ref{fig:15}). The cavity has a diameter of $\sim0\fdg25$, corresponding to $\sim$ 3.2 pc at a distance of 729 pc (Section \ref{subsubsec:distances}). An early-type star HD 262042 (B1.8 V) at 726 pc \citep{xu2021local} is just located in the gas cavity. 

\begin{figure*}[ht!]
  \gridline{\fig{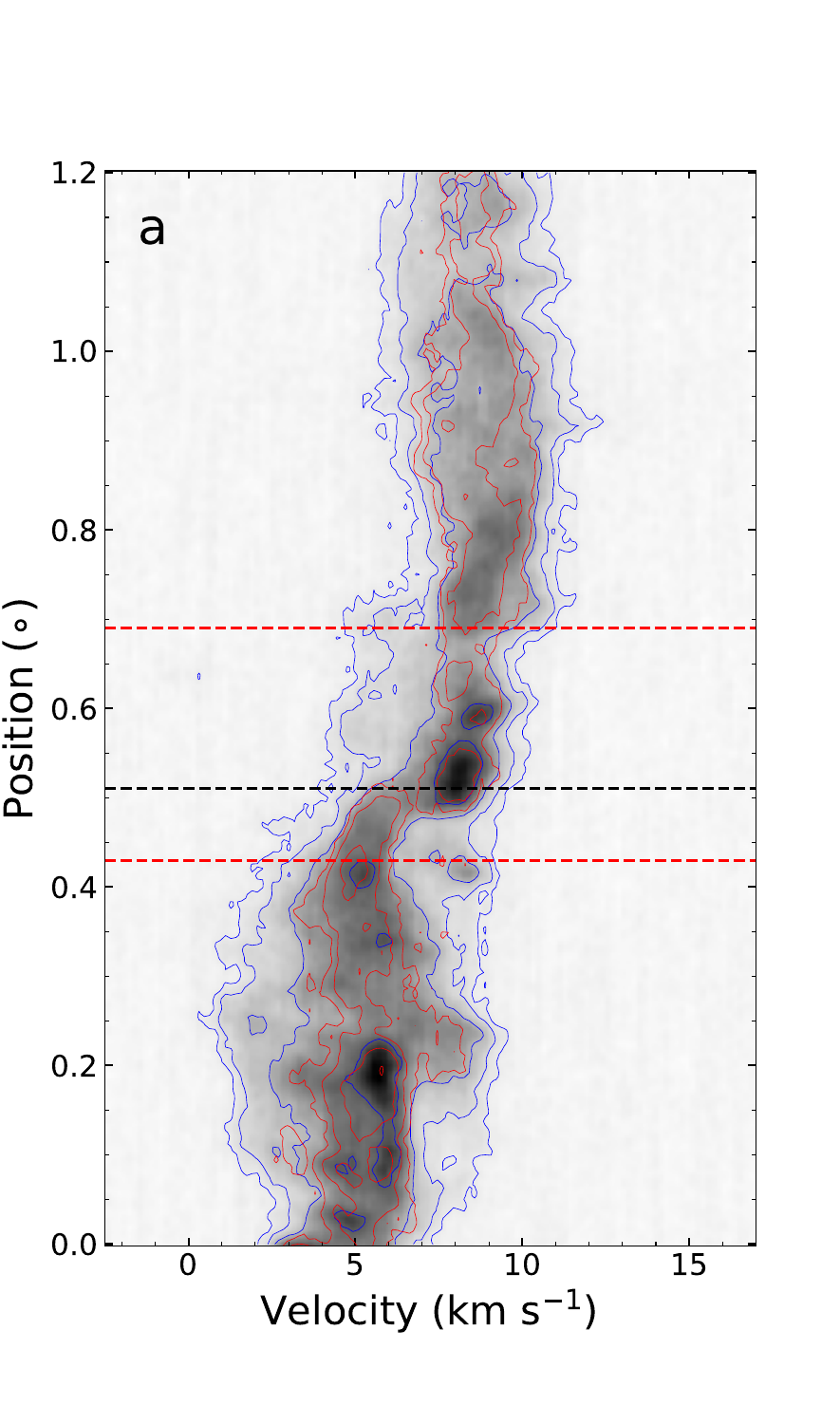}{0.32\textwidth}{}
            \fig{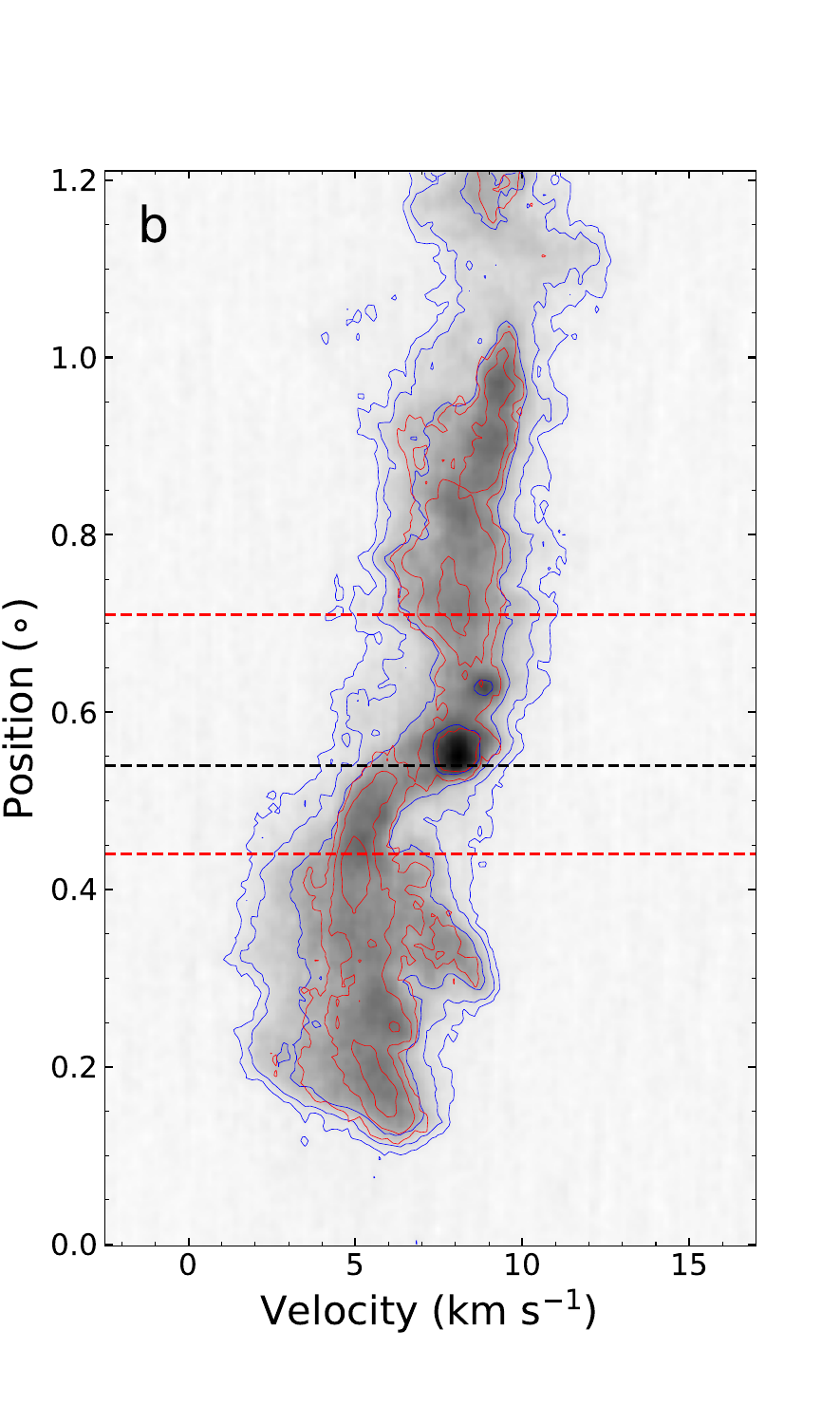}{0.32\textwidth}{}
            \fig{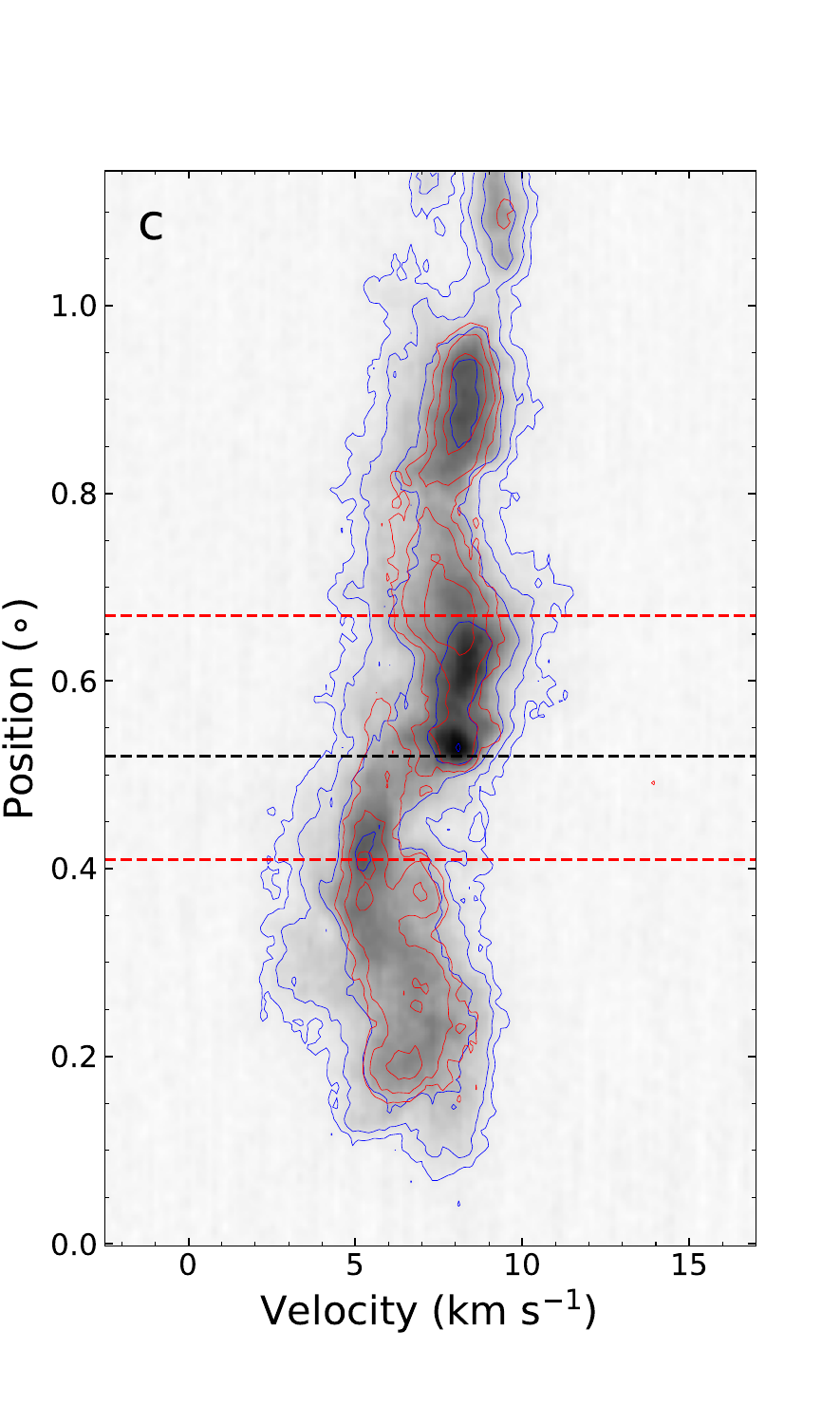}{0.32\textwidth}{}
            }
            \vspace{-20mm}
  \gridline{\fig{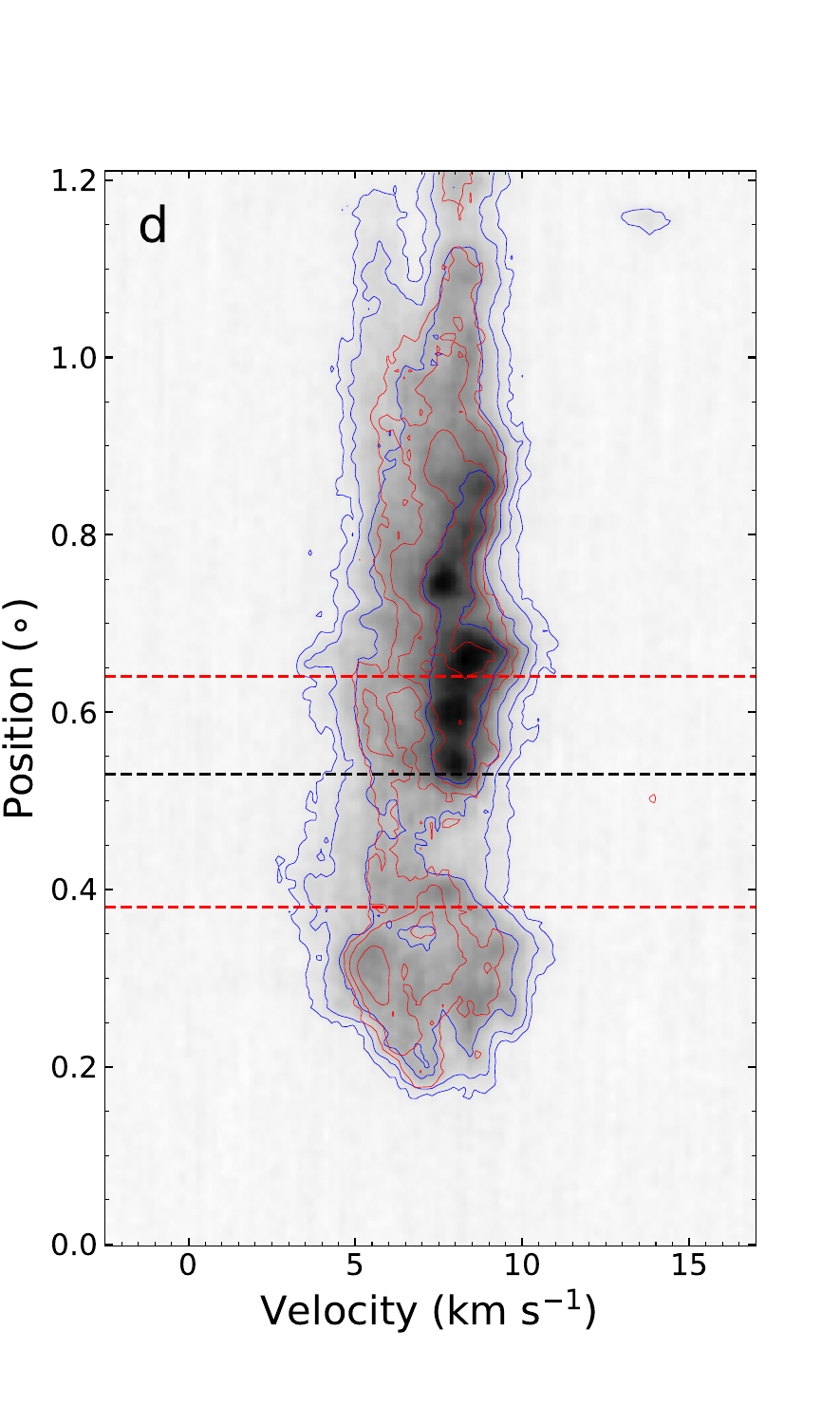}{0.32\textwidth}{}
            \fig{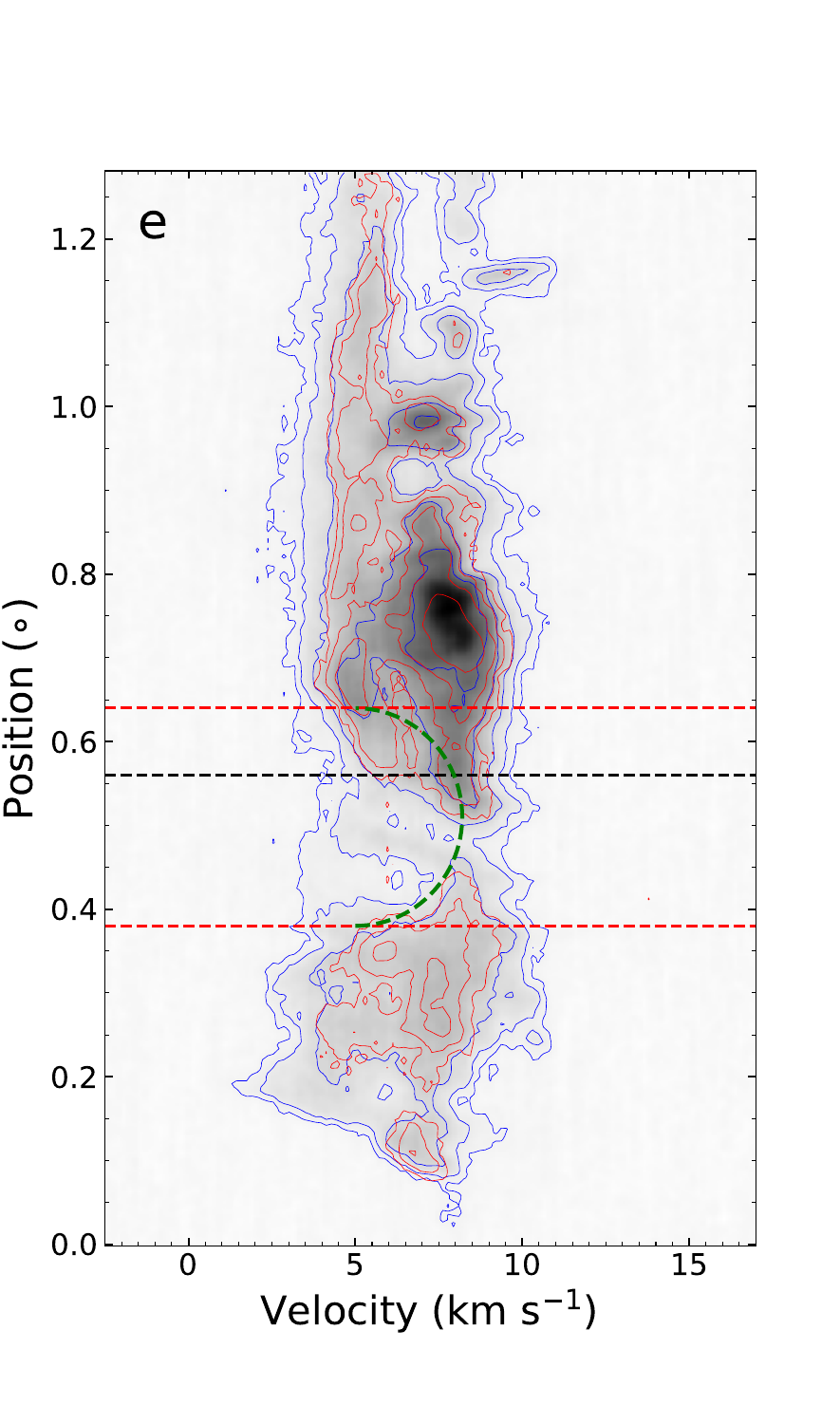}{0.32\textwidth}{}
            \fig{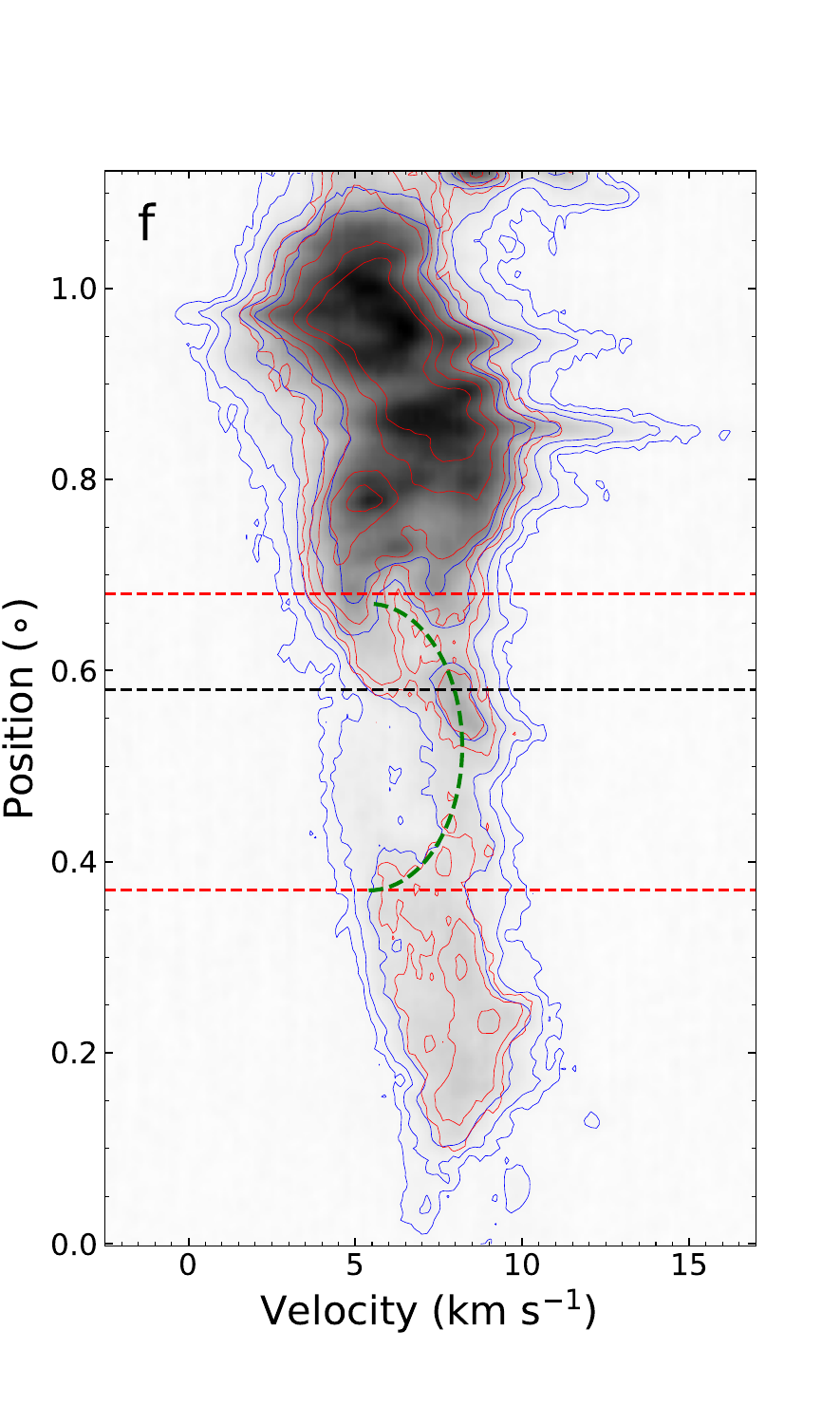}{0.32\textwidth}{}
            }
  \caption{PV maps of CO emission along the yellow arrows in Figure \ref{fig:15} with a width of $2\farcm5$. The red and blue contours indicate $\mathrm{^{13}CO}$ and $\mathrm{^{12}CO}$ emission, respectively. The area between two red dashed horizontal lines represents the extent of the cavity. The black dashed horizontal line indicates the position of the B1.8 V star HD 262042. The green dashed lines in panel (e) and panel (f) indicate manual fitting of the shell with an expanding velocity of $\sim3\mathrm{~km~s^{-1}}$.
  \label{fig:16}}
  \end{figure*}

We make PV maps across the gas cavity (yellow arrows in Figure \ref{fig:15}) to investigate the possible kinematic features of the surrounding molecular gas. Cavity-like PV patterns can be discerned in all panels of Figure \ref{fig:16}, indicating an expansion velocity of $\sim\mathrm{3~km~s^{-1}}$ (see esp. the green dashed line in Figure \ref{fig:16} (e) and (f)) for the gas structure. 
The expansion velocity is roughly comparable to the mean FWHM ($\sim\mathrm{2.1~km~s^{-1}}$) of $\rm ^{13}CO$ structures in Table \ref{table:phy-para}. We note that HD 262042 is not located at the center of the cavity but is somewhat near the dense gas wall in the west. A possible explanation is that the expanding shell structure seems to be stopped by the dense gas environment therein. This scenario is also supported by the fact that the expanding features are more prominent toward the direction away from the dense gas wall. Therefore, we suggest that the expansion features of the gas cavity are likely associated with the HD 262042.


The kinematic age of the cavity can be derived by

\begin{equation} 
  t_{\mathrm{K}} (\mathrm{Myr}) \sim\frac{R}{V_{\mathrm{exp}}},
\end{equation}
where $R$ is the radius of cavity in pc and $V_{\mathrm{exp}}$ is the expansion velocity in $\mathrm{km~s^{-1}}$ \citep{1987ApJ...317..190M}. Adopting a radius of $\sim$ 1.6 pc and an expansion velocity of $\sim$ $\mathrm{3~km~s^{-1}}$, the estimated kinematic age of the cavity is $\sim$ 0.5 Myr. The age is slightly small but is comparable to the age of nearby young clusters ($\sim$ 1 -- 2 Myr, \citealt{venuti2019deep}). 
Assuming thin-shell approximation (e.g., \citealt{Giuliani1982,Dale2009}), we expand the radius of the red ring by $1'$ ($\sim$ 10\% of the radius) to form a concentric ring to represent the gas shell structure formed by the stellar wind. Because not all molecular gas here is related to the cavity shell by considering the projection effect, we take the calculated mass of about 160 $M_\odot$ as an upper limit. Then, the kinetic energy of the gas cavity can be estimated as $E_{\mathrm{K}}(\mathrm{gas~cavity}) = \frac{1}{2}Mv^2 \lesssim 1.4\times10^{46}~\mathrm{erg}$, leading to a kinematic energy input rate of $\dot{E}_{\mathrm{K}}(\mathrm{gas~cavity})\lesssim 9.0\times10^{32}~\mathrm{erg~s^{-1}}$.

A typical B1 V star can produce an energy input rate of $\mathrm{\sim1\times10^{33}~erg~s^{-1}}$ (\citealt{chevalier1999supernova, chen2013linear}), which is comparable to the estimated value of $\dot{E}_\mathrm{K}(\mathrm{gas~cavity})$. This means that the B1.8 V star HD 262042 is likely capable of producing the gas cavity in the molecular gas environment. We find that early-type B1 V -- B2 V stars can produce wind bubbles with a maximum radius of $\sim$ 2.6 -- 5.3 pc \citep{chevalier1999supernova}. By considering that the gas cavity near NGC 2264 is still expanding with a velocity of $\sim\mathrm{3~km~s^{-1}}$, we propose that the gas cavity with a radius of $\sim$ 1.6 pc likely originates from the B1.8 V star HD 262042. In addition, \citet{Rosen2021} found that energy-driven feedback from the stellar wind of an early-type B star can drive roughly parsec-scale bubbles in dense MCs, which agrees with the gas cavity observed by CO emission here.

On the other hand, an SNR also can produce cavity structures near its natal MCs. 
But there is probably no supernova explosion toward the cluster in the past several megayears, by considering the young age of NGC 2264 and the lack of a massive star with spectral type earlier than S Mon.
Therefore, gas cavities with scales of several parsecs in this region may not be related to SNRs.
In addition, the accretion process of a protostar can input considerable energy and momentum to the surrounding gas environment by bipolar outflows. In principle, energetic activities of massive protostars can affect the distribution of the surrounding molecular gas. However, outflow activities in the NGC 2264 region are not enough to produce cavities on the scale of several parsecs \citep{buckle2012structure}. We thus suggest that the large-scale gas cavity structures in the east cloud should be related to the radiation pressure and wind from the surrounding early-type stars.

\clearpage
\section{Summary} \label{sec:summary}
We present a large-scale MWISP CO mapping observation toward Mon OB1 region of $198^{\circ}\le l \le 206^{\circ}$ and $-0\fdg5\le l \le 3\fdg5$ with velocity of $\mathrm{-5~km~s^{-1}}\le v \le\mathrm{20~km~s^{-1}}$. The main results are listed below:

(1). 
Molecular gas in the Mon OB1 region exhibits complicated hierarchical structures and morphology (e.g., shell-like, cavity-like, and filamentary structures). In the region of $l\sim[201\fdg0,203\fdg2]$, $v\sim[-5,12]~\mathrm{km~s^{-1}}$, we find a parallelogram-like PV structure between the east cloud and the west cloud (Figure \ref{fig:2}), representing a large cavity-like structure located at ($l\sim202\fdg1,~b\sim1\fdg2$) with a major axis of $\sim2\fdg6$.

(2). $\mathrm{^{12}CO}$ emission can well trace the distribution of weak and diffuse emission of molecular gas toward the region, while $\mathrm{^{13}CO}$ and $\mathrm{C^{18}O}$ lines trace denser skeleton structures within the extended $\mathrm{^{12}CO}$ emission. $\mathrm{^{13}CO}$ and $\mathrm{C^{18}O}$ emission covers 37.2\% and 3.2\% of $\mathrm{^{12}CO}$ emission area, respectively.
The overall mass of the MCs from the integrated emission (see the 2D distribution in Figure \ref{fig:5}) traced by $\mathrm{^{12}CO}$, $\mathrm{^{13}CO}$, and $\mathrm{C^{18}O}$, is $1.1\times10^5~M_\odot$, $5.4\times10^4~M_\odot$, and $9.1\times10^3~M_\odot$, respectively. 

(3). Based on coherent spatial and velocity structures of $\rm ^{13}CO$ emission, we identify a total of 263 $\rm ^{13}CO$ structures toward the Mon OB1 region. 
Molecular gas near NGC 2264 and the Mon R1 loop has a relatively high temperature and column density.
In addition, the $\mathrm{C^{18}O}$ emission is also prominent in these regions.
We also find that the regions with relatively high column density correspond to the area where multiple $\rm ^{13}CO$ structures are superimposed.

(4). Combining Gaia DR3 data, we obtain the distances of 32 $\rm ^{13}CO$ structures with angular size larger than $\sim10'$. 
Molecular gas in the Mon OB1 region consists of a GMC complex at an average distance of $729^{+45}_{-45}$ pc, which is in good agreement with previous studies. 
The flux of $\rm ^{13}CO$ structures with distance makes up 90.9\% of the total flux of identified $\rm ^{13}CO$ structures.

(5). $\rm ^{13}CO$ structures in the Mon R1 loop (in the velocity range of $\rm[-4,2]~km~s^{-1}$) show a slightly smaller distance compared to other $\rm ^{13}CO$ structures in the Mon OB1 region. 
On the other hand, the Mon OB1 GMC may span a large depth of $\sim$ 150 pc or more along the LOS, which is about twice its projection size.
CO emission with $v\sim$ 20 -- 35$~\mathrm{km~s^{-1}}$ is located at a distance of $>$ 2 kpc and is not related to the Mon OB1 GMC.


(6). Most of the $\rm ^{13}CO$ structures in Table \ref{table:phy-para} are in an approximate virial equilibrium state. The total mass of these identified $\rm ^{13}CO$ structures traced by $\mathrm{^{12}CO}$, $\mathrm{^{13}CO}$, and $\mathrm{C^{18}O}$ is $6.1\times10^4~M_\odot$, $4.3~\times10^4~M_\odot$, and $8.4\times10^3~M_\odot$, respectively. 
The molecular gas traced by $\mathrm{^{12}CO}$ but no $\mathrm{^{13}CO}$ accounts for about 44.5\% of the total gas mass.
The total mass from the identified $\rm ^{13}CO$ structures here is somewhat less than that from the whole 2D integrated emission.
In addition, there are significant differences in dense gas fraction between the east cloud (12.4\%) and the west cloud (3.3\%), although they have a similar total mass ($M_{\mathrm{east}}/M_{\mathrm{west}}\approx1.2$) and angular area ($A_{\mathrm{east}}/A_{\mathrm{west}}\approx0.84$).

(7). For identified $\rm ^{13}CO$ structures associated with NGC 2264, the CO gas shows a higher mean value in $T_{\mathrm{MB,12_{peak}}}$, $N_{\mathrm{H_2}}(\mathrm{LTE},13)$, velocity dispersion, and $M\mathrm{(C^{18}O)}/M\mathrm{(^{13}CO)}$ compared to the remaining $\rm ^{13}CO$ structures in the Mon OB1 region. 
The dense gas environment indicates ongoing and forthcoming star-forming activities.
On the other hand, star-forming activities in NGC 2264 have a large impact on the physical properties (e.g, larger velocity dispersion and higher temperature) of surrounding molecular gas. 
We also discuss the connection between star-forming activities and the distribution/kinematics of the surrounding molecular gas.

\begin{acknowledgments}
  We would like to thank the anonymous referee for the valuable comments and suggestions that have largely improved the manuscript.  
  This research made use of the data from the Milky Way Imaging Scroll Painting (MWISP) project, which is a multiline survey in $\rm ^{12}CO$/$\rm ^{13}CO$/$\rm C^{18}O$ along the northern Galactic plane with the PMO-13.7 m telescope. We are grateful to all the members of the MWISP working group, particularly the staff members at the PMO-13.7 m telescope, for their long-term support. MWISP was sponsored by the National Key R\&D Program of China with grants, 2023YFA1608000 and 2017YFA0402701 and the CAS Key Research Program of Frontier Sciences with grant QYZDJ-SSW-SLH047. This work is supported by NSFC grant 12173090. X.C. acknowledges the support by the CAS International Cooperation Program (grant No. 114332KYSB20190009). Y.S. acknowledges support from the Youth Innovation Promotion Association, CAS (Y2022085), and the “Light of West China” Program (No. xbzg-zdsys-202212). This work also made use of data from the European Space Agency (ESA) mission Gaia (https://www.cosmos.esa.int/gaia), processed by the Gaia Data Processing and Analysis Consortium (DPAC, https://www.cosmos.esa.int/web/gaia/dpac/consortium). Funding for the DPAC has been provided by national institutions, in particular the institutions participating in the Gaia Multilateral Agreement.
\end{acknowledgments}

%

\vspace{5mm}
\facilities{PMO:DLH, Gaia}




\bibliography{sample631}{}

\begin{thebibliography}{}
\expandafter\ifx\csname natexlab\endcsname\relax\def\natexlab#1{#1}\fi
\providecommand{\url}[1]{\href{#1}{#1}}
\providecommand{\dodoi}[1]{doi:~\href{http://doi.org/#1}{\nolinkurl{#1}}}
\providecommand{\doeprint}[1]{\href{http://ascl.net/#1}{\nolinkurl{http://ascl.net/#1}}}
\providecommand{\doarXiv}[1]{\href{https://arxiv.org/abs/#1}{\nolinkurl{https://arxiv.org/abs/#1}}}

\bibitem[{{Alina} {et~al.}(2022){Alina}, {Montillaud}, {Hu}, {Lazarian},
  {Ristorcelli}, {Abdikamalov}, {Sagynbayeva}, {Juvela}, {Liu}, \&
  {Carri{\`e}re}}]{alina2022large}
{Alina}, D., {Montillaud}, J., {Hu}, Y., {et~al.} 2022, \aap, 658, A90,
  \dodoi{10.1051/0004-6361/202039065}

\bibitem[{{Allen}(1972)}]{allen1972infrared}
{Allen}, D.~A. 1972, \apjl, 172, L55, \dodoi{10.1086/180890}

\bibitem[{{Anderson} {et~al.}(2014){Anderson}, {Bania}, {Balser}, {Cunningham},
  {Wenger}, {Johnstone}, \& {Armentrout}}]{anderson2014wise}
{Anderson}, L.~D., {Bania}, T.~M., {Balser}, D.~S., {et~al.} 2014, \apjs, 212,
  1, \dodoi{10.1088/0067-0049/212/1/1}

\bibitem[{{Andrae} {et~al.}(2023){Andrae}, {Fouesneau}, {Sordo},
  {Bailer-Jones}, {Dharmawardena}, {Rybizki}, {De Angeli}, {Lindstr{\o}m},
  {Marshall}, {Drimmel}, {Korn}, {Soubiran}, {Brouillet}, {Casamiquela}, {Rix},
  {Abreu Aramburu}, {{\'A}lvarez}, {Bakker}, {Bellas-Velidis}, {Bijaoui},
  {Brugaletta}, {Burlacu}, {Carballo}, {Chaoul}, {Chiavassa}, {Contursi},
  {Cooper}, {Creevey}, {Dafonte}, {Dapergolas}, {de Laverny}, {Delchambre},
  {Demouchy}, {Edvardsson}, {Fr{\'e}mat}, {Garabato}, {Garc{\'\i}a-Lario},
  {Garc{\'\i}a-Torres}, {Gavel}, {Gomez}, {Gonz{\'a}lez-Santamar{\'\i}a},
  {Hatzidimitriou}, {Heiter}, {Jean-Antoine Piccolo}, {Kontizas}, {Kordopatis},
  {Lanzafame}, {Lebreton}, {Licata}, {Livanou}, {Lobel}, {Lorca}, {Magdaleno
  Romeo}, {Manteiga}, {Marocco}, {Mary}, {Nicolas}, {Ordenovic}, {Pailler},
  {Palicio}, {Pallas-Quintela}, {Panem}, {Pichon}, {Poggio}, {Recio-Blanco},
  {Riclet}, {Robin}, {Santove{\~n}a}, {Sarro}, {Schultheis}, {Segol},
  {Silvelo}, {Slezak}, {Smart}, {S{\"u}veges}, {Th{\'e}venin}, {Torralba
  Elipe}, {Ulla}, {Utrilla}, {Vallenari}, {van Dillen}, {Zhao}, \&
  {Zorec}}]{2023A&A...674A..27A}
{Andrae}, R., {Fouesneau}, M., {Sordo}, R., {et~al.} 2023, \aap, 674, A27,
  \dodoi{10.1051/0004-6361/202243462}

\bibitem[{{Arun} {et~al.}(2019){Arun}, {Mathew}, {Manoj}, {Ujjwal}, {Kartha},
  {Viswanath}, {Narang}, \& {Paul}}]{arun2019mass}
{Arun}, R., {Mathew}, B., {Manoj}, P., {et~al.} 2019, \aj, 157, 159,
  \dodoi{10.3847/1538-3881/ab0ca1}

\bibitem[{{Bailer-Jones}(2015)}]{bailer2015est}
{Bailer-Jones}, C. A.~L. 2015, \pasp, 127, 994, \dodoi{10.1086/683116}

\bibitem[{{Bhadari} {et~al.}(2020){Bhadari}, {Dewangan}, {Pirogov}, \&
  {Ojha}}]{bhadari2020star}
{Bhadari}, N.~K., {Dewangan}, L.~K., {Pirogov}, L.~E., \& {Ojha}, D.~K. 2020,
  \apj, 899, 167, \dodoi{10.3847/1538-4357/aba2c6}

\bibitem[{{Bolatto} {et~al.}(2013){Bolatto}, {Wolfire}, \&
  {Leroy}}]{bolatto2013co}
{Bolatto}, A.~D., {Wolfire}, M., \& {Leroy}, A.~K. 2013, \araa, 51, 207,
  \dodoi{10.1146/annurev-astro-082812-140944}

\bibitem[{{Bourke} {et~al.}(1997){Bourke}, {Garay}, {Lehtinen},
  {K{\"o}hnenkamp}, {Launhardt}, {Nyman}, {May}, {Robinson}, \&
  {Hyland}}]{Bourke_1997}
{Bourke}, T.~L., {Garay}, G., {Lehtinen}, K.~K., {et~al.} 1997, \apj, 476, 781,
  \dodoi{10.1086/303642}

\bibitem[{{Buckle} {et~al.}(2012){Buckle}, {Richer}, \&
  {Davis}}]{buckle2012structure}
{Buckle}, J.~V., {Richer}, J.~S., \& {Davis}, C.~J. 2012, \mnras, 423, 1127,
  \dodoi{10.1111/j.1365-2966.2012.20941.x}

\bibitem[{{Burton} {et~al.}(2013){Burton}, {Braiding}, {Glueck}, {Goldsmith},
  {Hawkes}, {Hollenbach}, {Kulesa}, {Martin}, {Pineda}, {Rowell}, {Simon},
  {Stark}, {Stutzki}, {Tothill}, {Urquhart}, {Walker}, {Walsh}, \&
  {Wolfire}}]{burton2013mopra}
{Burton}, M.~G., {Braiding}, C., {Glueck}, C., {et~al.} 2013, \pasa, 30, e044,
  \dodoi{10.1017/pasa.2013.22}

\bibitem[{{Caratti o Garatti} {et~al.}(2006){Caratti o Garatti}, {Giannini},
  {Nisini}, \& {Lorenzetti}}]{o2006h2}
{Caratti o Garatti}, A., {Giannini}, T., {Nisini}, B., \& {Lorenzetti}, D.
  2006, \aap, 449, 1077, \dodoi{10.1051/0004-6361:20054313}

\bibitem[{{Chen} {et~al.}(2020){Chen}, {Li}, {Yuan}, {Huang}, {Tian}, {Wang},
  {Zhang}, {Wang}, \& {Liu}}]{chen2020large}
{Chen}, B.~Q., {Li}, G.~X., {Yuan}, H.~B., {et~al.} 2020, \mnras, 493, 351,
  \dodoi{10.1093/mnras/staa235}

\bibitem[{{Chen} {et~al.}(2013){Chen}, {Zhou}, \& {Chu}}]{chen2013linear}
{Chen}, Y., {Zhou}, P., \& {Chu}, Y.-H. 2013, \apjl, 769, L16,
  \dodoi{10.1088/2041-8205/769/1/L16}

\bibitem[{{Chevalier}(1999)}]{chevalier1999supernova}
{Chevalier}, R.~A. 1999, \apj, 511, 798, \dodoi{10.1086/306710}

\bibitem[{{Dahm}(2008)}]{2008hsf1.book..966D}
{Dahm}, S.~E. 2008, in Handbook of Star Forming Regions, Volume I, ed.
  B.~{Reipurth}, Vol.~4, 966, \dodoi{10.48550/arXiv.0808.3835}

\bibitem[{{Dale} {et~al.}(2009){Dale}, {W{\"u}nsch}, {Whitworth}, \&
  {Palou{\v{s}}}}]{Dale2009}
{Dale}, J.~E., {W{\"u}nsch}, R., {Whitworth}, A., \& {Palou{\v{s}}}, J. 2009,
  \mnras, 398, 1537, \dodoi{10.1111/j.1365-2966.2009.15213.x}

\bibitem[{{Dame} {et~al.}(2001){Dame}, {Hartmann}, \&
  {Thaddeus}}]{dame2001milky}
{Dame}, T.~M., {Hartmann}, D., \& {Thaddeus}, P. 2001, \apj, 547, 792,
  \dodoi{10.1086/318388}

\bibitem[{{de Zeeuw} {et~al.}(1999){de Zeeuw}, {Hoogerwerf}, {de Bruijne},
  {Brown}, \& {Blaauw}}]{de1999hipparcos}
{de Zeeuw}, P.~T., {Hoogerwerf}, R., {de Bruijne}, J.~H.~J., {Brown}, A.~G.~A.,
  \& {Blaauw}, A. 1999, \aj, 117, 354, \dodoi{10.1086/300682}

\bibitem[{{Edenhofer} {et~al.}(2023){Edenhofer}, {Zucker}, {Frank}, {Saydjari},
  {Speagle}, {Finkbeiner}, \& {En{\ss}lin}}]{Edenhofer2023}
{Edenhofer}, G., {Zucker}, C., {Frank}, P., {et~al.} 2023, arXiv e-prints,
  arXiv:2308.01295, \dodoi{10.48550/arXiv.2308.01295}

\bibitem[{{Flaccomio} {et~al.}(2023){Flaccomio}, {Micela}, {Peres},
  {Sciortino}, {Salvaggio}, {Prisinzano}, {Guarcello}, {Venuti}, {Bonito}, \&
  {Pillitteri}}]{flaccomio2023spatial}
{Flaccomio}, E., {Micela}, G., {Peres}, G., {et~al.} 2023, \aap, 670, A37,
  \dodoi{10.1051/0004-6361/202244872}

\bibitem[{{Foreman-Mackey} {et~al.}(2013){Foreman-Mackey}, {Hogg}, {Lang}, \&
  {Goodman}}]{Foreman2013emcee}
{Foreman-Mackey}, D., {Hogg}, D.~W., {Lang}, D., \& {Goodman}, J. 2013, \pasp,
  125, 306, \dodoi{10.1086/670067}

\bibitem[{{Frerking} {et~al.}(1982){Frerking}, {Langer}, \&
  {Wilson}}]{frerking1982relationship}
{Frerking}, M.~A., {Langer}, W.~D., \& {Wilson}, R.~W. 1982, \apj, 262, 590,
  \dodoi{10.1086/160451}

\bibitem[{{Gaia Collaboration} {et~al.}(2023){Gaia Collaboration}, {Vallenari},
  {Brown}, {Prusti}, {de Bruijne}, {Arenou}, {Babusiaux}, {Biermann},
  {Creevey}, {Ducourant}, {Evans}, {Eyer}, {Guerra}, {Hutton}, {Jordi},
  {Klioner}, {Lammers}, {Lindegren}, {Luri}, {Mignard}, {Panem}, {Pourbaix},
  {Randich}, {Sartoretti}, {Soubiran}, {Tanga}, {Walton}, {Bailer-Jones},
  {Bastian}, {Drimmel}, {Jansen}, {Katz}, {Lattanzi}, {van Leeuwen}, {Bakker},
  {Cacciari}, {Casta{\~n}eda}, {De Angeli}, {Fabricius}, {Fouesneau},
  {Fr{\'e}mat}, {Galluccio}, {Guerrier}, {Heiter}, {Masana}, {Messineo},
  {Mowlavi}, {Nicolas}, {Nienartowicz}, {Pailler}, {Panuzzo}, {Riclet}, {Roux},
  {Seabroke}, {Sordo}, {Th{\'e}venin}, {Gracia-Abril}, {Portell}, {Teyssier},
  {Altmann}, {Andrae}, {Audard}, {Bellas-Velidis}, {Benson}, {Berthier},
  {Blomme}, {Burgess}, {Busonero}, {Busso}, {C{\'a}novas}, {Carry}, {Cellino},
  {Cheek}, {Clementini}, {Damerdji}, {Davidson}, {de Teodoro}, {Nu{\~n}ez
  Campos}, {Delchambre}, {Dell'Oro}, {Esquej}, {Fern{\'a}ndez-Hern{\'a}ndez},
  {Fraile}, {Garabato}, {Garc{\'\i}a-Lario}, {Gosset}, {Haigron}, {Halbwachs},
  {Hambly}, {Harrison}, {Hern{\'a}ndez}, {Hestroffer}, {Hodgkin}, {Holl},
  {Jan{\ss}en}, {Jevardat de Fombelle}, {Jordan}, {Krone-Martins}, {Lanzafame},
  {L{\"o}ffler}, {Marchal}, {Marrese}, {Moitinho}, {Muinonen}, {Osborne},
  {Pancino}, {Pauwels}, {Recio-Blanco}, {Reyl{\'e}}, {Riello}, {Rimoldini},
  {Roegiers}, {Rybizki}, {Sarro}, {Siopis}, {Smith}, {Sozzetti}, {Utrilla},
  {van Leeuwen}, {Abbas}, {{\'A}brah{\'a}m}, {Abreu Aramburu}, {Aerts},
  {Aguado}, {Ajaj}, {Aldea-Montero}, {Altavilla}, {{\'A}lvarez}, {Alves},
  {Anders}, {Anderson}, {Anglada Varela}, {Antoja}, {Baines}, {Baker},
  {Balaguer-N{\'u}{\~n}ez}, {Balbinot}, {Balog}, {Barache}, {Barbato},
  {Barros}, {Barstow}, {Bartolom{\'e}}, {Bassilana}, {Bauchet}, {Becciani},
  {Bellazzini}, {Berihuete}, {Bernet}, {Bertone}, {Bianchi}, {Binnenfeld},
  {Blanco-Cuaresma}, {Blazere}, {Boch}, {Bombrun}, {Bossini}, {Bouquillon},
  {Bragaglia}, {Bramante}, {Breedt}, {Bressan}, {Brouillet}, {Brugaletta},
  {Bucciarelli}, {Burlacu}, {Butkevich}, {Buzzi}, {Caffau}, {Cancelliere},
  {Cantat-Gaudin}, {Carballo}, {Carlucci}, {Carnerero}, {Carrasco},
  {Casamiquela}, {Castellani}, {Castro-Ginard}, {Chaoul}, {Charlot}, {Chemin},
  {Chiaramida}, {Chiavassa}, {Chornay}, {Comoretto}, {Contursi}, {Cooper},
  {Cornez}, {Cowell}, {Crifo}, {Cropper}, {Crosta}, {Crowley}, {Dafonte},
  {Dapergolas}, {David}, {David}, {de Laverny}, {De Luise}, {De March}, {De
  Ridder}, {de Souza}, {de Torres}, {del Peloso}, {del Pozo}, {Delbo},
  {Delgado}, {Delisle}, {Demouchy}, {Dharmawardena}, {Di Matteo}, {Diakite},
  {Diener}, {Distefano}, {Dolding}, {Edvardsson}, {Enke}, {Fabre}, {Fabrizio},
  {Faigler}, {Fedorets}, {Fernique}, {Fienga}, {Figueras}, {Fournier},
  {Fouron}, {Fragkoudi}, {Gai}, {Garcia-Gutierrez}, {Garcia-Reinaldos},
  {Garc{\'\i}a-Torres}, {Garofalo}, {Gavel}, {Gavras}, {Gerlach}, {Geyer},
  {Giacobbe}, {Gilmore}, {Girona}, {Giuffrida}, {Gomel}, {Gomez},
  {Gonz{\'a}lez-N{\'u}{\~n}ez}, {Gonz{\'a}lez-Santamar{\'\i}a},
  {Gonz{\'a}lez-Vidal}, {Granvik}, {Guillout}, {Guiraud},
  {Guti{\'e}rrez-S{\'a}nchez}, {Guy}, {Hatzidimitriou}, {Hauser}, {Haywood},
  {Helmer}, {Helmi}, {Sarmiento}, {Hidalgo}, {Hilger}, {H{\l}adczuk}, {Hobbs},
  {Holland}, {Huckle}, {Jardine}, {Jasniewicz}, {Jean-Antoine Piccolo},
  {Jim{\'e}nez-Arranz}, {Jorissen}, {Juaristi Campillo}, {Julbe}, {Karbevska},
  {Kervella}, {Khanna}, {Kontizas}, {Kordopatis}, {Korn}, {K{\'o}sp{\'a}l},
  {Kostrzewa-Rutkowska}, {Kruszy{\'n}ska}, {Kun}, {Laizeau}, {Lambert},
  {Lanza}, {Lasne}, {Le Campion}, {Lebreton}, {Lebzelter}, {Leccia}, {Leclerc},
  {Lecoeur-Taibi}, {Liao}, {Licata}, {Lindstr{\o}m}, {Lister}, {Livanou},
  {Lobel}, {Lorca}, {Loup}, {Madrero Pardo}, {Magdaleno Romeo}, {Managau},
  {Mann}, {Manteiga}, {Marchant}, {Marconi}, {Marcos}, {Marcos Santos},
  {Mar{\'\i}n Pina}, {Marinoni}, {Marocco}, {Marshall}, {Martin Polo},
  {Mart{\'\i}n-Fleitas}, {Marton}, {Mary}, {Masip}, {Massari},
  {Mastrobuono-Battisti}, {Mazeh}, {McMillan}, {Messina}, {Michalik}, {Millar},
  {Mints}, {Molina}, {Molinaro}, {Moln{\'a}r}, {Monari}, {Mongui{\'o}},
  {Montegriffo}, {Montero}, {Mor}, {Mora}, {Morbidelli}, {Morel}, {Morris},
  {Muraveva}, {Murphy}, {Musella}, {Nagy}, {Noval}, {Oca{\~n}a}, {Ogden},
  {Ordenovic}, {Osinde}, {Pagani}, {Pagano}, {Palaversa}, {Palicio},
  {Pallas-Quintela}, {Panahi}, {Payne-Wardenaar}, {Pe{\~n}alosa Esteller},
  {Penttil{\"a}}, {Pichon}, {Piersimoni}, {Pineau}, {Plachy}, {Plum}, {Poggio},
  {Pr{\v{s}}a}, {Pulone}, {Racero}, {Ragaini}, {Rainer}, {Raiteri}, {Rambaux},
  {Ramos}, {Ramos-Lerate}, {Re Fiorentin}, {Regibo}, {Richards}, {Rios Diaz},
  {Ripepi}, {Riva}, {Rix}, {Rixon}, {Robichon}, {Robin}, {Robin}, {Roelens},
  {Rogues}, {Rohrbasser}, {Romero-G{\'o}mez}, {Rowell}, {Royer}, {Ruz Mieres},
  {Rybicki}, {Sadowski}, {S{\'a}ez N{\'u}{\~n}ez}, {Sagrist{\`a} Sell{\'e}s},
  {Sahlmann}, {Salguero}, {Samaras}, {Sanchez Gimenez}, {Sanna},
  {Santove{\~n}a}, {Sarasso}, {Schultheis}, {Sciacca}, {Segol}, {Segovia},
  {S{\'e}gransan}, {Semeux}, {Shahaf}, {Siddiqui}, {Siebert}, {Siltala},
  {Silvelo}, {Slezak}, {Slezak}, {Smart}, {Snaith}, {Solano}, {Solitro},
  {Souami}, {Souchay}, {Spagna}, {Spina}, {Spoto}, {Steele},
  {Steidelm{\"u}ller}, {Stephenson}, {S{\"u}veges}, {Surdej}, {Szabados},
  {Szegedi-Elek}, {Taris}, {Taylor}, {Teixeira}, {Tolomei}, {Tonello}, {Torra},
  {Torra}, {Torralba Elipe}, {Trabucchi}, {Tsounis}, {Turon}, {Ulla}, {Unger},
  {Vaillant}, {van Dillen}, {van Reeven}, {Vanel}, {Vecchiato}, {Viala},
  {Vicente}, {Voutsinas}, {Weiler}, {Wevers}, {Wyrzykowski}, {Yoldas}, {Yvard},
  {Zhao}, {Zorec}, {Zucker}, \& {Zwitter}}]{2023A&A...674A...1G}
{Gaia Collaboration}, {Vallenari}, A., {Brown}, A.~G.~A., {et~al.} 2023, \aap,
  674, A1, \dodoi{10.1051/0004-6361/202243940}

\bibitem[{{Giuliani}(1982)}]{Giuliani1982}
{Giuliani}, J.~L., J. 1982, \apj, 256, 624, \dodoi{10.1086/159939}

\bibitem[{{Goldsmith} {et~al.}(2008){Goldsmith}, {Heyer}, {Narayanan}, {Snell},
  {Li}, \& {Brunt}}]{2008ApJ...680..428G}
{Goldsmith}, P.~F., {Heyer}, M., {Narayanan}, G., {et~al.} 2008, \apj, 680,
  428, \dodoi{10.1086/587166}

\bibitem[{{Green}(2019)}]{green2019revised}
{Green}, D.~A. 2019, Journal of Astrophysics and Astronomy, 40, 36,
  \dodoi{10.1007/s12036-019-9601-6}

\bibitem[{{Gro{\ss}schedl} {et~al.}(2018){Gro{\ss}schedl}, {Alves}, {Meingast},
  {Ackerl}, {Ascenso}, {Bouy}, {Burkert}, {Forbrich}, {F{\"u}rnkranz},
  {Goodman}, {Hacar}, {Herbst-Kiss}, {Lada}, {Larreina}, {Leschinski},
  {Lombardi}, {Moitinho}, {Mortimer}, \& {Zari}}]{grossschedl20183d}
{Gro{\ss}schedl}, J.~E., {Alves}, J., {Meingast}, S., {et~al.} 2018, \aap, 619,
  A106, \dodoi{10.1051/0004-6361/201833901}

\bibitem[{{Henshaw} {et~al.}(2019){Henshaw}, {Ginsburg}, {Haworth}, {Longmore},
  {Kruijssen}, {Mills}, {Sokolov}, {Walker}, {Barnes}, {Contreras}, {Bally},
  {Battersby}, {Beuther}, {Butterfield}, {Dale}, {Henning}, {Jackson},
  {Kauffmann}, {Pillai}, {Ragan}, {Riener}, \& {Zhang}}]{henshaw2019brick}
{Henshaw}, J.~D., {Ginsburg}, A., {Haworth}, T.~J., {et~al.} 2019, \mnras, 485,
  2457, \dodoi{10.1093/mnras/stz471}

\bibitem[{{Kamezaki} {et~al.}(2014){Kamezaki}, {Imura}, {Omodaka}, {Handa},
  {Tsuboi}, {Nagayama}, {Hirota}, {Sunada}, {Kobayashi}, {Chibueze}, {Kawai},
  \& {Nakano}}]{kamezaki2014annual}
{Kamezaki}, T., {Imura}, K., {Omodaka}, T., {et~al.} 2014, \apjs, 211, 18,
  \dodoi{10.1088/0067-0049/211/2/18}

\bibitem[{{Kauffmann} {et~al.}(2008){Kauffmann}, {Bertoldi}, {Bourke}, {Evans},
  \& {Lee}}]{2008A&A...487..993K}
{Kauffmann}, J., {Bertoldi}, F., {Bourke}, T.~L., {Evans}, N.~J., I., \& {Lee},
  C.~W. 2008, \aap, 487, 993, \dodoi{10.1051/0004-6361:200809481}

\bibitem[{{Kauffmann} {et~al.}(2013){Kauffmann}, {Pillai}, \&
  {Goldsmith}}]{kauffmann2013low}
{Kauffmann}, J., {Pillai}, T., \& {Goldsmith}, P.~F. 2013, \apj, 779, 185,
  \dodoi{10.1088/0004-637X/779/2/185}

\bibitem[{{Kawamura} {et~al.}(1998){Kawamura}, {Onishi}, {Yonekura}, {Dobashi},
  {Mizuno}, {Ogawa}, \& {Fukui}}]{kawamura1998a13co}
{Kawamura}, A., {Onishi}, T., {Yonekura}, Y., {et~al.} 1998, \apjs, 117, 387,
  \dodoi{10.1086/313119}

\bibitem[{{Kounkel} {et~al.}(2017){Kounkel}, {Hartmann}, {Loinard},
  {Ortiz-Le{\'o}n}, {Mioduszewski}, {Rodr{\'\i}guez}, {Dzib}, {Torres}, {Pech},
  {Galli}, {Rivera}, {Boden}, {Evans}, {Brice{\~n}o}, \&
  {Tobin}}]{kounkel2017gould}
{Kounkel}, M., {Hartmann}, L., {Loinard}, L., {et~al.} 2017, \apj, 834, 142,
  \dodoi{10.3847/1538-4357/834/2/142}

\bibitem[{{Kutner} {et~al.}(1979){Kutner}, {Dickman}, {Tucker}, \&
  {Machnik}}]{kutner1979ring}
{Kutner}, M.~L., {Dickman}, R.~L., {Tucker}, K.~D., \& {Machnik}, D.~E. 1979,
  \apj, 232, 724, \dodoi{10.1086/157332}

\bibitem[{{Lada} \& {Fich}(1996)}]{lada1996structure}
{Lada}, C.~J., \& {Fich}, M. 1996, \apj, 459, 638, \dodoi{10.1086/176929}

\bibitem[{{Li} {et~al.}(2018{\natexlab{a}}){Li}, {Wang}, {Zhang}, {Ma}, {Fang},
  \& {Yang}}]{li2018large}
{Li}, C., {Wang}, H., {Zhang}, M., {et~al.} 2018{\natexlab{a}}, \apjs, 238, 10,
  \dodoi{10.3847/1538-4365/aad963}

\bibitem[{{Li} {et~al.}(2018{\natexlab{b}}){Li}, {Li}, {Xu}, {Wang}, {Du},
  {Yang}, \& {Yang}}]{2018ApJS..235...15L}
{Li}, Y., {Li}, F.-C., {Xu}, Y., {et~al.} 2018{\natexlab{b}}, \apjs, 235, 15,
  \dodoi{10.3847/1538-4365/aaab67}

\bibitem[{{Lim} {et~al.}(2022){Lim}, {Naz{\'e}}, {Hong}, {Yoon}, {Lee},
  {Hwang}, {Park}, \& {Lee}}]{lim2022gaia}
{Lim}, B., {Naz{\'e}}, Y., {Hong}, J., {et~al.} 2022, \aj, 163, 266,
  \dodoi{10.3847/1538-3881/ac63b6}

\bibitem[{{Lynds}(1965)}]{lynds1965catalogue}
{Lynds}, B.~T. 1965, \apjs, 12, 163, \dodoi{10.1086/190123}

\bibitem[{{Ma} {et~al.}(2022){Ma}, {Wang}, {Zhang}, {Wang}, {Zhang}, {Liu},
  {Li}, {Zheng}, {Yuan}, \& {Yang}}]{ma2022gas}
{Ma}, Y., {Wang}, H., {Zhang}, M., {et~al.} 2022, \apjs, 262, 16,
  \dodoi{10.3847/1538-4365/ac7797}

\bibitem[{{Magakian} {et~al.}(2022){Magakian}, {Tatarnikov}, {Movsessian}, \&
  {Andreasyan}}]{magakian2022near}
{Magakian}, T.~Y., {Tatarnikov}, A.~M., {Movsessian}, T.~A., \& {Andreasyan},
  H.~R. 2022, \mnras, 510, 2139, \dodoi{10.1093/mnras/stab3585}

\bibitem[{{Margulis} {et~al.}(1988){Margulis}, {Lada}, \&
  {Snell}}]{margulis1988molecular}
{Margulis}, M., {Lada}, C.~J., \& {Snell}, R.~L. 1988, \apj, 333, 316,
  \dodoi{10.1086/166748}

\bibitem[{{Margulis} {et~al.}(1989){Margulis}, {Lada}, \&
  {Young}}]{margulis1989young}
{Margulis}, M., {Lada}, C.~J., \& {Young}, E.~T. 1989, \apj, 345, 906,
  \dodoi{10.1086/167960}

\bibitem[{{McCray} \& {Kafatos}(1987)}]{1987ApJ...317..190M}
{McCray}, R., \& {Kafatos}, M. 1987, \apj, 317, 190, \dodoi{10.1086/165267}

\bibitem[{{Miville-Desch{\^e}nes} {et~al.}(2017){Miville-Desch{\^e}nes},
  {Murray}, \& {Lee}}]{miville2017physical}
{Miville-Desch{\^e}nes}, M.-A., {Murray}, N., \& {Lee}, E.~J. 2017, \apj, 834,
  57, \dodoi{10.3847/1538-4357/834/1/57}

\bibitem[{{Montillaud} {et~al.}(2019){Montillaud}, {Juvela}, {Vastel}, {He},
  {Liu}, {Ristorcelli}, {Eden}, {Kang}, {Kim}, {Koch}, {Lee}, {Rawlings},
  {Saajasto}, {Sanhueza}, {Soam}, {Zahorecz}, {Alina}, {B{\"o}gner}, {Cornu},
  {Doi}, {Malinen}, {Marshall}, {Micelotta}, {Pelkonen}, {Viktor T{\'o}th},
  {Traficante}, \& {Wang}}]{montillaud2019multib}
{Montillaud}, J., {Juvela}, M., {Vastel}, C., {et~al.} 2019, \aap, 631, A3,
  \dodoi{10.1051/0004-6361/201834903}

\bibitem[{{Movsessian} {et~al.}(2021){Movsessian}, {Magakian}, \&
  {Dodonov}}]{movsessian2021new}
{Movsessian}, T.~A., {Magakian}, T.~Y., \& {Dodonov}, S.~N. 2021, \mnras, 500,
  2440, \dodoi{10.1093/mnras/staa3302}

\bibitem[{{Naik} \& {Widmark}(2024)}]{naik2023missing}
{Naik}, A.~P., \& {Widmark}, A. 2024, \mnras, 527, 11559,
  \dodoi{10.1093/mnras/stad3822}

\bibitem[{{Nony} {et~al.}(2021){Nony}, {Robitaille}, {Motte}, {Gonzalez},
  {Joncour}, {Moraux}, {Men'shchikov}, {Didelon}, {Louvet}, {Buckner},
  {Schneider}, {Lumsden}, {Bontemps}, {Pouteau}, {Cunningham}, {Fiorellino},
  {Oudmaijer}, {Andr{\'e}}, \& {Thomasson}}]{2021A&A...645A..94N}
{Nony}, T., {Robitaille}, J.~F., {Motte}, F., {et~al.} 2021, \aap, 645, A94,
  \dodoi{10.1051/0004-6361/202039353}

\bibitem[{{Oliver} {et~al.}(1996){Oliver}, {Masheder}, \&
  {Thaddeus}}]{oliver1996new}
{Oliver}, R.~J., {Masheder}, M.~R.~W., \& {Thaddeus}, P. 1996, \aap, 315, 578

\bibitem[{{Onishi} {et~al.}(1996){Onishi}, {Mizuno}, {Kawamura}, {Ogawa}, \&
  {Fukui}}]{onishi1996ac}
{Onishi}, T., {Mizuno}, A., {Kawamura}, A., {Ogawa}, H., \& {Fukui}, Y. 1996,
  \apj, 465, 815, \dodoi{10.1086/177465}

\bibitem[{{Peek} {et~al.}(2022){Peek}, {Tchernyshyov}, \&
  {Miville-Deschenes}}]{peek2022burton}
{Peek}, J.~E.~G., {Tchernyshyov}, K., \& {Miville-Deschenes}, M.-A. 2022, \apj,
  925, 201, \dodoi{10.3847/1538-4357/ac3f34}

\bibitem[{{Peretto} {et~al.}(2006){Peretto}, {Andr{\'e}}, \&
  {Belloche}}]{peretto2006probing}
{Peretto}, N., {Andr{\'e}}, P., \& {Belloche}, A. 2006, \aap, 445, 979,
  \dodoi{10.1051/0004-6361:20053324}

\bibitem[{{Pineda} {et~al.}(2010){Pineda}, {Goldsmith}, {Chapman}, {Snell},
  {Li}, {Cambr{\'e}sy}, \& {Brunt}}]{pineda2010relation}
{Pineda}, J.~L., {Goldsmith}, P.~F., {Chapman}, N., {et~al.} 2010, \apj, 721,
  686, \dodoi{10.1088/0004-637X/721/1/686}

\bibitem[{{Rapson} {et~al.}(2014){Rapson}, {Pipher}, {Gutermuth}, {Megeath},
  {Allen}, {Myers}, \& {Allen}}]{rapson2014spitzer}
{Rapson}, V.~A., {Pipher}, J.~L., {Gutermuth}, R.~A., {et~al.} 2014, \apj, 794,
  124, \dodoi{10.1088/0004-637X/794/2/124}

\bibitem[{{Reid} {et~al.}(2019){Reid}, {Menten}, {Brunthaler}, {Zheng}, {Dame},
  {Xu}, {Li}, {Sakai}, {Wu}, {Immer}, {Zhang}, {Sanna}, {Moscadelli}, {Rygl},
  {Bartkiewicz}, {Hu}, {Quiroga-Nu{\~n}ez}, \& {van
  Langevelde}}]{reid2019trigonometric}
{Reid}, M.~J., {Menten}, K.~M., {Brunthaler}, A., {et~al.} 2019, \apj, 885,
  131, \dodoi{10.3847/1538-4357/ab4a11}

\bibitem[{{Reipurth} {et~al.}(2004{\natexlab{a}}){Reipurth}, {Pettersson},
  {Armond}, {Bally}, \& {Vaz}}]{2004hreipurthalpha}
{Reipurth}, B., {Pettersson}, B., {Armond}, T., {Bally}, J., \& {Vaz}, L. P.~R.
  2004{\natexlab{a}}, \aj, 127, 1117, \dodoi{10.1086/380935}

\bibitem[{{Reipurth} {et~al.}(2004{\natexlab{b}}){Reipurth}, {Yu},
  {Moriarty-Schieven}, {Bally}, {Aspin}, \& {Heathcote}}]{reipurth2004deep}
{Reipurth}, B., {Yu}, K.~C., {Moriarty-Schieven}, G., {et~al.}
  2004{\natexlab{b}}, \aj, 127, 1069, \dodoi{10.1086/380933}

\bibitem[{{Riener} {et~al.}(2020){Riener}, {Kainulainen}, {Beuther}, {Henshaw},
  {Orkisz}, \& {Wang}}]{riener2020autonomous}
{Riener}, M., {Kainulainen}, J., {Beuther}, H., {et~al.} 2020, \aap, 633, A14,
  \dodoi{10.1051/0004-6361/201936814}

\bibitem[{{Riener} {et~al.}(2019){Riener}, {Kainulainen}, {Henshaw}, {Orkisz},
  {Murray}, \& {Beuther}}]{riener2019gausspy+}
{Riener}, M., {Kainulainen}, J., {Henshaw}, J.~D., {et~al.} 2019, \aap, 628,
  A78, \dodoi{10.1051/0004-6361/201935519}

\bibitem[{{Rosen} {et~al.}(2021){Rosen}, {Offner}, {Foley}, \&
  {Lopez}}]{Rosen2021}
{Rosen}, A.~L., {Offner}, S. S.~R., {Foley}, M.~M., \& {Lopez}, L.~A. 2021,
  arXiv e-prints, arXiv:2107.12397, \dodoi{10.48550/arXiv.2107.12397}

\bibitem[{{Rosolowsky} {et~al.}(2008){Rosolowsky}, {Pineda}, {Kauffmann}, \&
  {Goodman}}]{rosolowsky2008structural}
{Rosolowsky}, E.~W., {Pineda}, J.~E., {Kauffmann}, J., \& {Goodman}, A.~A.
  2008, \apj, 679, 1338, \dodoi{10.1086/587685}

\bibitem[{{Schreyer} {et~al.}(1997){Schreyer}, {Helmich}, {van Dishoeck}, \&
  {Henning}}]{schreyer1997molecular}
{Schreyer}, K., {Helmich}, F.~P., {van Dishoeck}, E.~F., \& {Henning}, T. 1997,
  \aap, 326, 347

\bibitem[{{Shan} {et~al.}(2012){Shan}, {Yang}, {Shi}, {Yao}, {Zuo}, {Lin},
  {Chen}, {Zhang}, {Duan}, {Cao}, {Li}, {Li}, {Liu}, \&
  {Zhong}}]{shan2012development}
{Shan}, W., {Yang}, J., {Shi}, S., {et~al.} 2012, IEEE Transactions on
  Terahertz Science and Technology, 2, 593, \dodoi{10.1109/TTHZ.2012.2213818}

\bibitem[{{Su} {et~al.}(2017){Su}, {Zhou}, {Yang}, {Chen}, {Chen}, {Liu},
  {Wang}, {Li}, \& {Zhang}}]{su2017molecular}
{Su}, Y., {Zhou}, X., {Yang}, J., {et~al.} 2017, \apj, 836, 211,
  \dodoi{10.3847/1538-4357/aa5cb7}

\bibitem[{{Su} {et~al.}(2019){Su}, {Yang}, {Zhang}, {Gong}, {Wang}, {Zhou},
  {Wang}, {Chen}, {Sun}, {Chen}, {Xu}, \& {Jiang}}]{su2019milky}
{Su}, Y., {Yang}, J., {Zhang}, S., {et~al.} 2019, \apjs, 240, 9,
  \dodoi{10.3847/1538-4365/aaf1c8}

\bibitem[{{Su} {et~al.}(2020){Su}, {Yang}, {Yan}, {Gong}, {Chen}, {Zhang},
  {Sun}, {Zhang}, {Chen}, {Zhou}, {Wang}, {Wang}, {Xu}, \&
  {Jiang}}]{su2020local}
{Su}, Y., {Yang}, J., {Yan}, Q.-Z., {et~al.} 2020, \apj, 893, 91,
  \dodoi{10.3847/1538-4357/ab7fff}

\bibitem[{{Sun} {et~al.}(2021){Sun}, {Yang}, {Yan}, {Lin}, {Zhang}, {Su}, {Xu},
  {Chen}, {Wang}, \& {Zhou}}]{2021ApJS..256...32S}
{Sun}, Y., {Yang}, J., {Yan}, Q.-Z., {et~al.} 2021, \apjs, 256, 32,
  \dodoi{10.3847/1538-4365/ac11fe}

\bibitem[{{Sung} {et~al.}(2004){Sung}, {Bessell}, \& {Chun}}]{sung2004initial}
{Sung}, H., {Bessell}, M.~S., \& {Chun}, M.-Y. 2004, \aj, 128, 1684,
  \dodoi{10.1086/423440}

\bibitem[{{Teixeira} {et~al.}(2021){Teixeira}, {Alves}, {Sicilia-Aguilar},
  {Hacar}, \& {Scholz}}]{teixeira2021monoceros}
{Teixeira}, P.~S., {Alves}, J., {Sicilia-Aguilar}, A., {Hacar}, A., \&
  {Scholz}, A. 2021, \mnras, 504, L17, \dodoi{10.1093/mnrasl/slab029}

\bibitem[{{Tobin} {et~al.}(2015){Tobin}, {Hartmann}, {F{\H{u}}r{\'e}sz}, {Hsu},
  \& {Mateo}}]{tobin2015kinematic}
{Tobin}, J.~J., {Hartmann}, L., {F{\H{u}}r{\'e}sz}, G., {Hsu}, W.-H., \&
  {Mateo}, M. 2015, \aj, 149, 119, \dodoi{10.1088/0004-6256/149/4/119}

\bibitem[{{Torii} {et~al.}(2019){Torii}, {Fujita}, {Nishimura}, {Tokuda},
  {Kohno}, {Tachihara}, {Inutsuka}, {Matsuo}, {Kuriki}, {Tsuda}, {Minamidani},
  {Umemoto}, {Kuno}, \& {Miyamoto}}]{torii2019forest}
{Torii}, K., {Fujita}, S., {Nishimura}, A., {et~al.} 2019, \pasj, 71, S2,
  \dodoi{10.1093/pasj/psz033}

\bibitem[{{Venuti} {et~al.}(2019){Venuti}, {Damiani}, \&
  {Prisinzano}}]{venuti2019deep}
{Venuti}, L., {Damiani}, F., \& {Prisinzano}, L. 2019, \aap, 621, A14,
  \dodoi{10.1051/0004-6361/201833253}

\bibitem[{{Wang} {et~al.}(2019){Wang}, {Yang}, {Su}, {Du}, {Ma}, \&
  {Zhang}}]{wang2019molecular}
{Wang}, C., {Yang}, J., {Su}, Y., {et~al.} 2019, \apjs, 243, 25,
  \dodoi{10.3847/1538-4365/ab2d2e}

\bibitem[{{Wang} {et~al.}(2017){Wang}, {Yang}, {Xu}, {Li}, {Su}, \&
  {Zhang}}]{wang2017molecular}
{Wang}, C., {Yang}, J., {Xu}, Y., {et~al.} 2017, \apjs, 230, 5,
  \dodoi{10.3847/1538-4365/aa6c6b}

\bibitem[{{Wang} {et~al.}(2023{\natexlab{a}}){Wang}, {Feng}, {Yang}, {Chen},
  {Su}, {Yan}, {Du}, {Ma}, \& {Cai}}]{wang2023molecular}
{Wang}, C., {Feng}, H., {Yang}, J., {et~al.} 2023{\natexlab{a}}, \aj, 165, 106,
  \dodoi{10.3847/1538-3881/acafee}

\bibitem[{{Wang} {et~al.}(2023{\natexlab{b}}){Wang}, {Feng}, {Yang}, {Chen},
  {Su}, {Yan}, {Du}, {Ma}, \& {Cai}}]{2023AJ....166..121W}
---. 2023{\natexlab{b}}, \aj, 166, 121, \dodoi{10.3847/1538-3881/acebdd}

\bibitem[{{Wang} {et~al.}(2024){Wang}, {Koch}, {Clarke}, {Fuller}, {Peretto},
  {Tang}, {Yen}, {Lai}, {Ohashi}, {Arzoumanian}, {Johnstone}, {Furuya},
  {Inutsuka}, {Lee}, {Ward-Thompson}, {Le Gouellec}, {Liu}, {Fanciullo},
  {Hwang}, {Pattle}, {Poidevin}, {Tahani}, {Onaka}, {Rawlings}, {Chung}, {Liu},
  {Lyo}, {Priestley}, {Hoang}, {Tamura}, {Berry}, {Bastien}, {Ching},
  {Coud{\'e}}, {Kwon}, {Chen}, {Eswaraiah}, {Soam}, {Hasegawa}, {Qiu},
  {Bourke}, {Byun}, {Chen}, {Chen}, {Chen}, {Cho}, {Choi}, {Choi}, {Choi},
  {Chrysostomou}, {Dai}, {Di Francesco}, {Diep}, {Doi}, {Duan}, {Duan}, {Eden},
  {Fiege}, {Fissel}, {Franzmann}, {Friberg}, {Friesen}, {Gledhill}, {Graves},
  {Greaves}, {Griffin}, {Gu}, {Han}, {Hayashi}, {Houde}, {Inoue}, {Iwasaki},
  {Jeong}, {K{\"o}nyves}, {Kang}, {Kang}, {Karoly}, {Kataoka}, {Kawabata},
  {Khan}, {Kim}, {Kim}, {Kim}, {Kim}, {Kim}, {Kim}, {Kim}, {Kirchschlager},
  {Kirk}, {Kobayashi}, {Kusune}, {Kwon}, {Lacaille}, {Law}, {Lee}, {Lee},
  {Lee}, {Lee}, {Li}, {Li}, {Li}, {Li}, {Lin}, {Liu}, {Liu}, {Lu}, {Mairs},
  {Matsumura}, {Matthews}, {Moriarty-Schieven}, {Nagata}, {Nakamura},
  {Nakanishi}, {Ngoc}, {Park}, {Parsons}, {Pyo}, {Qian}, {Rao}, {Rawlings},
  {Retter}, {Richer}, {Rigby}, {Sadavoy}, {Saito}, {Savini}, {Seta}, {Sharma},
  {Shimajiri}, {Shinnaga}, {Tang}, {Thuong}, {Tomisaka}, {Tram}, {Tsukamoto},
  {Viti}, {Wang}, {Whitworth}, {Wu}, {Xie}, {Yang}, {Yoo}, {Yuan}, {Yun},
  {Zenko}, {Zhang}, {Zhang}, {Zhang}, {Zhou}, {Zhu}, {de Looze}, {Andr{\'e}},
  {Dowell}, {Eyres}, {Falle}, {Robitaille}, \& {van Loo}}]{wang2024filamentary}
{Wang}, J.-W., {Koch}, P.~M., {Clarke}, S.~D., {et~al.} 2024, \apj, 962, 136,
  \dodoi{10.3847/1538-4357/ad165b}

\bibitem[{{Wilson} {et~al.}(1970){Wilson}, {Jefferts}, \&
  {Penzias}}]{wilson1970carbon}
{Wilson}, R.~W., {Jefferts}, K.~B., \& {Penzias}, A.~A. 1970, \apjl, 161, L43,
  \dodoi{10.1086/180567}

\bibitem[{{Wilson} \& {Rood}(1994)}]{wilson1994abundances}
{Wilson}, T.~L., \& {Rood}, R. 1994, \araa, 32, 191,
  \dodoi{10.1146/annurev.aa.32.090194.001203}

\bibitem[{{Wolf-Chase} {et~al.}(2003){Wolf-Chase}, {Moriarty-Schieven}, {Fich},
  \& {Barsony}}]{wolf2003star}
{Wolf-Chase}, G., {Moriarty-Schieven}, G., {Fich}, M., \& {Barsony}, M. 2003,
  \mnras, 344, 809, \dodoi{10.1046/j.1365-8711.2003.06863.x}

\bibitem[{{Wright} {et~al.}(2023){Wright}, {Jeffries}, {Jackson}, {Sacco},
  {Arnold}, {Franciosini}, {Gilmore}, {Gonneau}, {Morbidelli}, {Prisinzano},
  {Randich}, \& {Worley}}]{wright2023gaia}
{Wright}, N.~J., {Jeffries}, R.~D., {Jackson}, R.~J., {et~al.} 2023, arXiv
  e-prints, arXiv:2311.08358, \dodoi{10.48550/arXiv.2311.08358}

\bibitem[{{Xu} {et~al.}(2021){Xu}, {Hou}, {Bian}, {Hao}, {Liu}, {Li}, \&
  {Li}}]{xu2021local}
{Xu}, Y., {Hou}, L.~G., {Bian}, S.~B., {et~al.} 2021, \aap, 645, L8,
  \dodoi{10.1051/0004-6361/202040103}

\bibitem[{{Yan} {et~al.}(2019){Yan}, {Yang}, {Sun}, {Su}, \&
  {Xu}}]{yan2019molecular}
{Yan}, Q.-Z., {Yang}, J., {Sun}, Y., {Su}, Y., \& {Xu}, Y. 2019, \apj, 885, 19,
  \dodoi{10.3847/1538-4357/ab458e}

\bibitem[{{Yuan} {et~al.}(2023){Yuan}, {Yang}, {Du}, {Liu}, {Su}, {Yan},
  {Chen}, {Sun}, {Zhang}, {Zhou}, \& {Ma}}]{yuan2023spatial}
{Yuan}, L., {Yang}, J., {Du}, F., {et~al.} 2023, \apj, 944, 91,
  \dodoi{10.3847/1538-4357/acac26}

\bibitem[{{Zhang}(2023)}]{zhang2023distances}
{Zhang}, M. 2023, \apjs, 265, 59, \dodoi{10.3847/1538-4365/acc1e8}

\bibitem[{{Zhang} {et~al.}(2024){Zhang}, {Su}, {Chen}, {Fang}, {Yan}, {Zhang},
  {Sun}, {Wang}, {Feng}, {Ma}, {Zhang}, {Zhuang}, {Zhou}, {Chen}, \&
  {Yang}}]{zhang2024multilayer}
{Zhang}, S., {Su}, Y., {Chen}, X., {et~al.} 2024, arXiv e-prints, accepted by
  AJ, arXiv:2403.00061, \dodoi{10.48550/arXiv.2403.00061}

\bibitem[{{Zucker} {et~al.}(2020){Zucker}, {Speagle}, {Schlafly}, {Green},
  {Finkbeiner}, {Goodman}, \& {Alves}}]{2020A&A...633A..51Z}
{Zucker}, C., {Speagle}, J.~S., {Schlafly}, E.~F., {et~al.} 2020, \aap, 633,
  A51, \dodoi{10.1051/0004-6361/201936145}

\end{thebibliography}
\bibliographystyle{aasjournal}



\end{document}